\documentclass[apj]{emulateapj}
\usepackage{graphicx}

\newcommand{\mum}{\ifmmode{\rm \mu m}\else{$\mu$m}\fi}

\newcommand{\figpath}{./}

\begin{document}

\title{Carbon-rich dust past the asymptotic giant branch:
aliphatics, aromatics, and fullerenes in the Magellanic Clouds}

\author{
G.~C.~Sloan\altaffilmark{1},
E.~Lagadec\altaffilmark{1},
A.~A.~Zijlstra\altaffilmark{2},
K.~E.~Kraemer\altaffilmark{3},
A.~P.~Weis\altaffilmark{4,5},
M.~Matsuura\altaffilmark{6},
K.~Volk\altaffilmark{7},
E.~Peeters\altaffilmark{8,9},
W.~W.~Duley\altaffilmark{10},
J.~Cami\altaffilmark{8,9},
J.~Bernard-Salas\altaffilmark{11},
F.~Kemper\altaffilmark{12},
\& R.~Sahai\altaffilmark{13}
}
\altaffiltext{1}{Center for Radiophysics \& Space Research, Cornell 
  Univ., Ithaca, NY 14853-6801, USA, sloan@isc.astro.cornell.edu}\
\altaffiltext{2}{Jodrell Bank Centre for Astrophysics, Univ.\ of 
  Manchester, Manchester M13 9PL, UK}
\altaffiltext{3}{Institute for Scientific Research, Boston College,
  140 Commonwealth Avenue, Chestnut Hill, MA 02467, USA}
\altaffiltext{4}{Department of Astronomy \& Astrophysics, Columbia
  Univ., 550 W.\ 120th St., New York, NY 10027, USA}
\altaffiltext{5}{Research Experience for Undergraduates, Department of
  Astronomy, Cornell University, Ithaca, NY 14853-6801, USA}
\altaffiltext{6}{Astrophysics Group, Department of Physics \& Astronomy,
  Univ.\ College London, Gower Street, London WC1E 6BT, UK}
\altaffiltext{7}{Space Telescope Science Institute, 3700 San Martin Dr.,
  Baltimore, MD 21218}
\altaffiltext{8}{Department of Physics \& Astronomy, University of
  Western Ontario, London, ON N6A 3K7, Canada}
\altaffiltext{9}{SETI Institute, 189 Bernardo Ave., Suite 100, Mountain
  View, CA 94043}
\altaffiltext{10}{Department of Physics \& Astronomy, Univ.\ of Waterloo,
  200 University Ave. W., Waterloo, ON N2L 3G1, Canada}
\altaffiltext{11}{Department of Physical Sciences,  The Open University,
  Walton Hall, Milton Keynes, MK7 6AA, UK}
\altaffiltext{12}{Academia Sinica, Institute of Astronomy and Astrophysics, 
  11F Astronomy-Mathematics Building, NTU/AS, No. 1, Sec. 4, Roosevelt Rd., 
  Taipei 10617, Taiwan, R.O.C.}
\altaffiltext{13}{Jet Propulsion Laboratory, MS 183-900, California
  Institute of Technology, Pasadena, CA 91109, USA} 



\begin{abstract}
Infrared spectra of carbon-rich objects which have evolved 
off the asymptotic giant branch reveal a range of dust
properties, including fullerenes, polycyclic aromatic 
hydrocarbons (PAHs), aliphatic hydrocarbons, and several
unidentified features, including the 21~\mum\ emission
feature.  To test for the presence of fullerenes, we used
the position and width of the feature at 18.7--18.9~\mum\ 
and examined other features at 17.4 and 6--9~\mum.  This 
method adds three new fullerene sources to the known sample, 
but it also calls into question three previous 
identifications.  We confirm that the strong 11~\mum\ 
features seen in some sources arise primarily from SiC, which 
may exist as a coating around carbonaceous cores and result 
from photo-processing.  Spectra showing the 21~\mum\ feature 
usually show the newly defined Class D PAH profile at 
7--9~\mum.  These spectra exhibit unusual PAH profiles at 
11--14~\mum, with weak contributions at 12.7~\mum, which we 
define as Class D1, or show features shifted to $\sim$11.4, 
12.4, and 13.2~\mum, which we define as Class D2.  Alkyne 
hydrocarbons match the 15.8~\mum\ feature associated with 
21~\mum\ emission.  Sources showing fullerene emission but no 
PAHs have blue colors in the optical, suggesting a clear
line of sight to the central source.  Spectra with 21~\mum\ 
features and Class D2 PAH emission also show photometric 
evidence for a relatively clear line of sight to the central 
source.  The multiple associations of the 21~\mum\ feature 
with aliphatic hydrocarbons suggest that the carrier is 
related to this material in some way.
\end{abstract}


\section{Introduction} 

Amorphous carbon dominates the circumstellar dust in 
carbon-rich stars on the asymptotic giant branch \citep[AGB;
e.g.][]{mr87,gro09}.  By the time an object has evolved to a 
mature planetary nebula (PN), the well-known series of 
infrared (IR) emission features at 3.29, 6.2, 7.7--7.9, 8.6, 
11.3, and 12.7~\mum\ dominate the dust emission.  Polycyclic 
aromatic hydrocarbons (PAHs) are commonly invoked as their
carrier.  

Carbon-rich objects between the AGB and mature PNe can 
display a rich variety of spectra.  One family of intriguing
spectral features is associated with the still unidentified 
21~\mum\ feature discovered by \cite{kwo89} in spectra of 
carbon-rich post-AGB objects.  Related features include those 
at 8, 11, 16, and 26--30~\mum\ \citep[e.g.][]{kra02,vol11}.  

Circumstellar hydrocarbons can appear in even more exotic 
forms, as demonstrated by the discovery of C$_{60}$ and 
C$_{70}$ (referred to here collectively as fullerenes) in the 
spectrum of a Galactic PN \citep{cam10}.  
\cite{gh10} discovered several more PNe showing fullerenes, 
including one in the Small Magellanic Cloud (SMC), and 
\cite{gh11} added 10 more PNe to the list of Magellanic 
fullerene candidates \citep[see][for references to recent 
fullerene detections]{jbs12}.  Previous detections of 
fullerenes in PNe indicate that they only appear when the 
central star is still relatively cool 
\citep[e.g.][]{gh12,ots13}.

The sensitivity of the Infrared Spectrograph \citep[IRS;][]
{hou04} on the {\it Spitzer Space Telescope} \citep{wer04a}
made it possible to obtain rich sets of infrared spectra
in the Magellanic Clouds, where we can overcome the
uncertainty in distances to Galactic sources and obtain
samples less biased by dust extinction in the Galactic plane.
The IRS observed over 40 carbon-rich post-AGB objects in
the Magellanic Clouds.  This paper examines those spectra 
and relates the character of the dust emission to the 
properties of the central star and its circumstellar 
environment.  In such a way we can add to our knowledge of 
the forms carbon-rich dust can take and how those forms 
respond to photo-processing and evolve as they move from 
circumstellar environments into the interstellar medium.  

Our focus is primarily observational and will rely on
classification and methods of analysis that minimize 
assumptions about the spectra.  By analyzing the IRS data and 
the photometric properties of the sources, we can uncover
clues about the nature of carbon-rich circumstellar dust.
Section 2 defines the sample and describes how it was 
observed with the IRS.  In Section 3 we explain how the 
sources are classified in spectroscopic groups and focus
in turn on how different forms of dust contribute to the
spectra.  Section 4 turns to ancillary data including 
optical and infrared photometry and optical spectroscopy.  
Section 5 discusses the accumulated evidence, and Section~6
summarizes the results.


\begin{deluxetable*}{llccrrl}] 
\tablecolumns{7}
\tablewidth{0pt}
\tablenum{1}
\tablecaption{The sample}
\label{t.sample}
\tablehead{
  \colhead{Source\tablenotemark{a}} & \colhead{Alias} & \colhead{RA (J2000)} & \colhead{Dec.\ (J2000)} & 
  \colhead{Program ID} & \colhead{AOR key} & \colhead{Group\tablenotemark{b}}
}
\startdata
SMP SMC 001               & LHA 115-N 1     & 00 23 58.63 & $-$73 38 04.0 &   103 &  4953088 & Mixed \\
IRAS 00350$-$7436         &                 & 00 36 59.58 & $-$74 19 50.3 & 50240 & 27517184 & PAH-like \\
SMP SMC 006               & LHA 115-N 6     & 00 41 27.75 & $-$73 47 06.5 &   103 &  4954112 & PAH-like \\
LIN 49                    &                 & 00 43 53.89 & $-$72 55 14.7 & 50240 & 27537664 & Fullerene \\
2MASS J00444111$-$7321361 &                 & 00 44 41.11 & $-$73 21 36.2 & 50240 & 27525120 & 21~\mum\ (D2)\\ 
SMP SMC 011               & LHA 115-N 29    & 00 48 36.52 & $-$72 58 00.9 &   103 & 15902976 & Red \\          
SMP SMC 013               & LHA 115-N 38    & 00 49 51.64 & $-$73 44 21.5 & 20443 & 14706176 & Mixed \\
SMP SMC 015               & LHA 115-N 43    & 00 51 07.37 & $-$73 57 37.6 & 20443 & 14706688 & Mixed \\
SMP SMC 016               & LHA 115-N 42    & 00 51 27.17 & $-$72 26 11.7 & 20443 & 14706944 & Fullerene \\
SMP SMC 018               & LHA 115-N 47    & 00 51 58.31 & $-$73 20 31.8 & 20443 & 14707456 & Mixed \\
SMP SMC 020               & LHA 115-N 54    & 00 56 05.38 & $-$70 19 25.9 & 20443 & 14707968 & Big-11 \\
SMP SMC 024               & LHA 115-N 70    & 00 59 16.11 & $-$72 01 59.9 &   103 & 15901952 & Fullerene \\
SMP SMC 025               & LIN 357         & 00 59 40.51 & $-$71 38 15.0 & 20443 & 14708480 & Red (SX) \\
2MASS J01054645$-$7147053 &                 & 01 05 46.46 & $-$71 47 05.3 & 50240 & 27518464 & 21~\mum\ (D2)\\ 
SMP SMC 027               & LHA 115-N 87    & 01 21 10.65 & $-$73 14 34.8 & 20443 & 14708992 & Big-11 \\
SMP LMC 002               &                 & 04 40 56.70 & $-$67 48 01.5 &   103 &  4946944 & Fullerene \\
SMP LMC 008               & LHA 120-N 78    & 04 50 13.14 & $-$69 33 56.9 &   103 & 15902464 & Mixed \\
IRAS 05063$-$6908         & MSX LMC 234     & 05 06 03.67 & $-$69 03 58.8 & 30788 & 19006720 & PAH-like \\
SMP LMC 025               & L63 124         & 05 06 23.89 & $-$69 03 19.3 & 20443 & 14701568 & Mixed \\
IRAS 05073$-$6752         & LI-LMC 463      & 05 07 13.93 & $-$67 48 46.7 & 40519 & 24317184 & PAH-like \\     
IRAS 05092$-$7121         & MSX LMC 156     & 05 08 35.92 & $-$71 17 30.7 & 50338 & 25992448 & 21~\mum\ (D1)\\
IRAS 05110$-$6616         & MSX LMC 287     & 05 11 10.65 & $-$66 12 53.7 & 50338 & 25992704 & 21~\mum\ (D2)\\
IRAS 05127$-$6911         & LI-LMC 611      & 05 12 28.21 & $-$69 07 55.7 & 40519 & 24316928 & PAH-like \\     
IRAS 05185$-$6806         & MSX LMC 346     & 05 18 28.15 & $-$68 04 04.0 & 30788 & 19011840 & 21~\mum\ (B)\\
IRAS F05192$-$7008        & MSX LMC 390     & 05 18 45.26 & $-$70 05 34.5 & 40519 & 24314624 & 21~\mum\ (D2)\\
SMP LMC 048               & LHA 120-N 1230  & 05 20 09.49 & $-$69 53 39.1 & 20443 & 14703104 & Mixed \\
2MASS J05204385$-$6923403 &                 & 05 20 43.86 & $-$69 23 40.2 & 50338 & 27985920 & 21~\mum\ (D2)\\
SMP LMC 051               & LHA 120-N 125   & 05 20 52.44 & $-$70 09 35.5 & 40519 & 22421504 & Big-11 \\
SMP LMC 056               & LHA 120-N 136   & 05 23 31.12 & $-$69 04 04.5 & 40519 & 22423808 & Fullerene \\
SMP LMC 058               & LHA 120-N 133   & 05 24 20.75 & $-$70 05 01.5 &   103 &  4950784 & Big-11 \\
IRAS Z05259$-$7052        & SAGE MC J052520 & 05 25 20.77 & $-$70 50 07.5 & 50338 & 25996032 & 21~\mum\ (D2)\\
NGC 1978 WBT 2665         & MSX LMC 580     & 05 29 02.41 & $-$66 15 27.8 &  3591 & 11239680 & 21~\mum\ (D1)\\
IRAS 05315$-$7145         &                 & 05 30 44.14 & $-$71 43 00.6 & 40519 & 22429440 & Red \\
SMP LMC 076               & LHA 120-N 60    & 05 33 56.10 & $-$67 53 09.0 &   103 &  4951296 & Big-11 \\
IRAS 05360$-$7121         & MSX LMC 741     & 05 35 25.85 & $-$71 19 56.6 & 30788 & 19008256 & 21~\mum\ (B)\\
IRAS 05370$-$7019         &                 & 05 36 32.49 & $-$70 17 37.8 & 40519 & 24315648 & Big-11 \\
SMP LMC 085               & LHA 120-N 69    & 05 40 30.80 & $-$66 17 37.4 &   103 &  4952320 & Big-11 \\
IRAS 05413$-$6934         &                 & 05 40 54.29 & $-$69 33 18.6 & 40650 & 23884032 & PAH-like \\
IRAS 05495$-$7034         & LI-LMC 1745     & 05 49 00.01 & $-$70 33 22.5 & 40519 & 22447104 & Red \\
IRAS 05537$-$7015         &                 & 05 53 11.98 & $-$70 15 22.5 & 40519 & 27084032 & Big-11 \\
IRAS 05588$-$6944         & MSX LMC 1601    & 05 58 25.98 & $-$69 44 25.8 & 50338 & 25992960 & Big-11 \\
IRAS 06111$-$7023         & MSX LMC 1710    & 06 10 32.00 & $-$70 24 40.8 & 30788 & 19013120 & 21~\mum\ (D1)\\
SMP LMC 099               & LHA 120-N 221   & 06 18 58.05 & $-$71 35 50.3 & 20443 & 14705664 & Red (PN) 
\enddata
\tablenotetext{a}{We will refer to the targets by these names, shortening the names based on 
2MASS and IRAS coordinates.}
\tablenotetext{b}{Defined in Section~\ref{s.groups}; 
subgroups of the 21~\mum\ group are defined in Section~\ref{s.pahclass}.}
\end{deluxetable*}

\section{The spectral sample} 

\subsection{Selection} 

Our sample is composed of spectra from several {\it Spitzer} 
programs.  We inspected all IRS spectra observed in the
Magellanic Clouds, focused on those showing carbon-rich
properties, and classified them following the criteria and
methods described by \cite{kra02}.  The carbon-rich sample 
was then culled, based on our spectral classifications, so 
that we could focus on the behavior of the dust in those 
carbon-rich sources in evolutionary phases between the AGB 
and mature PNe.  We did not consider the photometry at this 
stage because we wish to consider the photometric properties 
of the sample independently from their spectral properties 
(Sec.~\ref{s.ircolor}). 

The largest of the excluded carbon-rich groups consists of 
nearly 150 carbon stars in the LMC and 40 in the SMC.  These 
stars are on the AGB and have spectral energy distributions 
(SEDs) that peak at wavelengths below $\sim$20~\mum\ and show 
absorption from C$_2$H$_2$ at 7.5 and 13.7~\mum\ and/or 
emission from SiC dust at $\sim$11~\mum.  The brighter stars 
in this group have been the subject of several studies 
\citep[e.g.][]{slo06,zij06,lag07,lei08,gru08}.  The dust in 
their spectra is predominantly amorphous carbon, SiC, and 
MgS (assuming that MgS is the carrier of the 26--30~\mum\ 
feature; see Section~\ref{s.mgs}).

We also exclude the other evolutionary endpoint, mature
PNe, which have spectra dominated by PAHs and forbidden
lines.  Multiple papers have studied the Magellanic PNe
observed by the IRS \cite[e.g.][]{sta07,jbs09}.

R CrB candidates are also not included, because they 
represent an unusual evolutionary path outside our
scope \cite[e.g.][]{cla12}.   Their spectra typically 
resemble a Planck curve of a few hundred K, due to emission 
from warm amorphous carbon dust \citep[see][]{kra05}.  The 
recent work by \cite{gh13} reveals PAH-related structures in
the spectra of several Galactic R CrB stars observed by
the IRS. 

Also excluded are those spectra showing strong hydrocarbon 
absorption features in their spectra \citep{kra06,jbs06}. 
These sources appear to be special cases of edge-on disk 
geometry.

Table~\ref{t.sample} presents the resulting sample.  We added 
one source, the PN SMP LMC~099, which has the spectral
characteristics of a mature PN, but had previously been 
included in sources with fullerenes identified in their 
spectra \citep{gh11}.  The remaining sources should be 
intermediate in nature between the carbon-rich AGB and mature 
PNe, and they show a rich variety of carbon-rich dust 
emission features, including fullerene emission, the 
21~\mum\ features, other related features, and/or unusual 
PAH-like emission features.

The sources in our sample were observed in several IRS 
programs constructed to meet different scientific objectives.
Thus our sample must be affected by multiple biases and 
should not be considered to be complete.  It includes 
at least one source, SMP SMC~025, which turned out to be 
oxygen-rich, despite the presence of an 11~\mum\ feature 
similar to the SiC feature commonly seen in carbon-rich 
objects.  



Where sources have SMP designations, we have adopted them in
Table~\ref{t.sample} to remain consistent with previous 
publications and to indicate that they have been previously 
identified as PNe \citep{san78}.  Otherwise, we adopt the 
historical name for an object, although if a source received 
a coordinate-based name at roughly the same time it received 
a shorter designation, we preferred the latter.  

The coordinates in Table~\ref{t.sample} are based on the
positions of the sources in near-IR surveys.  Appendix~2
provides more detail on how these were determined.

\subsection{Observations and Reduction} \label{s.obsred} 

All observations were obtained by the IRS in standard staring
mode, which involved two pointings in each spectral aperture.
All but one source was observed with the Short-Low (SL) and 
Long-Low (LL) modules.  The exception, IRAS~05413, was 
observed using SL, along with the Short-High (SH) and 
Long-High (LH) modules.

We followed the standard data reduction scheme as described
in more detail by \cite{slo14}, starting with data produced
by the S18.18 version of the pipeline at the {\it Spitzer}
Science Center.  For SL, we used images with the source in 
one aperture as the background for when the source is in
the other aperture (i.e., SL order 1 $-$ SL order 2 and vice 
versa).  For LL, we removed the background by differencing
images with the source in the same aperture, but the other
nod position.  For the source observed in SH and LH, 
we subtracted a dedicated background image obtained in a 
nearby field.  Images were cleaned for bad pixels using the 
{\sc irsclean} package in IDL.

We deviated from previous reduction schemes by extracting
the spectra using the optimal extraction algorithm described
by \cite{leb10} for all spectra except IRAS~05413, which was
extracted using the older additive algorithm (also known as 
tapered-column extraction), due to the observational design
with SH and LH.  We resampled this spectrum to match the 
spectral resolution of the other observations in our sample.

Spectra from separate nods were compared to remove spikes and 
divots and then combined.  To remove discontinuities between
spectral segments, we normalized them upward to the segment
which was presumably best centered in the spectrographic
slits.

Appendix~1 describes our method of spectral analysis 
and includes tables presenting extracted strengths
and positions of the various spectral features considered in
the following sections.

\section{Spectral properties} \label{s.properties} 

\subsection{Polycyclic aromatic hydrocarbons (PAHs)} \label{s.pahs} 

\begin{figure} 
\includegraphics[width=3.4in]{\figpath 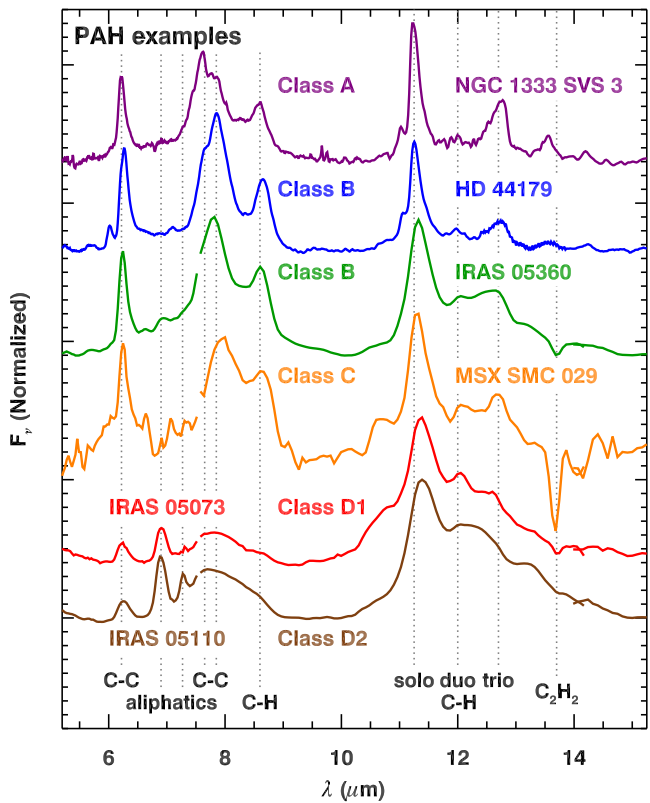}
\caption{Examples of spectra showing different classes of PAH
emission.  The top two spectra are based on {\it ISO}/SWS
spectra, the rest are IRS data.  Of the IRS sources, all but
MSX SMC 029 are in our sample.  Classes D1 and D2 are defined
in Section~\ref{s.pahclass}.  The vertical dotted lines
mark the wavelengths of aliphatic modes at 6.90 and 
7.27~\mum, aromatic modes at other wavelengths, and the
location of the Q branch from molecular
C$_2$H$_2$ absorption at 13.7~\mum.\label{f.pah}}
\end{figure}

When a warm radiation field illuminates carbon-rich dust,
the resulting spectrum shows emission from aromatic 
hydrocarbons, which are usually described as PAHs.  We will 
adopt this nomenclature, but it is important to keep in mind
that the particles responsible for the observed emission
features are not necessarily the free and primarily aromatic 
molecules usually assumed \citep[e.g.][]{atb89}.  An
alternative perspective is that the aromatic bonds responsible
for the well-known IR emission features exist in an amorphous 
mixture with a range of structures from aliphatic to
aromatic \citep[for recent work, see][]{kwo11,jon12a,jon12c}.

In aliphatics, all electrons are localized in specific bonds, 
while in aromatics they are not.  In aromatics, the energy
of an absorbed photon is quickly distributed over all 
available bonds, making the PAH structure much more resistant 
to photodissociation in a harsh radiative environment.  The 
electron cloud also protects aromatic structures from 
collisions with electrons, with greater protection for 
larger molecules \citep{mic10a,mic10b,boc12}.  Classic PAH 
emission arises from material which is dominated by aromatic 
bonds.

Figure~\ref{f.pah} 
presents spectra of the different classes of PAH emission.  
The top two spectra in the figure are from the 
Short-Wavelength Spectrometer (SWS) on the {\it Infrared 
Space Observatory} ({\it ISO}), as calibrated by \cite{slo03} 
and resampled to the IRS wavelength grid.  The remainder are 
IRS data, and they are from the current sample, with the 
exception of MSX~SMC~029, which was discussed by 
\cite{kra05}.  Its spectrum in Figure~\ref{f.pah} is 
processed identically to the spectra in the current sample.

In the PAH model, C--C modes produce the features observed
at 6.25, 7.65, and 7.85~\mum.   The 8.60~\mum\ feature 
arises from an in-plane C--H bending mode, while the 
out-of-plane C--H bending modes produce features at 11.25, 
12.0, and 12.7~\mum, with the wavelength determined by the
number of adjacent C--H bonds on the same aromatic ring.
In wavelength order, these are the solo, duo, and trio
modes \citep{atb89}.

\cite{pee02} grouped PAH spectra into Classes A, B, and C, 
based primarily on the position of the 6.2~\mum\ feature
and the shape of the 7.7--7.9~\mum\ complex, with the peaks 
shifting from blue to red in the sequence A--B--C.  They
noted that the different features do not always shift in
unison.  We will concentrate on the 7.7--7.9~\mum\ region, 
which generally consists of two components with peaks 
$\sim$7.65 and 7.85~\mum.  The former dominates in Class A 
spectra, while the latter dominates in Class B.  \cite{pee02} 
found that most, but not all, PNe showed Class B PAH emission,
while Class A PAH emission sources were more likely to be
associated with young stellar objects.  

For Class C PAH emission, the peak is further to the red.  In
the {\it ISO}/SWS sample examined by \cite{pee02}, only two 
Class C sources were present, and both were post-AGB objects.  
Observations with the IRS added several more Class C sources, 
many of which are not post-AGB objects \citep{slo07,ack10,
smo10,eva10}.  Generally Class C PAH spectra have a central 
wavelength beyond a limit of $\sim$7.9--8.0~\mum\ 
\citep{kel08}.  The PAH features at 6.2 and 11.3~\mum\ also 
shift to longer wavelengths in this class.

Laboratory data show that these wavelengths shifts can arise
from hydrocarbon mixtures with higher ratios of aliphatics to 
aromatics \citep{pin08}.  As this ratio increases, the 
mixture becomes better described as hydrogenated amorphous 
carbon \citep[HAC; e.g.][]{dw81,jon90}.  Class C PAHs appear 
to be a special case seen only when hydrocarbon material has 
not been as heavily photo-processed as the more typical Class 
A or B PAHs, either because the radiation field is less 
harsh, or because they have been processed for shorter 
periods of time \citep[see][for an example of the latter 
case]{eva10}.  

\cite{mat14} added a new class of PAH emission seen in some
post-AGB spectra, Class D, based primarily on the unusual 
shape to the emission in the 6--9~\mum\ region \citep[see 
also][]{mat11}.  While typical PAH spectra show emission 
peaks in the 7.5--8~\mum\ region and at 8.6~\mum, with a
definite minimum between them, in Class D PAH spectra, the
region from 8 to 8.5~\mum\ is filled in with additional
emission.  Class D differs from Class C, which shows a
clear peak to the red of 8~\mum.  In 
Section~\ref{s.pahclass}, we show that Class D can be
broken into two separate classes.

In addition to the features from aromatic hydrocarbons, 
aliphatic features can also appear in our spectra, most
notably at 6.90 and 7.27~\mum\ (these wavelengths are 
determined below in Section~\ref{s.spec69}).  Methyl 
(CH$_3$) groups produce features from asymmetric and 
symmetric C--H bending modes at $\sim$6.8 and $\sim$7.2~\mum,
respectively.  Methylene (CH$_2$) groups can contribute at
$\sim$6.8~\mum\ with a C--H scissoring mode. 
Additional methylene C--H wagging and twisting modes 
appear at 7.6--7.7~\mum, but these are buried by the strong 
C--C PAH modes at these wavelengths.  


\subsection{Post-AGB spectral groups} \label{s.groups} 

\begin{figure} 
\includegraphics[width=3.4in]{\figpath 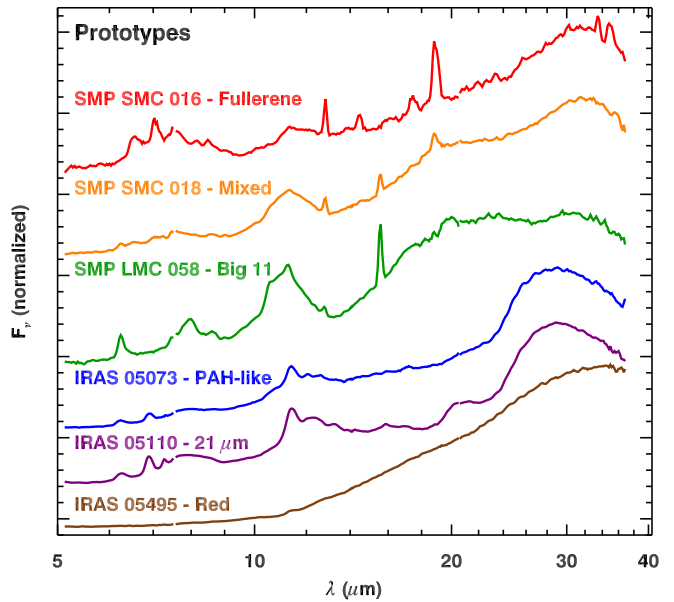}
\caption{Examples of the spectral groups into which we have
divided the sample of carbon-rich Magellanic post-AGB 
spectra.\label{f.proto}}
\end{figure}

\begin{figure} 
\includegraphics[width=3.4in]{\figpath 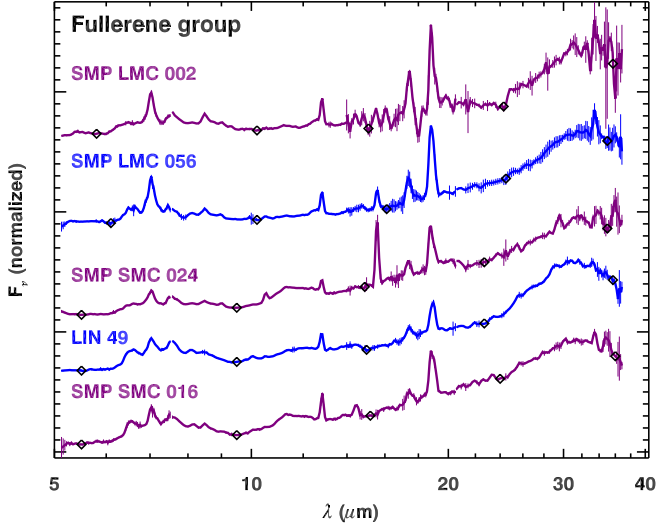}
\caption{The fullerene spectra.  The black diamonds mark
the wavelengths through which splines were fit.
\label{f.spfull}}
\end{figure}

\begin{figure} 
\includegraphics[width=3.4in]{\figpath 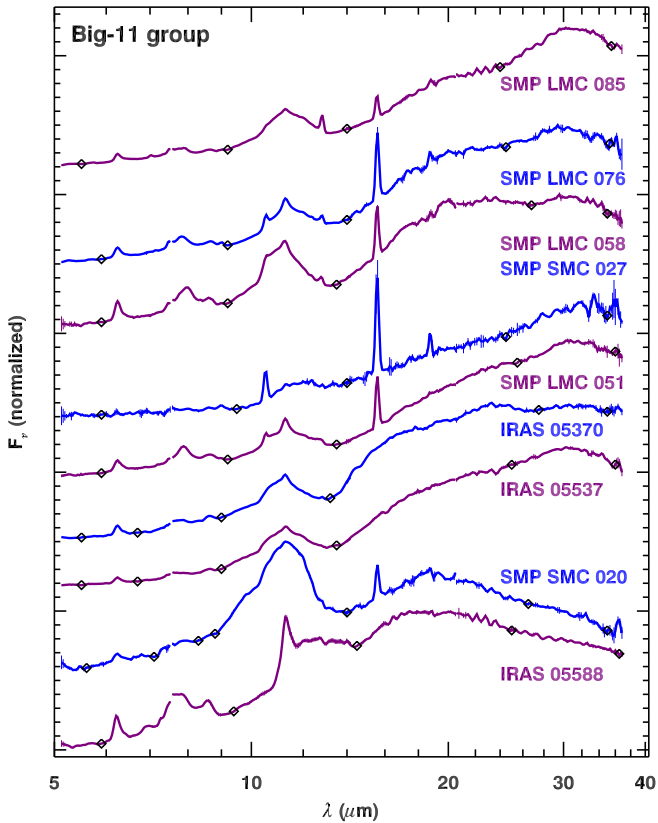}
\caption{The spectra showing the big-11 feature at 
$\sim$11.3~\mum\ and other associated features, including
emission from PAHs at 6--9~\mum, a shoulder at 18~\mum,
and often strong a strong emission at $\sim$30~\mum.
The spline points appear as black diamonds.\label{f.spbig11}}
\end{figure}

\begin{figure} 
\includegraphics[width=3.4in]{\figpath 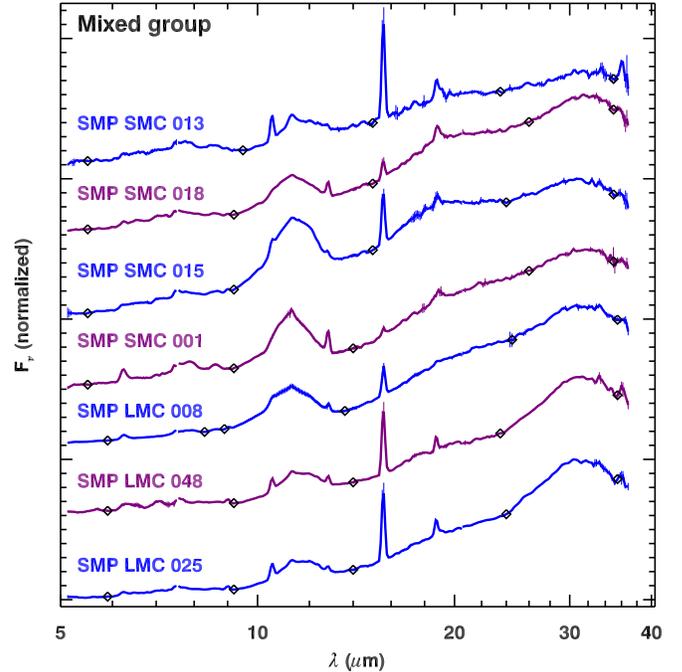}
\caption{The spectra showing a mixture of fullerene features
and features associated with the big-11 feature.  The 
boundaries between these groups are unavoidably arbitrary.
The black diamonds mark the spline points.
\label{f.spmixed}}
\end{figure}

\begin{figure} 
\includegraphics[width=3.4in]{\figpath 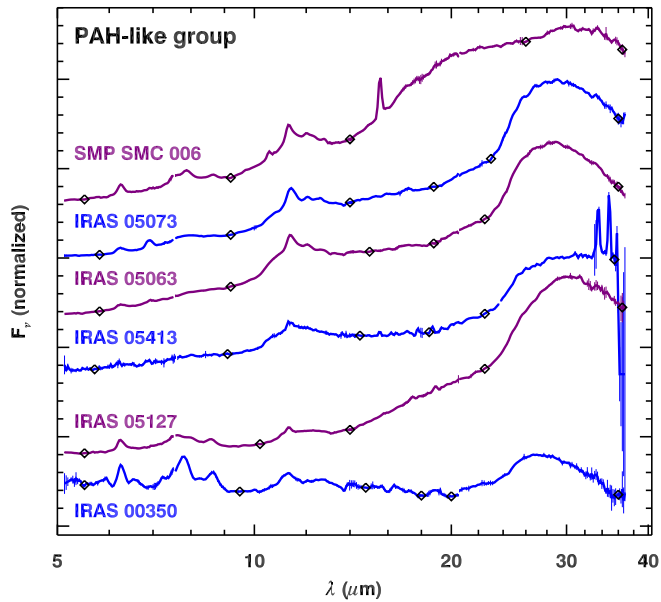}
\caption{Spectra showing PAH emission at 6--9~\mum\ and a
PAH-like spectrum at 11--14~\mum.  The black diamonds
mark the spline points.\label{f.sppah}}
\end{figure}

\begin{figure} 
\includegraphics[width=3.4in]{\figpath 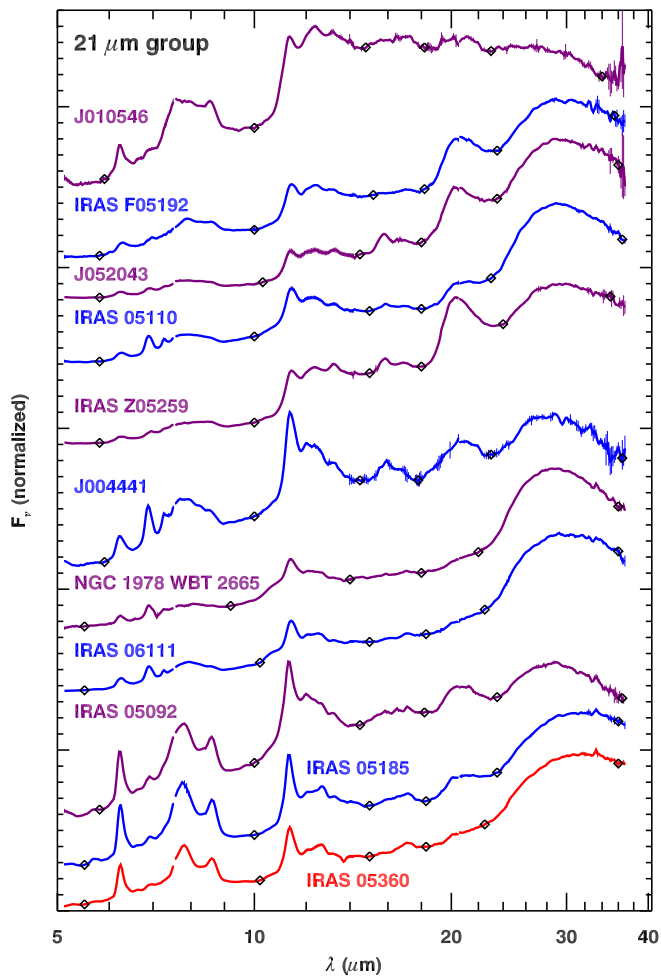}
\caption{Spectra showing the 21~\mum\ feature and associated 
features at 8, 11, 16, and 30~\mum.  Black diamonds mark the 
spline points. IRAS~05360 is plotted in red because it is
the comparison Class B PAH spectrum in several of the
following figures.\label{f.sp21}}
\end{figure}

\begin{figure} 
\includegraphics[width=3.4in]{\figpath 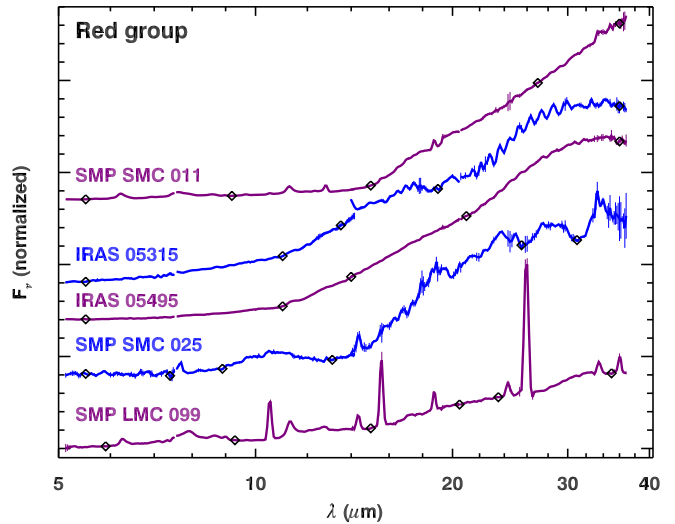}
\caption{The reddest spectra in our sample.  Two show little 
recognizable spectral structure, while one shows normal PAH 
emission and one shows crystalline silicate emission.  The 
fifth spectrum is a source with fullerenes identified in its 
spectrum by others but not confirmed here.  The spline points 
are marked with black diamonds.\label{f.spred}}
\end{figure}

We have assigned each spectrum to one of several groups
based on the criteria described below and explained more 
fully in the following sections.  Figure~\ref{f.proto} 
presents the prototypes for the groups, and 
Figures~\ref{f.spfull} to \ref{f.spred} plot the individual
spectra in each group.  We will capitalize the names of the 
spectral groups to avoid confusion with the use of the same 
words to describe individual spectral features.\footnote{For 
example, ``big-11'' refers to the spectral feature, while 
``Big-11'' refers to the spectra in the group dominated by 
this feature.}

The {\bf Fullerene group} shows emission features at 17.4 and
18.9~\mum.  These spectra have a distinctive structure in the 
6--9~\mum\ region which includes the previously identified 
fullerene features at 7.03 and 8.50~\mum\ \citep{cam10} and 
additional features which differ from the usual 6--9~\mum\ 
PAH emission profile.  
LIN~49 is a new identification of a fullerene source.

The {\bf Big-11 group} shows a strong emission feature at
$\sim$11~\mum, which we refer to throughout this paper as
the ``big-11'' feature.  Spectra in this group also show
PAH-like emission features, but no clear evidence of 
fullerenes.  They usually also show PAH features in the 
6--9~\mum\ region and a shoulder at 18~\mum\ from an 
unidentified carrier.  The key discriminants for this group 
are (1) a lack of fullerene emission and (2) {\it either} a 
clear big-11 feature {\it or} a combination of a more 
ambiguous big-11 feature and the 18~\mum\ shoulder.

The {\bf Mixed group} consists of positively identified 
fullerene sources that show additional structure in the 
6--9~\mum\ region usually associated with PAHs.  These 
spectra show clear distinctions from the Fullerene and Big-11 
groups to either side, but all three groups represent a 
continuum and where one draws the line between them is 
somewhat arbitrary.  These spectra typically show the big-11 
feature and the associated shoulder at 18~\mum\ (as do some 
of the Fullerene spectra), but the clear presence of 
fullerenes distinguishes them.  Two sources, SMP LMC~008 and 
SMP SMC~001, are new fullerene identifications.

The {\bf PAH-like group} shows typical PAH emission features 
in the 6--9~\mum\ region and PAH-like structure in the 
11--14~\mum\ region.  Our classification as PAH-like is based 
primarily on the unusual structure at 11--14~\mum, with the 
relative strengths of the 12.0 and 12.7~\mum\ features 
reversed, producing a triangular shape for the 11--14~\mum\
complex.  The PAH-like spectra also tend to show a broader
blue wing to the 11.3~\mum\ feature.


The {\bf 21~\mum\ group} typically shows the family of
associated emission features at 8, 11, 16, 21, and 
26--30~\mum, although not all spectra show all of the
features.  \cite{vol11} analyzed many of the same spectra 
discussed here.  Their classification of what belongs in the 
21~\mum\ group and ours do not coincide exactly, due to a 
combination of the improved optimal spectral extraction used 
here and different constraints on the classification.  In 
some of our spectra, the 21~\mum\ feature is ambiguous, but 
other related features are present (most notably NGC~1978 
WBT~2665 and IRAS~0611).  Table~\ref{t.sample}
includes subclassifications for the 21~\mum\ group based on 
the nature of the PAH-like emission at 11--14~\mum; these are 
defined in Section~\ref{s.pahclass} below.

The {\bf Red group} is the final group.  The only requirement 
is that the spectrum show a red continuum.  Most show only
low-contrast features, but this group also serves as the 
depository for spectra which cannot be placed in other 
categories.  One source, SMP SMC~025, shows crystalline
silicate emission.  SMP LMC~099 is a fullerene candidate
\citep{gh11}, although we do not find fullerenes in its 
spectrum.

\subsection{Separating the Fullerene, Mixed, and Big-11 groups} 

The C$_{60}$ molecule produces four strong emission features, 
at 7.0, 8.5, 17.4 and 18.9~\mum\ \citep{cam10}.  The spectra 
containing fullerenes and the big-11 feature represent a 
continuum, with many spectra exhibiting characteristics of 
both, making it necessary to develop a straightforward means 
of recognizing when fullerenes are present.  Appendix~1 
describes in detail our analysis of the spectral features.  
In this section, we focus on the results.

\subsubsection{Fullerene at 17--19~\mum} 
\label{s.full18}

\begin{figure} 
\includegraphics[width=3.4in]{\figpath 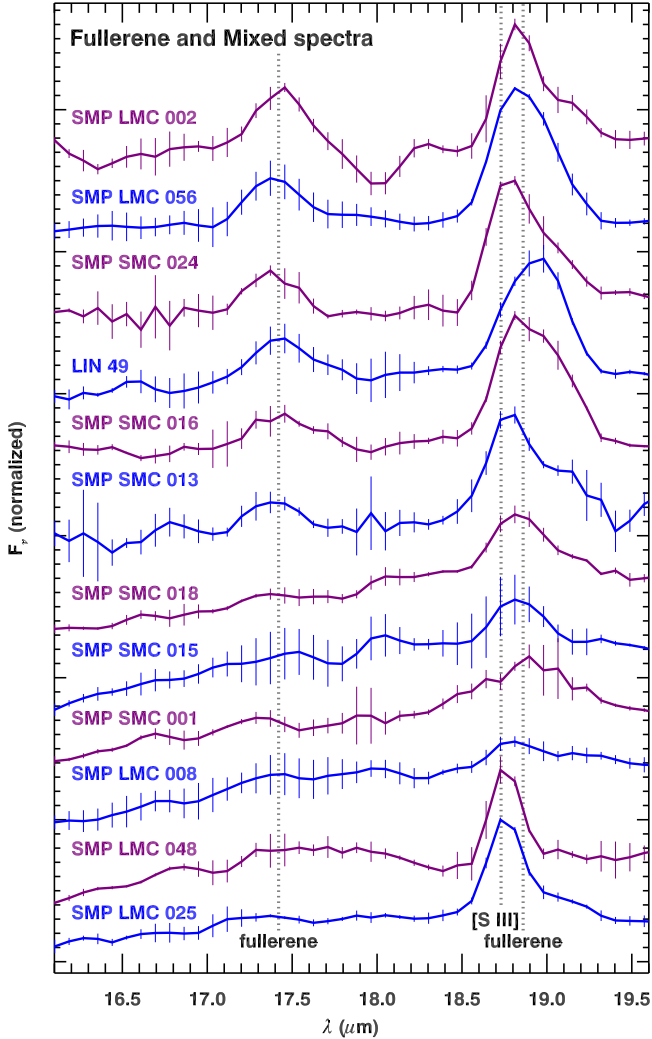}
\caption{The 16--20~\mum\ region of the spectra showing 
evidence for fullerene features.  The vertical dotted lines 
are at 17.41 and 18.88, the nominal positions of the 
fullerene features in our data, and 18.71~\mum, the nominal 
position of the forbidden lines from [S III].  These spectra
have had splines fitted and removed.
\label{f.sp19a}}
\end{figure}

\begin{figure} 
\includegraphics[width=3.4in]{\figpath 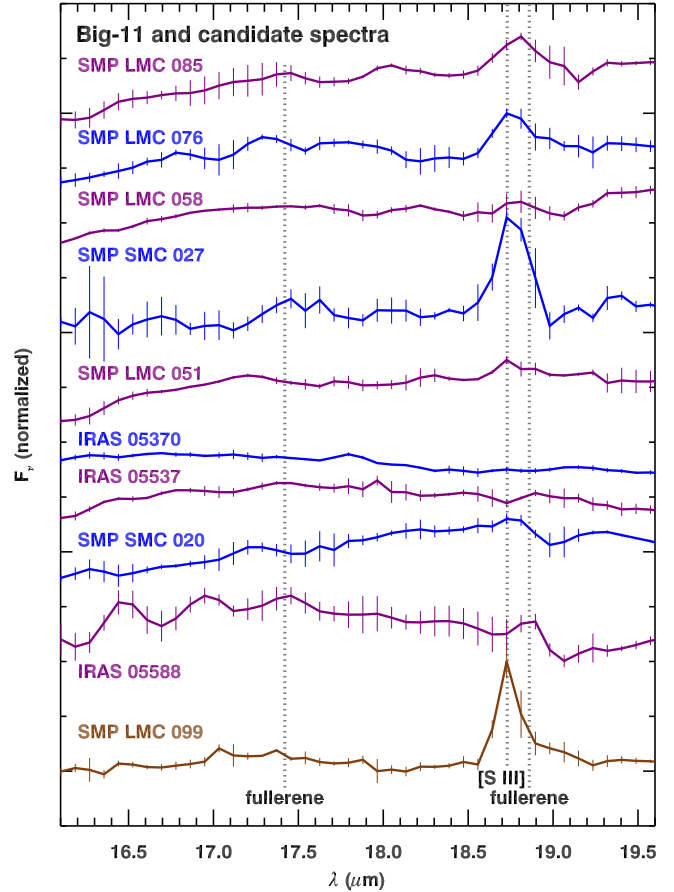}
\caption{The 16--20~\mum\ region of the spectra in the Big-11 
group, along with SMP LMC~099, which shows [S III] but no 
fullerenes.  The vertical dotted lines are as defined for 
Fig.~\ref{f.sp19a}.  These spectra have had splines fitted 
and removed.
\label{f.sp19b}}
\end{figure}

\begin{figure} 
\includegraphics[width=3.4in]{\figpath 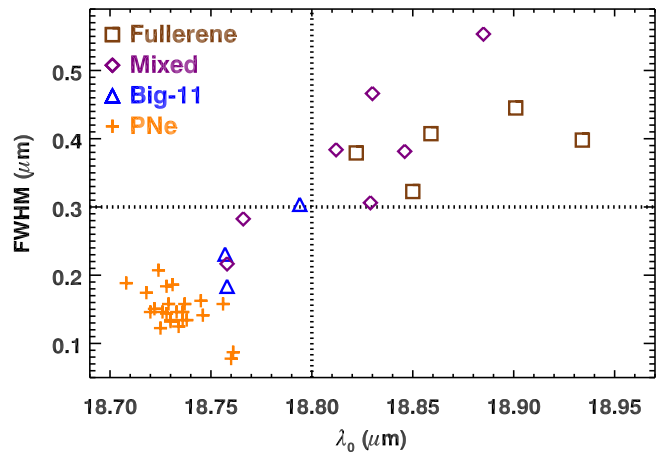}
\caption{Width of the 18.7--18.9~\mum\ feature vs.\ 
wavelength for the Fullerene, Mixed, and Big-11 spectra,
as well as a control sample of planetary nebulae.  The data
plotted are based on fitted gaussians and differ from the
data in Tables~\ref{t.fulldat} and \ref{t.siiidat}.  Data
in the lower left quadrant are from [S III] lines, while the
upper right arises from fullerenes.\label{f.pos19}}
\end{figure}

Figures~\ref{f.sp19a} and \ref{f.sp19b} present the
(spline-removed) 17--19~\mum\ spectra of sources in the
Fullerene, Mixed, and Big-11 groups, as well as the fullerene 
candidate SMP LMC~099.  This region includes the fullerene 
features at 17.4 and 18.9~\mum.  Many of these spectra have
the [S III] line at 18.71~\mum, which, at the low resolution
of LL2 ($\lambda$/$\Delta$$\lambda$ = 130 at 19~\mum), is 
blended with the 18.9~\mum\ fullerene feature.  The 
17.4~\mum\ feature is weaker, and for spectra with low 
signal/noise ratios (SNRs), the 18.9~\mum\ feature is the 
better test, if we can address the blending problem.  

Appendix~1.1 describes our analysis of this spectral 
region.  Fitting gaussians to the 18.8~\mum\ blend separates 
those spectra dominated by [S III] from those dominated by 
fullerenes, as Figure~\ref{f.pos19} shows.  The widths and 
positions fall into two quadrants.  For most cases, it is not 
necessary to rely on data from SH or other higher-resolution 
spectrometers.  All of the spectra with $\lambda_0 > $ 
18.80~\mum\ have a FWHM $>$ 0.3~\mum, and in each case, we 
positively identify fullerenes in the spectrum.  The spectra 
with $\lambda_0 < $ 18.80~\mum\ have FWHM $<$ 0.3~\mum.  
Spectra in this quadrant include some Mixed and Big-11 
spectra, but all of these are shifted up and to the right of 
the PNe, indicating that some fullerenes may be blended with 
the [S III] line.

The weaker 17.4~\mum\ feature can help validate the
presence of fullerenes in all of the sources in the Fullerene
group.  PAHs can also produce a feature at 17.4~\mum\ 
\citep{wer04b}, but it is usually accompanied by a stronger
PAH feature at 16.4~\mum\ \citep{sel10} which is not apparent
in Figures~\ref{f.sp19a} and \ref{f.sp19b}.  A greater 
problem is the general weakness of the 17.4~\mum\ feature; it 
is detected unambiguously in only one third of the Mixed and 
Big-11 spectra.  Fortunately, the 6--9~\mum\ region can 
clarify the nature of the spectra.

\subsubsection{Fullerene and PAHs in the 6--9~\mum\ region} 
\label{s.spec69}

\begin{figure} 
\includegraphics[width=3.4in]{\figpath 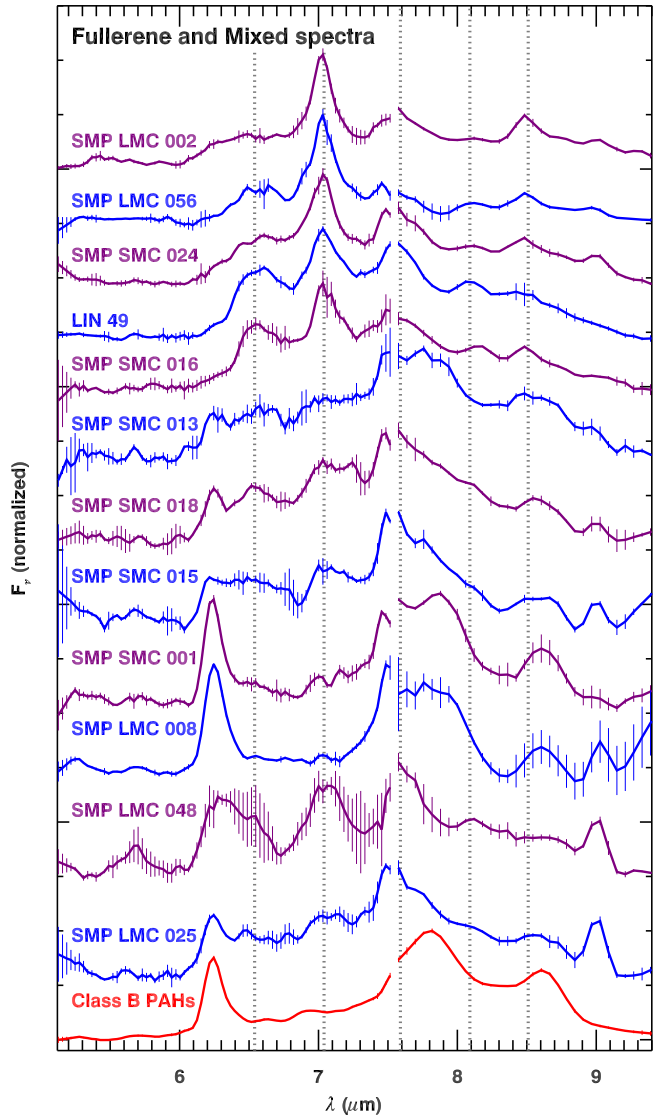}
\caption{The 6--9~\mum\ emission complex in the spectra 
showing contributions from fullerene features.  The first
five spectra are in the Fullerene group, and SMP SMC~013
is the first of the Mixed group.  The bottom comparison 
spectrum of Class B PAHs is the source IRAS~05360.  The 
vertical dotted lines are at 6.54, 7.04, 7.59, 8.09, and 
8.51~\mum, marking features associated with fullerenes.  All 
spectra have had splines fitted and removed.\label{f.sp69a}}
\end{figure}

\begin{figure} 
\includegraphics[width=3.4in]{\figpath 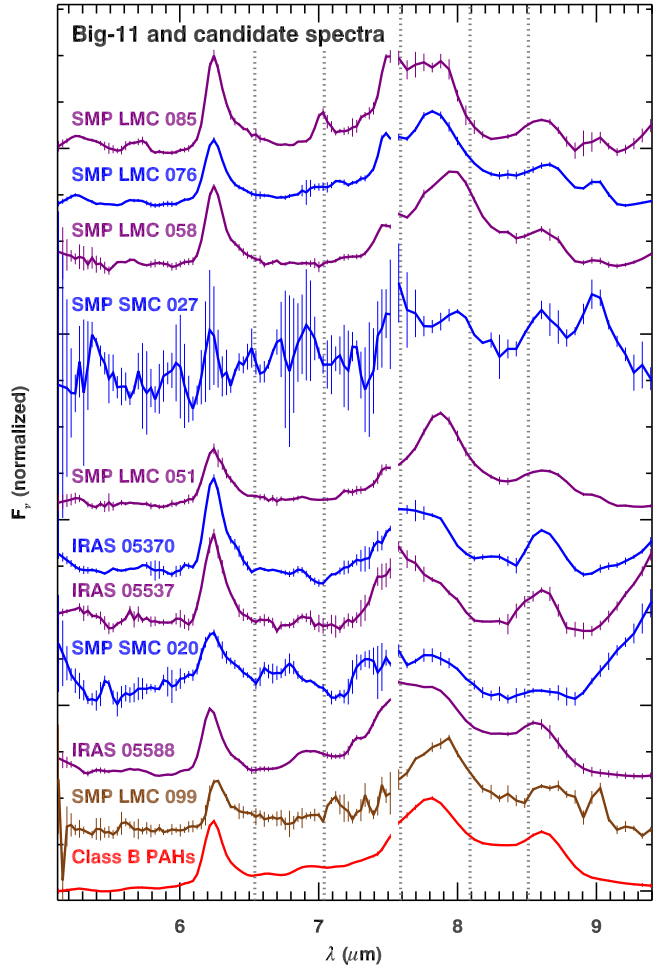}
\caption{The 6--9~\mum\ emission complex of the Big-11 
spectra and SMP LMC~099.  The comparison spectrum and the 
vertical dotted lines are as defined for Fig.~\ref{f.sp69a}.  
The spectra have had splines fitted and removed.  Some of the
spectra are climbing at the red edge due to the big-11 
feature.\label{f.sp69b}}
\end{figure}

Figure~\ref{f.sp69a} presents the 6--9~\mum\ spectra of the 
fullerene candidates identified in the previous section.  
This spectral region allows an independent assessment of the 
fullerene properties.  \cite{cam10} identified two features 
from C$_{60}$ in this wavelength region, at 7.0 and 8.5~\mum.  
Unfortunately, the former overlaps the [Ar II] line at 
6.99~\mum, and the latter is close to the in-plane C--H 
bending mode from PAHs at 8.6~\mum.  Our spectra also include 
features centered at 6.5 and 7.6~\mum, and in some cases, a 
weaker feature $\sim$8.1~\mum.

\cite{ber13} recently identified a series of related features
in spectra from the reflection nebula NGC~7023 at 6.4, 7.1, 
and 8.2~\mum, which they attributed to fullerene cations 
(C$_{60}^+$).  The cation features at 6.4 and 8.2~\mum\ are 
close to the 6.5 and 8.1~\mum\ features seen in our 
Magellanic sample.  The 7.1~\mum\ cation feature, if it 
appears in our spectra, is blended with the 7.0~\mum\ feature 
from neutral fullerenes (and with [Ar II]).

We also see a feature at 7.6~\mum\ in the fullerene spectra,
but it requires more caution because it is near the complex 
of hydrogen recombination lines which includes Pfund~$\alpha$ 
at 7.46~\mum\ and the Humphreys~$\beta$ and 11--7 transitions 
at 7.50~\mum.  These sources can show forbidden-line emission 
from the ionized region, as the presence of the [Ar III] line 
at 8.99~\mum\ in some spectra attests, and it follows that 
hydrogen emission might also be present.  Nonetheless, we 
suspect that the 7.6~\mum\ feature may be related to the 
6.5~\mum\ feature, because both appear together in spectra 
where [Ar III] and other emission lines are weak.  

The first five spectra in Figure~\ref{f.sp69a} show 
fullerenes, but no PAH emission.  The remaining spectra show 
some evidence of PAH emission, as do all of the spectra in
Figure~\ref{f.sp69b}, which includes the Big-11 spectra and
the fullerene candidate SMP LMC~099.  To aid our efforts to 
disentangle PAHs and fullerenes and characterize the features 
observed, we have extracted central wavelengths and 
integrated feature strengths as described in Appendix~1.2.

\begin{deluxetable}{lc} 
\tablecolumns{2}
\tablewidth{0pt}
\tablenum{2}
\tablecaption{Central wavelengths of 6--9~\mum\ features}
\label{t.lamc}
\tablehead{
  \colhead{Feature} & \colhead{$<$$\lambda_C$$>$ (\mum)}
}
\startdata
6.2~\mum\ PAH               &  6.25 $\pm$ 0.02 \\
6.9~\mum\ aliphatic         &  6.90 $\pm$ 0.03 \\
7.3~\mum\ aliphatic         &  7.27 $\pm$ 0.02 \\
8.6~\mum\ PAH               &  8.60 $\pm$ 0.02 \\
7--9 \mum\ base PAH feature &  7.78 $\pm$ 0.12 \\
\\
6.5~\mum\ fullerene-related &  6.54 $\pm$ 0.01 \\
7.0~\mum\ fullerene         &  7.04 $\pm$ 0.01 \\
7.6~\mum\ fullerene-related &  7.59 $\pm$ 0.02 \\
8.1~\mum\ fullerene-related &  8.10 $\pm$ 0.02 \\
8.5~\mum\ fullerene         &  8.51 $\pm$ 0.01 
\enddata
\end{deluxetable}

Table~\ref{t.lamc} gives the mean central wavelengths 
($\lambda_C$) for the various fullerene-related, PAH, and 
aliphatic features, based on the spectra which showed a 
given feature at a strength of 2.5~$\sigma$ or more.  For the 
PAH features, we excluded the Fullerene spectra, but 
considered all other detections.  For the aliphatic features, 
we considered the detections listed in Appendix~1.2.

Three of the fullerene-related features are more easily
separable from other features.  We detected the 6.5~\mum\ 
feature unambiguously in only five spectra:  four of the 
Fullerene group and one of the Mixed group.  Nine spectra 
contain an emission feature at 7.0~\mum\ that could arise 
from fullerenes or [Ar II]:  five of the 
Fullerene group, and four of the Mixed.  We only detect the 
8.1~\mum\ feature at 2.5~$\sigma$ or better in three spectra.  
The fullerene-related features at 7.6 and 8.5~\mum\ require 
some care, due to overlap with PAH features.

Most of the spectra show a tentative detection of a feature 
at 7.6~\mum, but the majority are really PAHs.  We find that 
four of the Fullerene spectra, five of the Mixed spectra, 
and even two of the Big-11 spectra contain a 7.6~\mum\ 
feature with consistent central wavelengths.  These spectra 
give $<\lambda_C>$ = 7.59 $\pm$ 0.02~\mum.  A central 
wavelength of 7.62~\mum\ appears to be the boundary between a 
fullerene-related and a PAH feature.  A central wavelength of 
7.59~\mum\ is 0.14~\mum\ from the position of the 
Pfund-$\alpha$ line, giving us some confidence that we are 
observing a distinct feature.

The 8.5~\mum\ fullerene feature can be blended with the 
8.6~\mum\ in-plane C--H bending mode from PAHs, but the
central wavelength can distinguish them.  The five 
Fullerene spectra with a feature centered at $\sim$8.5~\mum\ 
give $<\lambda_C>$ = 8.51 $\pm$ 0.01~\mum, while the 25 
other spectra with a feature in this vicinity have 
$<\lambda_C>$ = 8.61 $\pm$ 0.02~\mum.  The latter feature
is from PAHs.


In our spectra, the 7.0~\mum\ fullerene feature appears at 
7.04~\mum, compared to 6.99~\mum\ for [Ar II].  Most of the
spectra with a 7.0~\mum\ feature also have [Ne II] lines at
12.8~\mum, indicating that some contamination from [Ar II]
is likely, but the shift away from 6.99~\mum\ shows that 
fullerenes must be contributing.  The shift in fullerene
emission from 7.0~\mum\ to a longer wavelength could be a
temperature effect.  From 0 K to 1000 K, the C$_{60}$ 
feature moves from 6.97 to 7.11~\mum, and the C$_{70}$ 
feature moves from 6.98 to 7.08~\mum\ \citep{nem94}.  The 
observed shift might also arise from the presence of some 
fullerene cations.  However, the central wavelengths of the 
other putative cation features do not align particularly 
well with what \cite{ber13} reported for them:  6.54 vs.\ 
6.40~\mum\ and 8.10 vs.\ 8.23~\mum.  

\subsubsection{Fullerene classification} 

Sources in both the Fullerene and Mixed groups should be
treated as positive fullerene identifications.  Three of
these identifications are new:  LIN~49 in the Fullerene
group and SMP LMC~008 and SMP SMC~001 in the Mixed group.

The first five spectra in Figure~\ref{f.sp69a} show 
unambiguous fullerene features and no PAH emission and 
therefore belong to the Fullerene group.  The remaining 
spectra in the figure show a mixture of fullerene-related 
and PAH features which puts them in the Mixed group.  SMP 
SMC~013 is the transition between the two groups.  It
clearly has fullerene features at 17.4 and 18.9~\mum, but
the 6--9~\mum\ structure is more ambiguous, with definite 
PAH features at 6.2~\mum\ and $\sim$7.8~\mum\ alongside 
fullerene-related features at 6.5 and 7.0~\mum.

The bottom spectrum (of IRAS 05360) in Figure~\ref{f.sp69a} 
is an example of a Class B PAH spectrum typical for more 
evolved PNe and a good comparison for the features visible in 
the Mixed spectra.  All of the Mixed spectra show 
fullerene-like structure at 7.0 and 7.6~\mum, and all but 
SMP SMC~001 and SMP LMC~008 have a 6.5~\mum\ feature.  While 
the individual features are hard to discern in SMP LMC~025, 
the spectrum shows a clear excess between the 6.2 and 
7.7--7.9~\mum\ PAH features compared to a typical PAH 
spectrum.

Figure~\ref{f.sp69b} plots spectra from the Big-11 group as
well as SMP LMC~099.  None show convincing evidence of 
fullerenes.  The classification of the top spectrum, SMP 
LMC~085, is arguable, as it does show some evidence of 
features at 7.0 and 7.6~\mum.  However, it is lacking the 
plateau from 6.2 to 7.6~\mum\ and a clear detection of 
fullerenes at 17.4 or 18.9~\mum.  The spectrum of SMP LMC~076 
might show a 7.6~\mum\ feature, but it shows none of the 
other fullerene features expected in this wavelength region.  
The line at $\sim$18.8~\mum\ looks more like [S III] than 
fullerenes.  

We are unable to confirm previous identifications of 
fullerenes in the spectra of SMP SMC~020, 027, and SMP 
LMC~099 \citep{gh11}.  The structure in the 6--9~\mum\ region 
is not consistent with fullerenes, with SMP SMC~020 showing
spectral structure close to PAHs.  SMP SMC~020 also shows no 
clear feature at either 17.4 or 18.9~\mum.  SMP SMC~027 has 
what appears to be a fullerene feature at 17.4~\mum, but the 
structure at $\sim$18.8~\mum\ matches [S III], not 
fullerenes.  In SMP LMC~099, the 6--9~\mum\ spectrum looks
like Class B PAHs, the feature at $\sim$18.8~\mum\ is clearly 
[S III], and no recognizable feature appears at 17.4~\mum.

\subsubsection{The nature of the big-11 feature} 
\label{s.big11nature}

\begin{figure} 
\includegraphics[width=3.4in]{\figpath 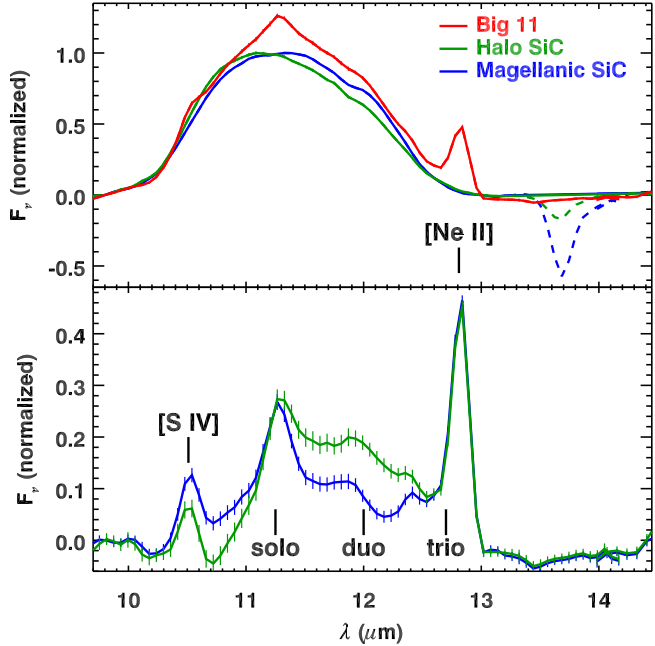}
\caption{The average big-11 feature compared to SiC profiles
from Magellanic carbon stars and carbon stars in the Galactic
Halo (top panel).  The dashed lines show the comparison 
profiles before removing the C$_2$H$_2$ absorption band at 
13.7~\mum\ by interpolating a line from the data to either
side.  The bottom panel shows the residual spectra after 
subtracting the two comparison profiles from the big-11 
feature.\label{f.sic}}
\end{figure}

The big-11 feature dominates the spectra not just of the
Big-11 group, but also the Mixed group.  It is clearly 
present in some of the Fullerene spectra as well.  This 
feature is usually accompanied by a broad feature at 
$\sim$18~\mum, which in the redder spectra appears as an 
inflection or shoulder on a continuum rising to longer 
wavelengths.  Many of these spectra also show a feature in 
the 26--30~\mum\ vicinity.

\cite{jbs09} attributed the big-11 feature primarily to SiC.
We investigated by comparing the big-11 feature to two 
coadded SiC profiles from separate control groups.  The first 
group consists of 26 spectra of carbon stars from the LMC and 
five from the SMC, all with good SNRs and well-defined SiC 
features.\footnote{The sample includes four spectra published 
by \cite{slo06}, 12 by \cite{zij06}, four by \cite{buc06}, 
one each by \cite{lag07} and \cite{lei08}, and nine 
unpublished spectra from program 30788 (P.I.\ R.\ Sahai).}  
The second group includes six carbon stars in the Galactic 
Halo.\footnote{Two in the line of sight toward the Sgr Dwarf 
Spheroidal \citep{lag09} and four in the nearby Halo 
\citep{lag12}.}  All of these spectra were observed and 
reduced as described in Section~\ref{s.obsred}, except that 
the brightness of the four Halo spectra made optimal 
extraction unnecessary.

To generate an average big-11 profile, we combined six 
spectra: SMP SMC~001, 015, 018, and 020, along with SMP 
LMC~008 and 085.  These sources include the four Mixed 
spectra with the least contamination from [S IV] at 
10.52~\mum, and the two spectra from the Big-11 group with 
the least structure from [S IV] and PAHs.  For the average 
big-11 profile and the two SiC profiles, a continuum was 
spline-fitted and removed to isolate the profiles, which were 
then normalized and averaged.  Both of the comparison SiC 
profiles also show absorption at 13.7~\mum\ from C$_2$H$_2$, 
which we removed from the data by replacing the affected 
wavelengths with a line interpolated from data to either 
side.  Our post-AGB spectra tend to show weak C$_2$H$_2$ 
absorption or none at all.

Figure~\ref{f.sic} compares the three profiles (top panel).  
We have normalized the big-11 profile to 1.0 at 10.95~\mum\ 
to align with the other two.  Both of the comparison SiC
profiles have a FWHM of 0.66~\mum.  The central wavelength is 
11.29~\mum\ for the Magellanic profile and 11.22~\mum\ for 
the Halo profile.  The differences in the shape of the
profile between the two are consistent with the wide range
of observed positions and profiles in published samples
\citep[e.g.][]{spe05,slo06}.

Figure~\ref{f.sic} also plots the residuals after subtracting
the two comparison SiC profiles from the average big-11 
feature (bottom panel).  The residual spectrum is a 
combination of [S IV] and [Ne II] lines at 10.52 and 
12.81~\mum, respectively, and PAH-like emission in between, 
with features or structure apparent at $\sim$11.3, 
$\sim$11.9, and$\sim$12.4~\mum.  The first two are close to 
the positions of the solo and duo out-of-plane C--H bending 
modes.  The position of the feature $\sim$11.9~\mum\
depends on the assumed underlying shape of the SiC feature,
and it is reasonable to assume that this is in fact the 
12.0~\mum\ duo mode.  The third is in the vicinity of the 
H$_2$ 0--0 S(2) transition at 12.28~\mum, but H$_2$ emission
does not appear at other wavelengths in these spectra and can
be ruled out.  A more likely culprit is the 
Humphreys~$\alpha$ transition at 12.37~\mum.  The presence of 
a strong [Ne II] line prevents us from making any statements 
about the trio mode at 12.7~\mum, but the duo mode appears to 
be unusually strong compared to the solo mode, making this 
residual emission look much like the PAH emission in the 
PAH-like group.  

If we ignore the [S IV] and [Ne II] lines (by replacing the
affected data as we did with the C$_2$H$_2$ line), then SiC 
emission contributes 88\% of the total big-11 feature.  This 
analysis confirms the description by \cite{jbs09} of the 
feature as primarily from SiC, and it contradicts a recent 
claim to the contrary by \cite{slo12}, who based their
preliminary conclusion on a comparison of too few spectra.  
We will address abundance issues raised by such a strong SiC 
feature in Section~\ref{s.coatings} below.

\subsection{PAH classification} 
\label{s.pahclass}

\begin{deluxetable}{llcc} 
\tablecolumns{4}
\tablewidth{0pt}
\tablenum{3}
\tablecaption{PAH classifications}
\label{t.pahclass}
\tablehead{
  \colhead{Target} & \colhead{Group} & \colhead{6--9~\mum} & 
  \colhead{11--14~\mum\tablenotemark{a}}
}
\startdata
SMP SMC 024       & Fullerene & \nodata & (D1) \\
SMP SMC 016       & Fullerene & \nodata & (D1) \\
\\
SMP SMC 013       & Mixed     & \nodata & (D1) \\
SMP SMC 018       & Mixed     & \nodata & (D1) \\
SMP SMC 015       & Mixed     & \nodata & (D1) \\
SMP SMC 001       & Mixed     & \nodata & (D1) \\
SMP SMC 008       & Mixed     & \nodata & (D1) \\
SMP SMC 048       & Mixed     & \nodata & (D1) \\
SMP SMC 025       & Mixed     & \nodata & (D1) \\
\\
SMP LMC 085       & Big-11    & A:      & (D1) \\
SMP LMC 076       & Big-11    & B       & (D1) \\
SMP LMC 058       & Big-11    & B       & (D1) \\
SMP SMC 027       & Big-11    & ?       & (D1) \\
SMP LMC 051       & Big-11    & B       & (D1) \\
IRAS 05370        & Big-11    & A       & (D1) \\
IRAS 05537        & Big-11    & A       & (D1) \\
SMP SMC 020       & Big-11    & B       & (D1) \\
IRAS 05588        & Big-11    & A       &  A   \\
\\
SMP SMC 006       & PAH-like  & B       &  D1 \\
IRAS 05073        & PAH-like  & D       &  D1 \\
IRAS 05063        & PAH-like  & D       &  D1 \\
IRAS 05413        & PAH-like  & ?       &  D1 \\
IRAS 05127        & PAH-like  & A       &  D1 \\
IRAS 00350        & PAH-like  & B       &  D1 \\
\\
J010546           & 21~\mum   & D       &  D2 \\
IRAS F05192       & 21~\mum   & D       &  D2 \\
J052043           & 21~\mum   & D       &  D2 \\
IRAS 05110        & 21~\mum   & D       &  D2 \\
IRAS Z05259       & 21~\mum   & D       &  D2 \\
J004441           & 21~\mum   & D       &  D2 \\
NGC 1978 WBT 2665 & 21~\mum   & D       &  D1 \\
IRAS 06111        & 21~\mum   & D       &  D1 \\
IRAS 05092        & 21~\mum   & B       &  D1 \\
IRAS 05185        & 21~\mum   & B       &  B  \\
IRAS 05360        & 21~\mum   & B       &  B  \\
\\
SMP LMC 099       & Red       & B       &  B  \\
SMP SMC 011       & Red       & A       &  A 
\enddata
\tablenotetext{a}{Classifications in parentheses are residuals
of the big-11 feature and assumed to be D1.}
\end{deluxetable}

To classify the spectra in our sample into the PAH classes
described in Section~\ref{s.pahs}, we have examined both the 
6--9 and 11--14~\mum\ regions.  Table~\ref{t.pahclass} 
presents the results for both wavelength regimes.  Where the 
emission at 11--14~\mum\ is not the new Classes D1 or D2
(defined below in Section~\ref{s.pah11-14}), we have 
assigned a class consistent with the shape at 6--9~\mum, 
because Class A and B PAH emission are indistinguishable at 
11--14~\mum.  The following two sections describe the 
treatment of the two wavelength regimes.

\subsubsection{Spectral properties at 6--9~\mum} 
\label{s.pah6-9}

\begin{figure} 
\includegraphics[width=3.4in]{\figpath 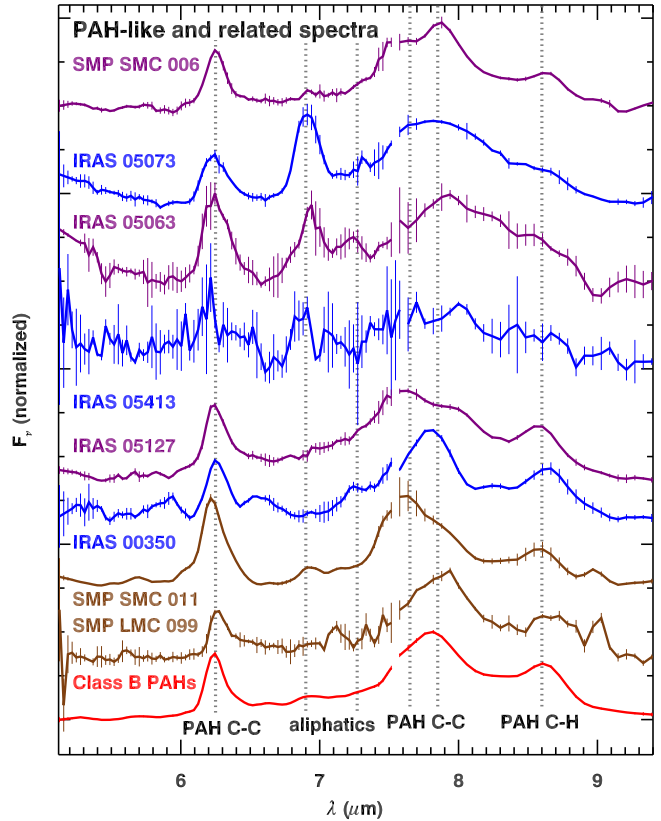}
\caption{The 6--9~\mum\ emission complex in the six spectra 
showing PAH-like features, two of the red spectra, and a PAH 
comparison spectrum (IRAS~05360).  The vertical dotted lines
mark the PAH and aliphatic features at 6.25, 6.90, 7.27, 
7.65, 7.85, and 8.60~\mum.  All spectra have had splines 
fitted and removed.\label{f.sp69c}}
\end{figure}

\begin{figure} 
\includegraphics[width=3.4in]{\figpath 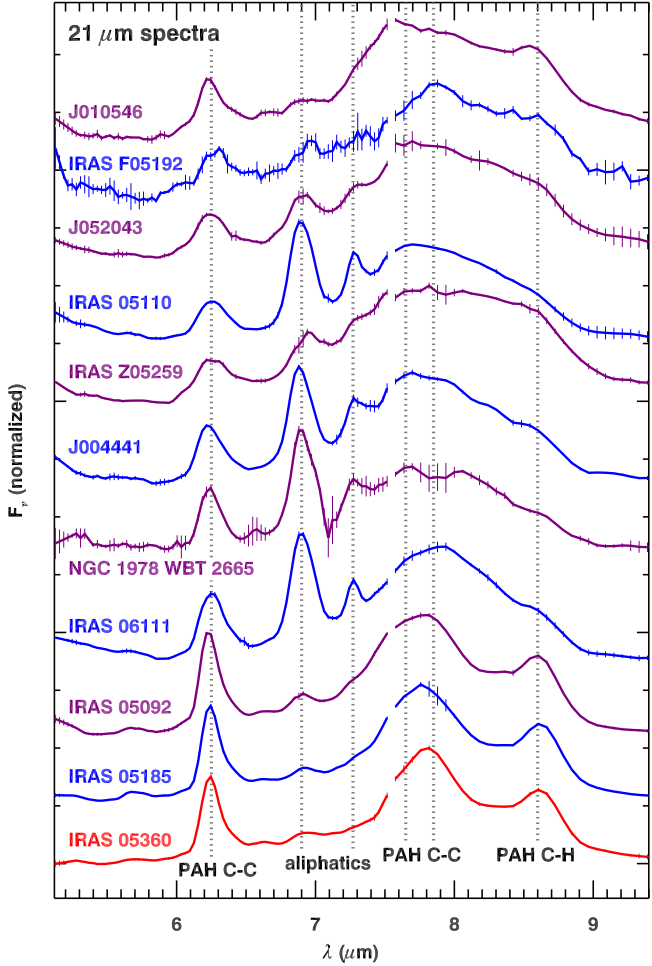}
\caption{The 6--9~\mum\ region of the spectra with 21~\mum\
and related emission features.  The vertical dotted lines 
and the comparison spectrum are as defined in 
Figure~\ref{f.sp69c}.\label{f.sp69d}}
\end{figure}

Variations in the spectra in the 6--9~\mum\ region can
distinguish the different classes of PAH emission.  Class A 
PAHs are stonger at 7.65~\mum, Class B PAHs are stronger at 
7.85~\mum, and Class C PAHs have feature with a central 
wavelength to the red of $\sim$8.0~\mum.  Only two spectra in 
our sample are clearly past that limit, both Red spectra with 
low-contrast spectral structure that may not be PAH emission.  
A third source, SMP LMC~058, is at the limit, and it 
resembles more the transitional ``B/C'' class defined by 
\cite{kel08} and seen in many Herbig AeBe stars.  It is 
interesting that even though the present sample is composed 
of carbon-rich post-AGB objects, and the two original 
Class C PAH spectra were both carbon-rich post-AGB objects 
\citep{pee02}, none of the spectra in our sample are good 
examples of Class C PAHs.

We excluded the Fullerene and Mixed groups from the PAH
classification, because the presence of the fullerene-related 
feature at 7.6~\mum\ can mimic the central wavelength of the
Class A spectrum (see Figure~\ref{f.sp69a}).  While most of
the spectra in the Big-11 group appear to be Class A, we 
cannot rule out low levels of fullerenes (see 
Figure~\ref{f.sp69b}).  Only three of the Big-11 spectra are 
unambiguously Class A.

Figure~\ref{f.sp69c} presents the 6--9~\mum\ region of the
PAH-like spectra, along with three other spectra.  Two
spectra are missing the usual minimum between the 7.85 and 
8.6~\mum\ features and therefore belong to Class D as defined 
by \cite{mat14}.  These two spectra also have the strongest
aliphatic features at 6.9~\mum\ in this group.

Figure~\ref{f.sp69d} shows the 21~\mum\ sources in the 
6--9~\mum\ region.  All except the bottom three show 
the Class D PAH profile.  The remaining spectra are 
consistent with Class B PAH emission.  All of the spectra 
with a strong aliphatic feature at 6.9~\mum\ are associated 
with Class D profiles at 6--9~\mum, although the inverse is 
not true.  All of the 21~\mum\ spectra show some aliphatic
6.9~\mum\ emission, even the Class B spectra.

The aliphatic features can appear in spectra with seeming 
disregard for the class of the PAH emission.  For example, 
the 6.9~\mum\ feature is clearly present (though weak) in 
the spectrum of SMP SMC~011, a Class A source 
(Figure~\ref{f.sp69c}).  It also appears in the spectrum of 
IRAS~05588, a Big-11 source, and possibly in SMP SMC~020 as 
well (Figure~\ref{f.sp69b}).  

\subsubsection{Spectral properties at 11--14~\mum} 
\label{s.pah11-14}

\begin{figure} 
\includegraphics[width=3.4in]{\figpath 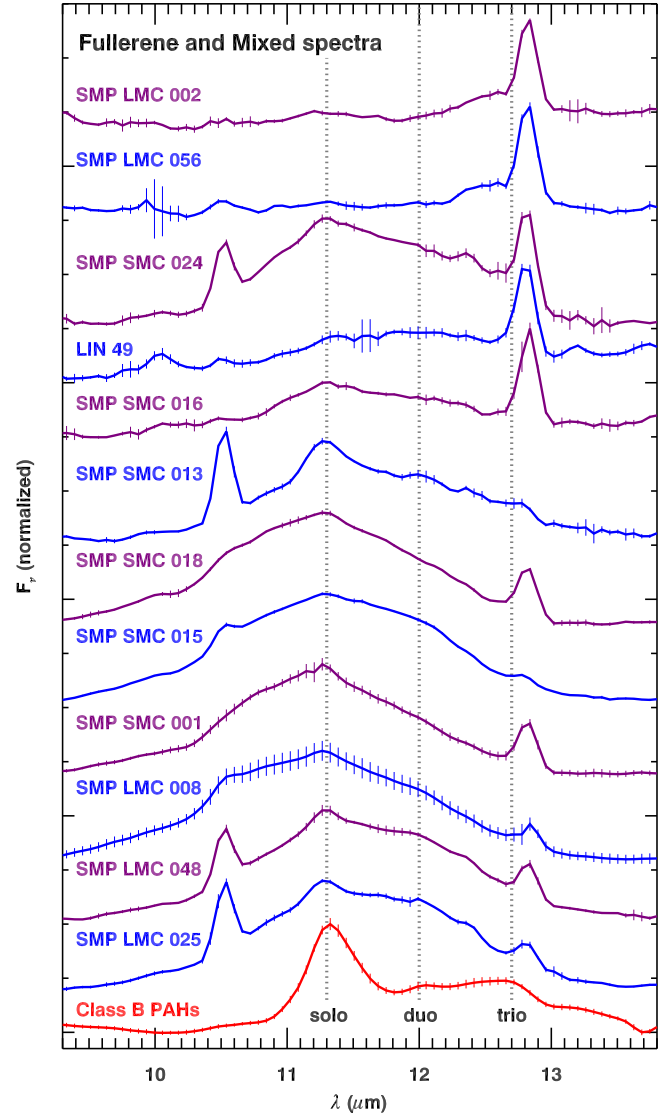}
\caption{The 11--14~\mum\ region of the spectra classified
as Fullerene or Mixed.  The vertical dotted lines mark the
nominal positions of the PAH modes from solo, duo, and trio
out-of-plane C--H bending at 11.3, 12.0, and 12.7~\mum,
respectively.  The comparison Class B PAH spectrum is of 
IRAS~05360.  Some of the spectra have emission lines from 
[S IV] and [Ne II] at 10.52 and 12.81~\mum, respectively.
\label{f.sp12a}}
\end{figure}

\begin{figure} 
\includegraphics[width=3.4in]{\figpath 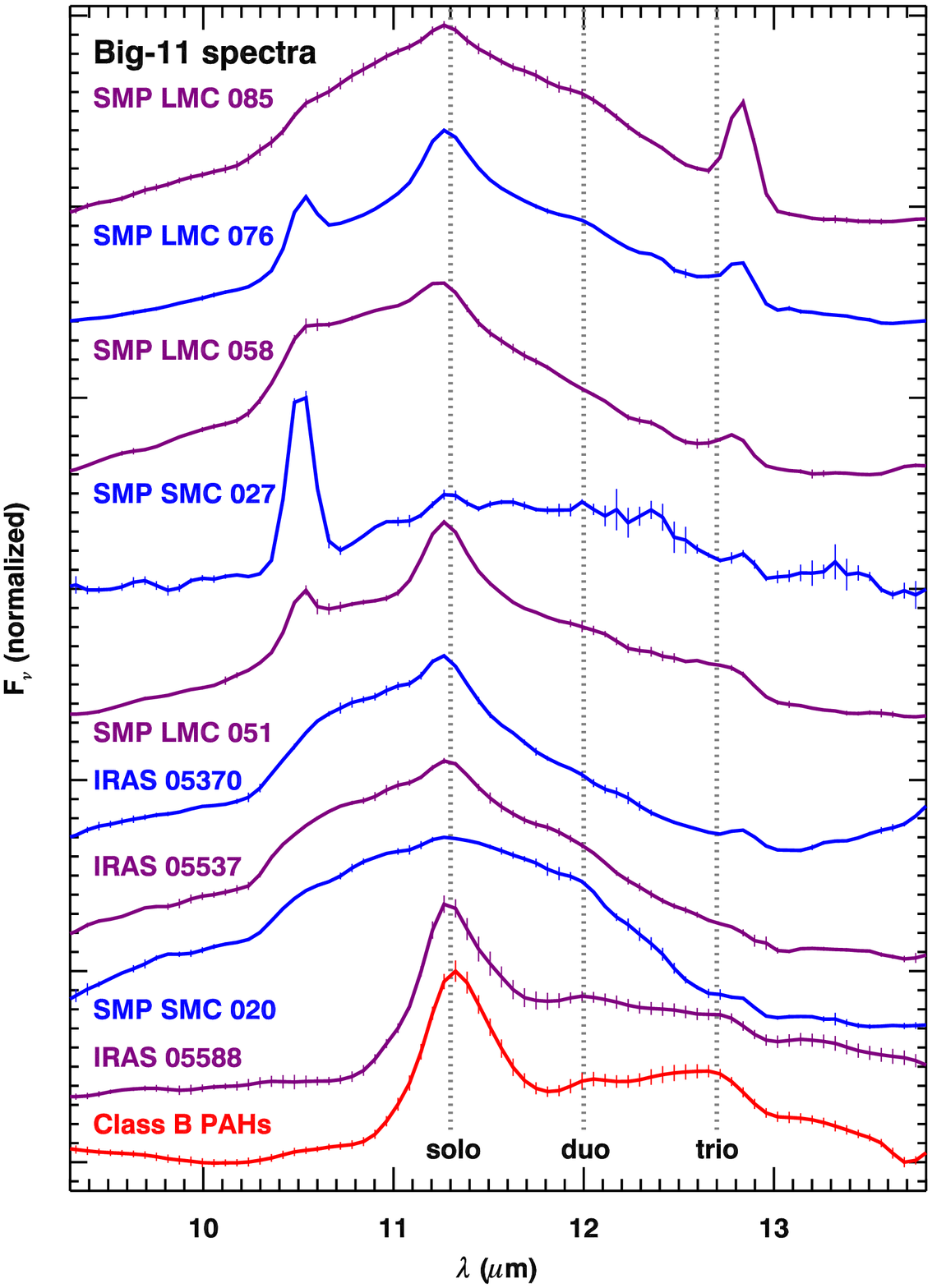}
\caption{The 11--14~\mum\ region for the Big-11 spectra.
The vertical lines and comparison data are as described in
Figure~\ref{f.sp12a}.\label{f.sp12b}}
\end{figure}

Figures~\ref{f.sp12a} and \ref{f.sp12b} focus on the 
11--14~\mum\ region in the spectra of the Fullerene, Mixed,
and Big-11 groups.  The big-11 feature usually appears in
spectra in the Fullerene and Mixed groups, because assignment
to those groups was based on the 17--19 and 6--9~\mum\ 
regions, not the 11--14~\mum\ region.  Some spectra, 
most notably SMP SMC~013, show inflections at 11.3 and 
12.0~\mum\ from a stronger contribution from PAHs.

Figure~\ref{f.sp12b} reveals the wide variation present in
the profile of the big-11 feature, with a range of relative
contributions between SiC and PAHs.  IRAS~05588, near the 
bottom, shows the least contaminated PAH spectrum, but with 
a slight reversal in the relative strength of 12.0 and 
12.7~\mum\ features.  This spectrum belongs to the Big-11 
group because of the strong shoulder at 18~\mum\ 
(Figure~\ref{f.spbig11}).

As noted in Section~\ref{s.big11nature}, the SiC dust feature 
dominates the big-11 feature, and the position and shape of 
the SiC feature can vary significantly from one carbon star
to the next.  Because the structure of the small PAH
residual depends on the assumed SiC profile, we did not 
classify the PAH contributions in the individual spectra 
with big-11 features.  Instead, we assume that the PAH 
contributions resemble the residual in Figure~\ref{f.sic}.  

\begin{figure} 
\includegraphics[width=3.4in]{\figpath 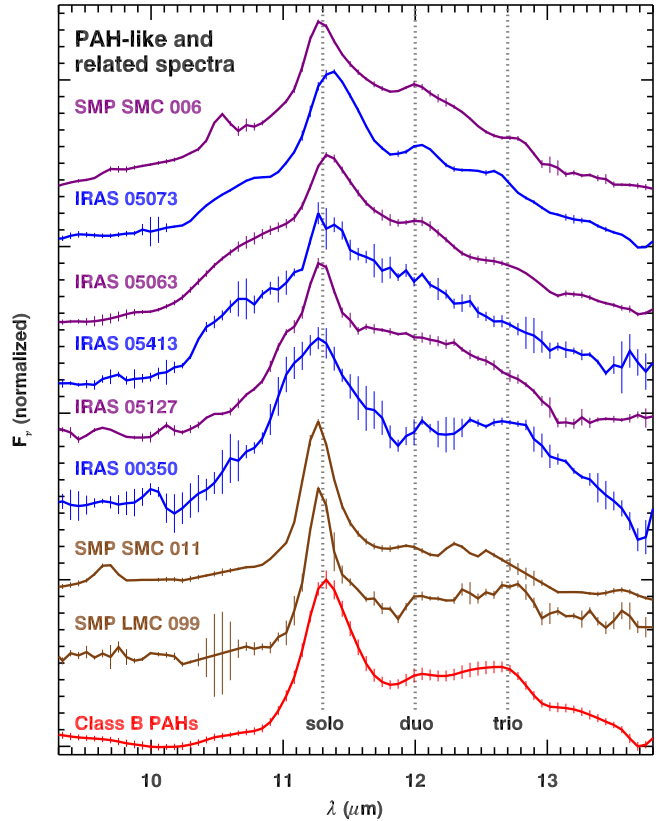}
\caption{The 11--14~\mum\ region for the spectra defined
as PAH-like, along with two Red spectra showing more typical
PAH features:  SMP SMC~011 and SMP LMC~099.  For clarity, the 
[S IV] line at 10.52~\mum\ has been removed from the spectra 
of both of these sources, and for SMP SMC~011, the [Ne II] 
line at 12.81~\mum\ has been removed as well.  The vertical 
lines and PAH comparison spectrum are as described in 
Figure~\ref{f.sp12a}.\label{f.sp12c}}
\end{figure}

\begin{figure} 
\includegraphics[width=3.4in]{\figpath 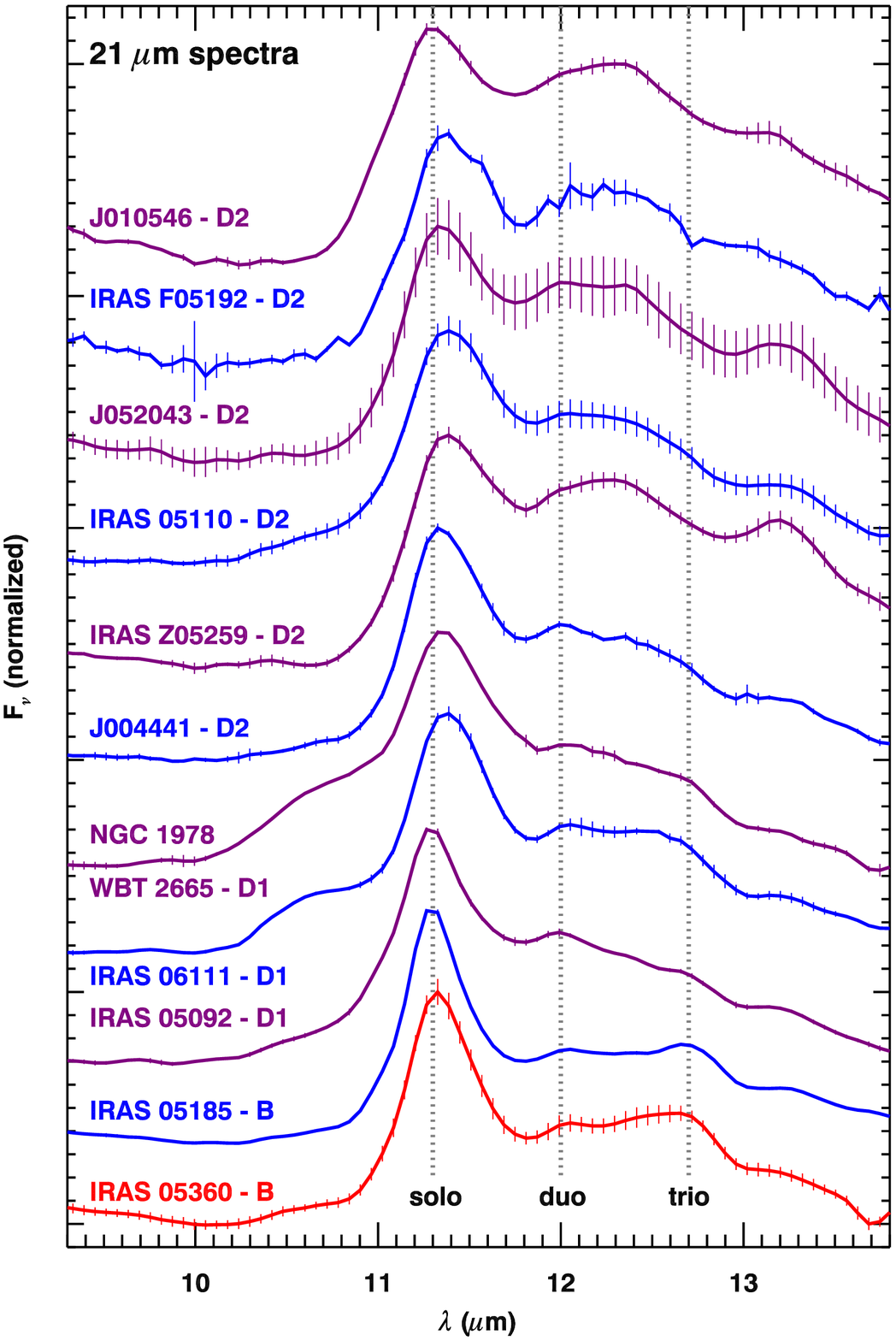}
\caption{The 11--14~\mum\ region of the spectra with 21~\mum\
and related emission features.  The vertical dotted lines are
as defined in Figure~\ref{f.sp12a}.  The spectrum of 
IRAS~05360 is plotted in red because it has served as the
prototypical PAH spectrum in several other figures.
\label{f.sp12d}}
\end{figure}

Figure~\ref{f.sp12c} illustrates how the PAH-like spectra
usually show an enhanced 12.0~\mum\ feature, like the 
residual PAH spectrum in the big-11 profile.  We identify
this as the Class D1 PAH profile.  It also has a broad blue 
wing to the 11.3~\mum\ feature not apparent in the three 
comparison spectra plotted at the bottom.  This blue excess 
at 10.5--11.0~\mum\ could arise from SiC.  IRAS~00350 is the 
only PAH-like spectrum with a 12.0~\mum\ feature that is
weaker than the 12.7~\mum\ feature, but it has a strong
blue excess.

Figure~\ref{f.sp12d} also reveals deviations from the usual
PAH profile at 11--14~\mum\ in most of the 21~\mum\ sources.
Only the bottom two spectra show a normal profile.
The remaining spectra can be separated into two distinct 
groups.  In the top six, the positions of the components at 
11--14~\mum\ have shifted to longer wavelengths, with broad 
emission components at $\sim$12.4 and 13.2~\mum, and the
shoulder at 10.5~\mum\ is missing.  We label this profile
as Class D2.  The remaining three sources show intermediate
spectra with PAH-like profiles with a stronger feature at 
12.0~\mum\ compared to 12.7~\mum\ and, in two of the three 
cases, a blue shoulder (i.e.\ Class D1).

Thus we have two unusual profiles apparent in the PAH-like
and 21~\mum\ groups.  Class D1 has components at 11.3, 12.0, 
and 12.7~\mum\ and a blue shoulder, and Class D2 has 
features at $\sim$11.4, $\sim$12.4, and 13.2~\mum.
In Figure~\ref{f.sp12d}, the spectrum of J004441 appears 
to be a blend of the D1 and D2 profiles.

In the D2 profile, the 12.4~\mum\ feature cannot be the H$_2$ 
0--0 S(2) transition at 12.28~\mum\ because it is too broad, 
and no H$_2$ lines could explain the 13.2~\mum\ feature.  In 
most of these spectra, the 11.3~\mum\ solo mode appears to be 
slightly shifted $\sim$0.1~\mum\ to the red, which might be 
expected from Class C PAH features typically seen in 
environments cooler than those responsible for more normal 
PAH emission.  Shifts of $\sim$0.4~\mum\ at 
12.0~\mum\ and $\sim$0.5~\mum\ at 12.7~\mum\ are a bit harder 
to explain.  While it is possible that the 12.4 and 
13.2~\mum\ components could be related to the duo and trio 
modes in some kind of modified PAH molecule, it seems equally 
reasonable that they could arise from some other unidentified 
bond in a hydrocarbon-related material.


The classifications applied by \cite{mat14} differ somewhat
from ours.  At 11--14~\mum, they describe Class A and B
profiles together as Class $\alpha$.  Their Class $\beta$ is
a mixture of Class A, B, and D1.  They describe our Class D2
as Class $\gamma$, and our big-11 feature as Class $\delta$.
Our classifications and theirs map together reasonably well,
but there are exceptions.  Most notably, IRAS~05063 is D1 in
our scheme but Class $\delta$ in theirs.  At 6--9~\mum, our
classifications generally agree.  One exception is 
IRAS~F05192, which we place in Class D and they place in
Class B.

\subsection{Spectra with the 21~\mum\ and related features} 

\begin{figure} 
\includegraphics[width=3.4in]{\figpath 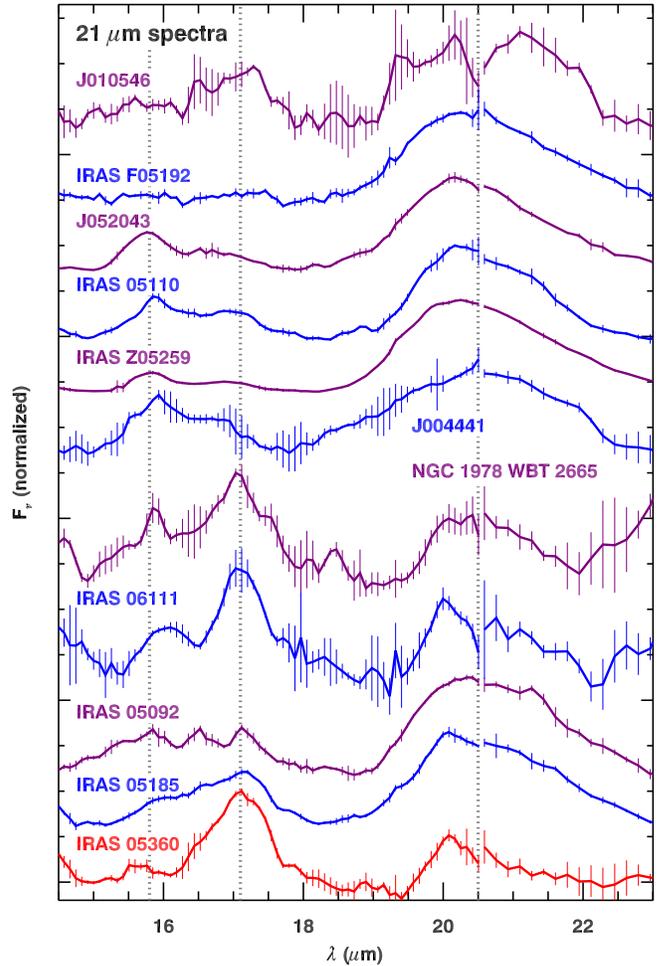}
\caption{The 15--23~\mum\ region of the spectra with 21~\mum\
features.  The spectra show three features, centered roughly
at 15.8, 17.1, and 20.5~\mum\ (the wavelengths of the dotted
vertical lines).  While the spectra show a 21~\mum\ feature, 
the variation in the relative strengths and positions is 
substantial.  The spectrum of IRAS~05360 is plotted in red 
because it has served as the prototype for Class B PAHs in 
several other figures.\label{f.sp15d}}
\end{figure}

Galactic sources with a 21~\mum\ feature in their spectra
typically show associated features at $\sim$8, 11, 16, and 
26--30~\mum, but not all features appear in all spectra 
\citep{kra02}.  Even the 21~\mum\ feature could be absent,
because the classification is based on the total spectrum,
not just the presence of one feature.  The Magellanic sample 
behave similarly (see Figure~\ref{f.sp21}).  The features do 
not appear in unison, and they can shift noticeably in 
wavelength from source to source, as Figure~\ref{f.sp15d}
shows.

We find the central wavelength of the 21~\mum\ feature, based
on all of the identified 21~\mum\ sources in our sample, to 
be 20.47 $\pm$ 0.10~\mum.  The standard deviation is 
surprisingly small given the variation in width of the 
feature.  Our analysis confirms the presence of the 21~\mum\ 
feature in all 11 spectra in this group, after their initial 
assignment based on visual inspection.  \cite{vol11} 
described the 21~\mum\ features as very weak in the spectra 
of IRAS~06111 and IRAS~05360 and as weak in the spectra of 
J010546 and NGC~1978~WBT~2665.  These are the detections with 
the lowest SNRs, but they still appear to be solid 
detections, and the features are visible in 
Figure~\ref{f.sp15d}.  The SNR gained from optimal extraction 
has helped here.

Our spectra show an association of 21~\mum\ spectra with
aliphatics and a further association with unusual spectral
structure in the 6--9 and 11--14~\mum\ regions.  The
aliphatic feature at 6.90 appears in all 11 spectra,
and most show Class D1 or D2 emission at 11--14~\mum.
Table~\ref{t.pahclass} reveals more trends.  All of the 
spectra with Class D2 profiles at 11--14~\mum\ have Class D 
profiles at 6--9~\mum, and two of the three Class D1 spectra 
are Class D at 6--9~\mum.  Aliphatics and the Class D 
profiles, especially D2, are linked.

\subsubsection{The 16~\mum\ emission complex} 

\begin{figure} 
\includegraphics[width=3.4in]{\figpath 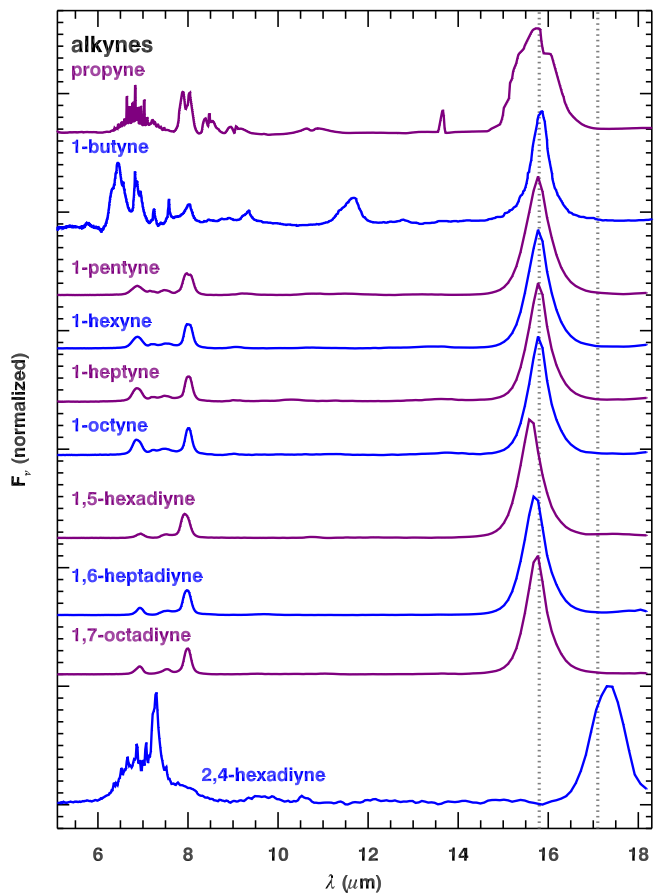}
\caption{Infrared laboratory spectra of several alkyne
chains. All except the last molecule have an emission feature 
at or near 15.8~\mum; the bottom alkyne shows a feature 
shifted to the vicinity of the 17.1~\mum\ feature.  The data 
are from NIST Chemistry WebBook, and Table~\ref{t.alkynes} 
gives the chemical formula for each molecule.
\label{f.alkynes}}
\end{figure}

\begin{deluxetable}{ll} 
\tablecolumns{2}
\tablewidth{0pt}
\tablenum{4}
\tablecaption{Alkynes}
\label{t.alkynes}
\tablehead{
  \colhead{Molecule} & \colhead{Formula}
}
\startdata
propyne         & CH$\equiv$C---CH$_3$ \\
1-butyne        & CH$\equiv$C---CH$_2$---CH$_3$ \\
1-pentyne       & CH$\equiv$C---CH$_2$---CH$_2$---CH$_3$ \\
1-hexyne        & CH$\equiv$C---CH$_2$---CH$_2$---CH$_2$---CH$_3$ \\
1-heptyne       & CH$\equiv$C---CH$_2$---CH$_2$---CH$_2$---CH$_2$---CH$_3$ \\
1-octyne        & CH$\equiv$C---CH$_2$---CH$_2$---CH$_2$---CH$_2$---CH$_2$---CH$_3$ \\
1,5-hexadiyne   & CH$\equiv$C---CH$_2$---CH$_2$--C$\equiv$CH \\
1,6-heptadiyne  & CH$\equiv$C---CH$_2$---CH$_2$---CH$_2$--C$\equiv$CH \\
1,7-octadiyne   & CH$\equiv$C---CH$_2$---CH$_2$---CH$_2$---CH$_2$--C$\equiv$CH \\
2,4-hexadiyne   & CH$_3$---C$\equiv$C---C$\equiv$C---CH$_3$
\enddata
\end{deluxetable}

The central wavelength of the 16~\mum\ complex is 16.51 $\pm$ 
0.39~\mum.  The larger spread in the apparent position of 
this feature results from its more complex structure.  In 
most of the spectra, it separates into two components, 
centered at $\sim$15.8 and 17.1~\mum, although both the
position and width of the component near 15.8~\mum\ can shift 
considerably, as Figure~\ref{f.sp15d} shows.

The position of the feature, 15.8~\mum, matches the C--H 
bending mode adjacent to a triple, or alkyne, C$\equiv$C 
bond.  Figure~\ref{f.alkynes} illustrates the laboratory 
spectra of several alkynes, all of which show a feature near 
15.8~\mum, using data from the NIST Chemistry 
WebBook\footnote{The National Institute of Standards and 
Technology maintains the WebBook at 
http://webbook.nist.gov/chemistry.}.  Table~\ref{t.alkynes} 
gives the chemical formulae for the molecules considered.  
Propyne, which can be considered the prototype, shows a broad 
feature, due to the interactions of the C--H bending mode with 
numerous modes in the adjacent methyl (CH$_3$) group.  
\cite{mal12} detected propyne in absorption in the spectrum of
SMP~LMC~011.  In longer aliphatic molecules, the intervening 
methylene (CH$_2$) insulates the modes on one side of the 
molecule from the other, and the resulting feature at 
15.8~\mum\ is narrower.  In molecules with a triple bond at 
both ends, the C--H mode is shifted to higher energies, and 
again, with more intervening methylene groups, this effect 
grows smaller.

If a long alkyne chain were attached to a larger molecule,
such as a PAH, the intervening methylene groups should
insulate the terminal C--H bending mode, much as it does in 
the longer aliphatic molecules shown in 
Figure~\ref{f.alkynes}.  Given other evidence for aliphatics 
in the spectra of 21~\mum\ sources (Figures~\ref{f.sp69c} and 
\ref{f.sp69d}), it is reasonable to expect that some of the 
aliphatic chains might be terminated by alkyne bonds.  Their 
presence may be enhanced by photo-processing as the hot 
stellar core grows more exposed.  Unfortunately, we are not 
aware of any laboratory spectra of such molecules to test 
this hypothesis.  Different combinations of PAH, alkyne 
sidegroups, and alkyne chains might be able to reproduce the 
variation in the shapes of the observed 15.8~\mum\ features.

The other component of the 16~\mum\ feature is centered at
$\sim$17.1~\mum.  Where it is strong, this component is
clearly distinct from the 17.4~\mum\ features associated
with either PAHs or fullerenes.  Its carrier may be related
to the alkyne chains which produce the 15.8~\mum\ feature.
If the alkyne bond is shifted one notch away from the end
of the chain, the terminal carbon becomes part of a methyl
group, and the energies of the C--H bending modes shift to
longer wavelengths.  In 2,4-hexadiyne, the shift is close
to the right amount to account for the 17.1~\mum\ feature.
Unfortunately, the NIST Chemistry WebBook does not contain
spectral data for longer analogues, and the laboratory 
spectra for related molecules with a methyl group at just 
one end (2-butyne, 2-pentyne, etc.), show little or no 
spectral structure at 15--18~\mum.  We conclude that the 
carrier of the 17.1~\mum\ feature is not yet confirmed, but
it is likely to be related to the alkyne chains responsible 
for the 15.8~\mum\ feature.

The combined strength of the two components of the 16~\mum\ 
complex does not exceed 5.6\% of the total emission from the
various features associated with the 21~\mum\ feature.  In
most casess it is $\la$2\%.  The laboratory spectra in 
Figure~\ref{f.alkynes} also show structure at 6--9~\mum,
but that structure would be difficult to detect, given its 
relative strength to the already weak 15.8 and 17.1~\mum\ 
features.

\subsubsection{Additional 21~\mum\ candidates} 
\label{s.new21}

\begin{figure} 
\includegraphics[width=3.4in]{\figpath 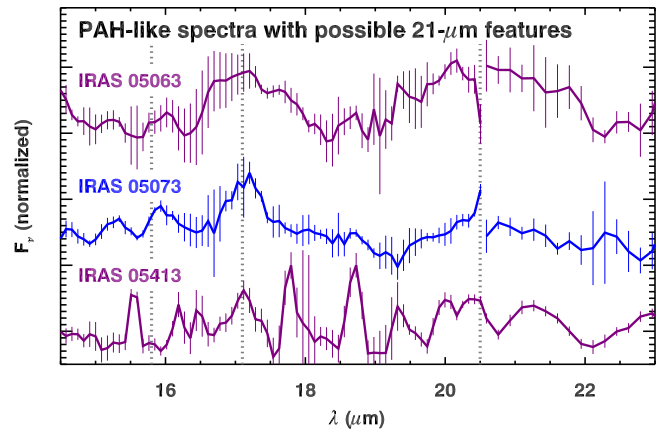}
\caption{The 15--23~\mum\ spectral region of the three
candidate 21~\mum\ sources among the PAH-like spectra.  The 
vertical dotted lines are as defined in Figure~\ref{f.sp15d}.
\label{f.sp15c}}
\end{figure}

The 6--9~\mum\ spectra of the PAH-like sources
(Figure~\ref{f.sp69c}) reveal two more Class D profiles,
for IRAS~05063 and IRAS~05073.  Both also show strong
aliphatics.  A third spectrum, IRAS~05413, is much noisier
but consistent with these two.  All three spectra are 
included in the analysis in Appendix~1.3 and are
plotted in Figure~\ref{f.sp15c}.  

Only IRAS~05063 has a reasonable detection at 21~\mum, with a 
SNR of 7.5.  Because the extraction algorithm used for these 
features shifts the wavelengths to match minima in the 
spectra, it will actually fit to the noise, so the extracted 
SNR should be taken as an upper limit.  The central 
wavelength of the extracted feature is within 1~$\sigma$ of 
the mean for this feature, which adds some confidence.  
Furthermore, the 17.1~\mum\ feature looks reasonable and is 
extracted with a SNR of 5, so this spectrum remains a viable 
candidate as a 21~\mum\ source.  \cite{vol11} describe its 
21~\mum\ feature as ``vW'' (very weak).  While we conclude 
that this source is a 21~\mum\ candidate, we cannot confirm 
it as a definite 21~\mum\ source.

IRAS~05073 does not have a measurable 21~\mum\ feature, and 
\cite{vol11} describe it possibly very weak (which they 
described as ``vW?'').  The structure in its spectrum at 15.8 
and 17.1~\mum\ suggests that it should also be considered as 
a candidate.  

While our analysis extracted a 21~\mum\ feature for 
IRAS~05413, it is only at 3~$\sigma$, and the central 
wavelength is about 3~$\sigma$ from the mean.  This source is 
the least likely of the three to be a 21~\mum\ source.

\subsection{The 26--30~\mum\ feature} 
\label{s.mgs}

\begin{deluxetable}{lcccc} 
\tablecolumns{5}
\tablewidth{0pt}
\tablenum{5}
\tablecaption{Mean wavelengths and MgS temperatures for the 26--30~\mum\ feature}
\label{t.mgswave}
\tablehead{
  \colhead{Group} & \colhead{Number} & \colhead{Out of} & 
  \colhead{$\lambda_C$ (\mum)} & \colhead{$T_{MgS}$ (K)} 
}
\startdata
Fullerene     &  4 &  5 & 29.88 $\pm$ 0.38 & 250 \\
Mixed         &  6 &  7 & 30.34 $\pm$ 0.59 & 220 \\
Big 11        &  5 & 10 & 30.65 $\pm$ 0.57 & 200 \\
Combined      & 15 & 22 & 30.32 $\pm$ 0.59 & 220 \\
\\
PAH-like      &  6 &  6 & 28.62 $\pm$ 1.49 & 360 \\
21~\mum\ (B)  &  2 &  2 & 28.73 $\pm$ 0.21 & 350 \\
21~\mum\ (D2) &  6 &  6 & 28.52 $\pm$ 0.30 & 370 \\
21~\mum\ (D1) &  3 &  3 & 28.26 $\pm$ 0.31 & 400 \\
Red           &  2 &  5 & 27.92 $\pm$ 0.57 & 450 \\
Combined      & 18 & 21 & 28.47 $\pm$ 0.86 & 380
\enddata
\end{deluxetable}

Table~\ref{t.mgswave} presents the mean wavelengths of the
26--30~\mum\ feature in the different spectral groups, based
on the analysis in Appendix~1.3.  The data show a shift in 
the position of the 26--30~\mum\ feature, which is apparent 
even in Figure~\ref{f.proto}.  The central wavelength moves 
from the red to the blue in the sequence Big-11, Mixed, 
Fullerene, PAH-like, 21~\mum, and Red.  A gap of over 
1~\mum\ separates the Fullerene and PAH-like groups, 
suggesting that we really have two larger groups.  

\cite{vol02} reported that in the {\it ISO}/SWS spectra of
several Galactic 21~\mum\ sources, the 26--30~\mum\ feature
appeared to consist of two components, with a narrower 
component centered at $\sim$26~\mum\ and a broader component
centered at $\sim$33~\mum.  These two components would 
explain a notch apparent in the feature at $\sim$27~\mum, but
that structure is more likely to be an artifact in the 
spectra due to a light-leak at the long-wavelength end of
Band 3D (27.3~\mum) and the poor quality of Band 3E 
(27.3--27.7~\mum) \citep{slo03}.  In our Magellanic sample,
the 26--30~\mum\ feature appears to shift smoothly from one 
spectrum to the next.  None of the individual spectra show 
evidence for two separate components to the feature.  The
difference between the Magellanic and Galactic samples is 
more likely due to the differences in the spectrometers and
not differences in the dust chemistry.

\cite{for81} discovered the 26--30~\mum\ feature in spectra 
from evolved carbon-rich objects, and \cite{gm85} proposed 
MgS dust as the carrier.  One of the major objections to MgS 
as the carrier of the 26--30~\mum\ feature has been the 
abundance problem created by the strength of the feature.  It 
can account for up to about a third of the total luminosity 
of some sources, which would require more sulfur than is 
available from the central star \citep{zha09b}.  If only the 
outer layers of the grains were MgS, then this would solve 
the problem.  Observational evidence from spectral 
studies of Magellanic carbon stars supports this scenario.  
As the dust shells around the stars grow increasingly red, 
the MgS feature strength climbs just as the SiC strength 
drops \citep[e.g.][]{slo06,zij06}.  This behavior is 
consistent with the coating of dust grains with MgS as 
they move away from the central star and cool 
\citep{lag07,lei08}.

Models of condensation of MgS in carbon-rich outflows
support this scenario, and the optical properties of 
MgS-coated grains reproduce the observed feature 
\citep{zhu08}.  A more recent study of the optical properties
of MgS-coated grains reveals that coatings thin enough to
stay within the abundance limits can produce the observed
feature \citep{lom12}.  

Assuming that MgS is the carrier of the 26--30~\mum\ feature, 
we estimate the dust temperature from its position, following 
\cite{hon02}.  They could fit the central wavelength by 
varying either the grain shape distribution (with spherical 
grains producing the bluest features) or the grain 
temperature.  If the latter is the cause of the shift, then 
their power law can be used:
$T_{MgS} = 5.1 \times 10^{14} \lambda_C^{-8.34}$.

Table~\ref{t.mgswave} includes a column giving the MgS dust
temperatures using the above equation.  The first three 
groups have MgS dust temperatures of $\sim$200--250~K, vs.\ 
the remaining groups, at $\sim$350--450~K.  Thus, if MgS 
produces the 26--30~\mum\ feature, then a shift in dust 
temperature can explain the observed shift in the position 
of the feature among the spectral groups we have defined,
with the Big-11 spectra having the coolest MgS dust and
the 21~\mum\ sources having the warmest.  


\section{Supporting photometry and spectroscopy} 

\subsection{Infrared colors} 
\label{s.ircolor}

\begin{figure} 
\includegraphics[width=3.4in]{\figpath 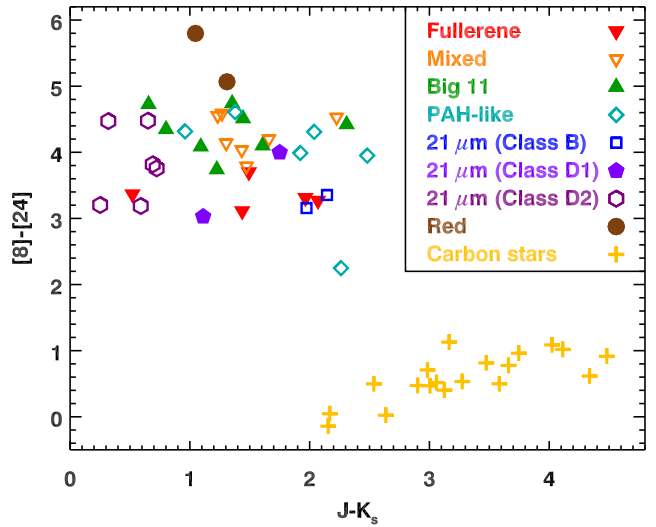}
\caption{A color-color plot showing the locations of the
spectral groups defined in this paper.  The 21~\mum\
sources are distinguished by the nature of the PAH emission
in their spectra at 6--9 and 11--14~\mum.  The comparison
sample of carbon stars are the SMC sample originally 
considered by \cite{slo06}.  The PAH-like source closest to
the carbon stars is IRAS~00350.\label{f.cc1}}
\end{figure}

\begin{figure} 
\includegraphics[width=3.4in]{\figpath 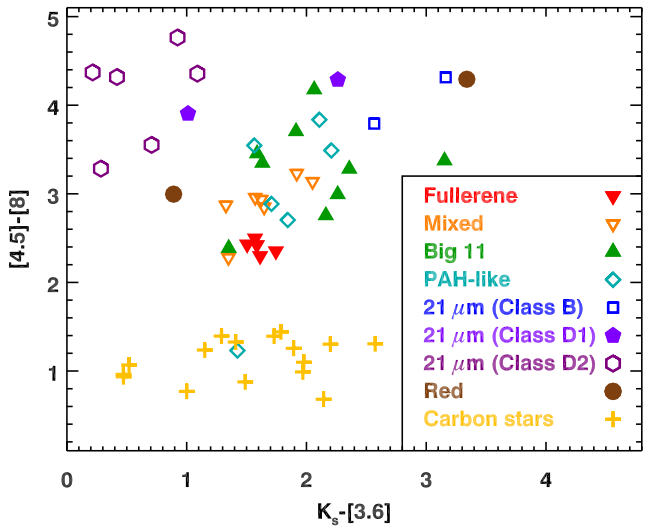}
\caption{A color-color plot with alternative axes showing 
the locations of the spectral groups defined in this 
paper.  The comparison sample is the same as in 
Fig.~\ref{f.cc1}.  The location of the Fullerene spectra are 
more tightly constrained in this plot.  IRAS~00350 is the
PAH-like source amongst the carbon stars, and NGC~1978
WBT~2665 is the Class D1 21~\mum\ source near to the D2
sources.\label{f.cc2}}
\end{figure}

\begin{deluxetable*}{lccccc} 
\tablecolumns{6}
\tablewidth{0pt}
\tablenum{6}
\tablecaption{Infrared colors}
\label{t.ircol}
\tablehead{
  \colhead{Group} & \colhead{$<J-K_s>$} & \colhead{$<K_s-[3.6]>$} & 
  \colhead{$<$[3.6]$-$[4.5]$>$} & \colhead{$<$[4.5]$-$[5.8]$>$} & 
  \colhead{$<$[8]$-$[24]$>$}
}
\startdata
Fullerene                 & 1.50 $\pm$ 0.61 & 1.60 $\pm$ 0.09 & 0.88 $\pm$ 0.09 & 2.40 $\pm$ 0.08 & 3.35 $\pm$ 0.22 \\
Mixed                     & 1.51 $\pm$ 0.35 & 1.64 $\pm$ 0.27 & 0.97 $\pm$ 0.23 & 2.89 $\pm$ 0.30 & 4.26 $\pm$ 0.30 \\
Big 11                    & 1.26 $\pm$ 0.51 & 2.05 $\pm$ 0.53 & 1.11 $\pm$ 0.16 & 3.28 $\pm$ 0.52 & 4.33 $\pm$ 0.34 \\
PAH-like                  & 1.84 $\pm$ 0.57 & 1.81 $\pm$ 0.31 & 0.93 $\pm$ 0.19 & 2.95 $\pm$ 0.94 & 3.90 $\pm$ 0.85 \\
21~\mum\ (Class B)        & 2.06 $\pm$ 0.12 & 2.86 $\pm$ 0.42 & 0.90 $\pm$ 0.30 & 4.05 $\pm$ 0.37 & 3.26 $\pm$ 0.14 \\
21~\mum\ (Class D1)       & 1.43 $\pm$ 0.55 & 1.65 $\pm$ 0.63 & 0.92 $\pm$ 0.37 & 4.10 $\pm$ 0.27 & 3.51 $\pm$ 0.69 \\
21~\mum\ (Class D2)       & 0.54 $\pm$ 0.20 & 0.61 $\pm$ 0.36 & 0.51 $\pm$ 0.45 & 4.11 $\pm$ 0.56 & 3.82 $\pm$ 0.57 \\
Red                       & 1.16 $\pm$ 0.14 & 1.89 $\pm$ 1.28 & 1.49 $\pm$ 0.41 & 4.16 $\pm$ 1.10 & 5.63 $\pm$ 0.50
\enddata
\end{deluxetable*}

For the sources in the sample, we have collected photometry 
from several infrared and optical catalogs.  Appendix~2
explains how we compiled the photometry and presents the
results for each source in the sample.  In this section we
draw from those data to better understand the photometric
properties of the spectral groups we have defined.

Figures~\ref{f.cc1} and \ref{f.cc2} show how the spectral 
groups are distributed in infrared color-color space.  We
have included a control sample of carbon stars defined by
\cite{slo06}, but with updated photometry and applying the
techniques described in Appendix~2.  The carbon stars trace 
a well-defined sequence from the bottom center of the diagram 
up and to the right.  The carbon-rich post-AGB objects occupy 
a different part of the diagram, with blue near-IR colors, 
because the central star is partially revealed, and red 
mid-IR colors, because the dust is moving away from the 
central star and cooling.  Table~\ref{t.ircol} gives 
mean colors for the spectral groups.

The Fullerene sources have bluer [8]$-$[24] colors than the 
Big-11 and Mixed groups.  Figure~\ref{f.cc2} reveals that
the colors of the Fullerene sources are tightly constrained 
in both $K-[3.6]$ and [4.5]$-$[8].

The 21~\mum\ sources with Class D2 PAH spectra stand out
from the other 21~\mum\ sources at shorter wavelengths, 
showing bluer colors in both $J-K_s$ and $K_s-[3.6]$.  The
21~\mum\ sources with Class D1 PAH spectra are bluer than
the Class B spectra in both of these colors, but only two 
spectra plotted in each figure is not enough to draw
conclusions.

\subsection{Temperatures} 

\begin{deluxetable}{lcll} 
\tablecolumns{4}
\tablewidth{0pt}
\tablenum{7}
\tablecaption{Temperatures in PNe}
\label{t.temp}
\tablehead{
  \colhead{Group} & \colhead{Num.} & \colhead{Electron T (K)} &
  \colhead{Effective T (K)}
}
\startdata
Fullerene                 &  4 & 12,430 $\pm$   590 & 34,900 $\pm$  5,100 \\
Mixed                     &  7 & 11,630 $\pm$ 1,340 & 31,300 $\pm$  2,000 \\
Big 11                    &  4 & 12,580 $\pm$   730 & 42,800 $\pm$ 23,400 \\
control PNe---C-rich      & 17 & 13,600 $\pm$ 1,330 & 64,300 $\pm$ 26,100 \\
control PNe---featureless & 21 & 14,280 $\pm$ 2,260 & 65,500 $\pm$ 43,400
\enddata
\end{deluxetable}

\begin{figure} 
\includegraphics[width=3.4in]{\figpath 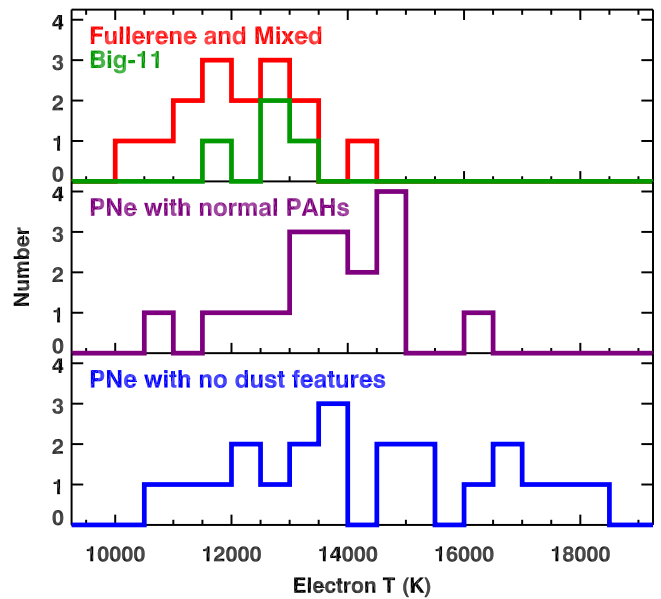}
\caption{The distribution of electron temperatures within
the sources in our sample classified as Fullerene, Mixed,
or Big 11 (top panel), compared to the control samples of
dusty carbon-rich and featureless PNe.  The three groups 
plotted in the top panel have similar distributions, and 
these differ from both of the control samples.\label{f.histexc}}
\end{figure}

\begin{figure} 
\includegraphics[width=3.4in]{\figpath 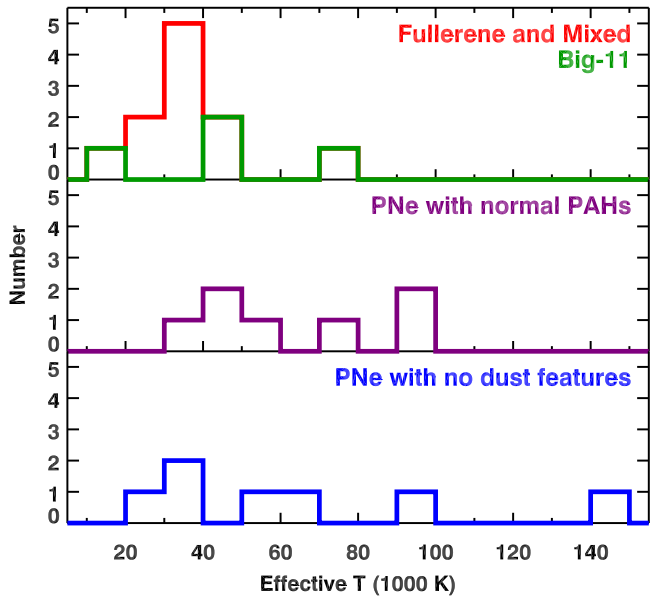}
\caption{The distribution of the effective temperatures of
the central stars of the PNe in our sample (top panel),
compared to the control samples of PNe (bottom two panel).  
\label{f.histeff}}
\end{figure}

The central stars of PNe exhibiting fullerene features 
generally appear to be cooler than in more typical PNe 
\citep[e.g.][]{gh12, ots13}.  To investigate the temperatures 
of the systems in our sample, we have examined the electron 
temperature of the gas, the effective temperature of the 
central star, and the optical colors of the nebulae.  Our 
control sample is based on the PNe observed by the IRS and 
examined by \cite{sta07} and \cite{jbs09}.  We include the 
spectra not observed by us and classified by them either as 
carbon-rich, based on the presence of normal PAH emission in 
their spectra (Class A or B), or featureless, based on the 
absence of any solid-state features.

\cite{lei06} measured the electron temperature in a large
sample of Magellanic PNe, using the [O III] and [N II] lines.
Where these produce different temperatures, we took an 
average.  Table~\ref{t.temp} presents the means for the
Fullerene, Mixed and Big-11 groups, along with control
samples.  The uncertainties are the standard deviations of 
the samples.  Figure~\ref{f.histexc} compares the temperature 
distribution of these groups with the two PNe control 
samples.  The Fullerene, Mixed, and Big-11 groups are 
indistinguishable.  The ionized gas in the PNe producing 
fullerene, mixed, or big-11 features tends to be cooler than 
the gas in PNe associated with typical PAH emission.  The 
ionized gas in featureless PNe spans a range of temperatures 
covering the PNe in our sample and PNe with normal PAHs and 
continuing to significantly higher temperatures.

Several observers have estimated the effective temperature of 
the central stars in Magellanic PNe.  We rely on the studies 
of the LMC and SMC by \cite{vil03,vil04}.
They based their estimates on He II lines and H$\beta$,
which often lead to different results.  For comparative 
purposes, we will concentrate on the estimates from H$\beta$ 
because they cover more of our sample.  We have supplemented
these temperatures with measurements of SMP LMC~002 and 085
by \cite{dop94} and \cite{her04} and a rough estimate of the 
effective temperature of IRAS~05588 by \cite{vol11}.

The mean effective temperatures for the different groups
appear in Table~\ref{t.temp}.  As with the electron
temperatures, the Fullerene and Mixed groups have overlapping
distributions.  The Big-11 sources show a wider spread, with
three sources shifting to higher temperatures.  The exception 
is IRAS~05588, which has $T_{eff} \approx$ 14,000 K and is 
better classified as a proto-PN \citep{vol11}.  Its presence 
in the Big-11 group is due primarily to a strong 18~\mum\ 
feature (Section~\ref{s.big11nature}), but its 
spectrum has the strongest PAH-like contribution at 11~\mum\ 
in the group, suggesting that it may differ from the rest.  
Excluding it raises $<T_{eff}>$ for the three remaining 
Big-11 sources to 52,400 $\pm$ 16,500 K, clearly hotter than 
the Fullerene and Mixed groups.

Figure~\ref{f.histeff} illustrates the differences between
the effective temperatures of the PNe in our sample and 
the control samples.  The results are similar to those for
the electron temperatures, although the number of control
sources for which we have effective temperatures is smaller.
The dusty PNe from the control group, which only show PAH
emission and no other dust, tend to have hotter central 
sources than the PNe in our sample, while the dust-free PNe 
show a wide range of temperatures.  

The temperature analysis confirms previous conclusions
about the fullerene sources.  Fullerenes appear in a subset
of PNe with lower electron temperatures and cooler central 
stars than the general population of PNe.

\subsection{Optical colors} 

\begin{deluxetable}{lrrr} 
\tablecolumns{4}
\tablewidth{0pt}
\tablenum{8}
\tablecaption{Optical colors}
\label{t.colopt}
\tablehead{
  \colhead{Group} & \colhead{$<U-B>$} & \colhead{$<B-V>$} & \colhead{$<V-I>$}
}
\startdata
Fullerene &  $-$1.07 $\pm$ 0.18    & 0.23 $\pm$ 0.33    & $-$0.32 $\pm$ 0.49 \\
Mixed     &  $-$0.58 $\pm$ 0.32    & 0.99 $\pm$ 0.22    & $-$1.04 $\pm$ 0.48 \\
Big 11    &  $-$0.34 $\pm$ 0.45    & 1.15 $\pm$ 0.68    & $-$0.46 $\pm$ 0.89 \\
PAH-like  &     0.03 $\pm$ 1.04    & 0.90 $\pm$ 0.55    &    0.20 $\pm$ 1.36 \\
21 \mum   &     1.08 $\pm$ 0.85    & 1.29 $\pm$ 0.50    &    1.29 $\pm$ 0.73 \\
Red       &  $-$0.49 $\pm$ \nodata & 0.63 $\pm$ \nodata & $-$0.94 $\pm$ \nodata
\enddata
\end{deluxetable}

\begin{figure} 
\includegraphics[width=3.4in]{\figpath 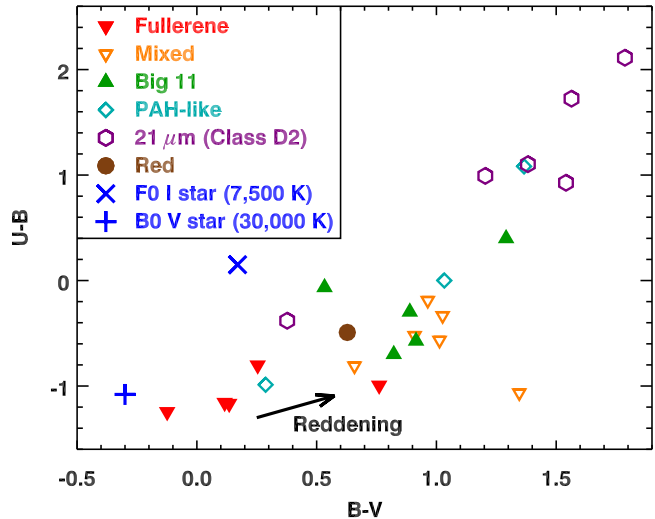}
\caption{An optical color-color plot of the carbon-rich
post-AGB sample, showing that the colors are consistent with
dust extinction.  The positions of an F0 supergiant and a 
main sequence B0 star (T$_{eff}$ = 30,000 K and 7,500 K,
respectively) are marked with blue crosses \citep{cox00}.
The reddening vector depicts one magnitude of extinction as 
determined by \cite{rie85}.  The two non-fullerene sources 
with $B-V < 0.5$ are IRAS~05127 (PAH-like) and J010546 
(21~\mum/D2).\label{f.ccubv}}
\end{figure}

An examination of optical colors reveals clear differences 
among the spectroscopic groups in our sample.  
Figure~\ref{f.ccubv} plots $U-B$ versus $B-V$.  The Fullerene 
sources occupy the blue end of the diagram, while the 
21~\mum\ sources for which we have optical colors occupy the 
red end.  The Mixed, Big-11, and PAH-like spectra are between 
the two extremes.  Many sources are missing from this diagram 
because they are too embedded for optical photometry at the 
distances involved.  Generally, sources with $J-K \ga 1$, are 
absent, which explains why for the 21~\mum\ sources, only 
those associated with Class D2 PAH emission appear.  The 
Fullerene and Mixed sources are an exception.  Despite their 
red $J-K$ colors, they are surprisingly blue in the optical.

Figure~\ref{f.ccubv} also shows the nominal positions of an
F0 supergiant, which is analogous to the central star in a 
post-AGB object with 21~\mum\ emission, a 30,000 K star to 
approximate the central star in PNe with fullerene emission 
\citep{cox00}, and a vector depicting one magnitude of 
interstellar extinction \citep{rie85}.  While the extinction 
from carbon-rich circumstellar dust will be different in 
detail from typical interstellar extinction, the reddening 
arrow still gives a reasonable impression.  The optical 
colors in our sample are consistent with extinction by 
circumstellar dust.

The blue colors of the Fullerene sources indicate that we 
have a clear line of sight to the central star in these 
systems, or if not that, that we at least can see a region 
scattering direct radiation from the central star.  
Polarimetry would settle that question, but in the meantime 
it is fair to say that the optical colors of the Fullerene 
sources are consistent with a reasonably clear line of sight 
to a radiation source of $\sim$30,000 K at their center.  The
colors of the Mixed and Big-11 sources indicate more dust
extinction.  In the other groups, the line of sight to the 
central region is more obscured.

Table~\ref{t.colopt} presents mean optical colors for the 
spectral groups in our sample and provides some quantitative
support for Figure~\ref{f.ccubv}.  The $V-I$ colors tell a
different story.  Once we move to wavelengths longer than 
$V$, the Fullerene sources cease to be the bluest.  The
near- and mid-IR colors plotted in Figures~\ref{f.cc1} and 
\ref{f.cc2} show that the Fullerene sources do have 
significant amounts of circumstellar dust; it is just not
distributed in the line of sight to the central star.

\subsection{Spectral energy distributions} 

\begin{figure} 
\includegraphics[width=3.4in]{\figpath 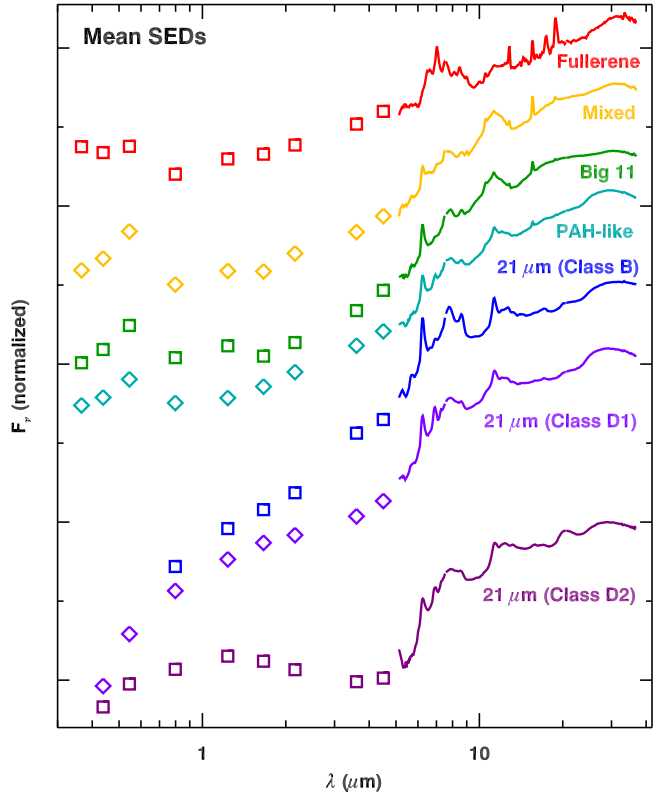}
\caption{Mean SEDs for each spectral group.\label{f.sed}}
\end{figure}

Figure~\ref{f.sed} plots the mean SED for each spectral 
group, separating the 21~\mum\ sources by the class of PAH 
emission.  We have not included the Red group because of the 
disparate nature of its members.  For each remaining group or 
subgroup, the spectra and photometric data were normalized at 
3.6~\mum\ and averaged.

The Fullerene spectra diverge from the other groups
at the shortest wavelengths, showing bluer SEDs.  This 
behavior is more obvious in Figure~\ref{f.ccubv}.

Within the 21~\mum\ group, the Class D2 sources differ from
the others, showing a secondary maximum peaking at $J$ while
the rest drop monotonically from [4.5] to $V$ and $U$, due to
extinction from dust.  The dust around the Class D2 sources 
is optically thin enough to reveal a secondary peak at $J$,
corresponding to a blackbody temperature of $\sim$4000 K.
If the peak were due to hot amorphous carbon dust, its
temperature would be $\sim$2500 K, well above its
condensation temperature.  The inflection at $\sim$2~\mum\
in the Class D1 SED may be of the same nature.

A better explanation of the differences between the D2
sources and the other 21~\mum\ sources comes from the
models of their SEDs by \cite{vol11}.  The optical depth
that they fitted at 0.55~\mum\ ranged from 1.1 to 5.6 for
the six Class D2 sources, with a mean of 2.5 $\pm$ 1.9.
For the five other 21~\mum\ sources, the minimum optical
depth was 9.4, and the mean was 18.8 $\pm$ 8.3, with no 
obvious distinction between the Class D2 and Class B
sources.  The distribution of temperatures of the central 
stars showed no difference between the different subgroups
of 21~\mum\ sources.  The mean temperature was 7,400 $\pm$ 
2,000 K, quite cool compared to the central temperatures of 
the Fullerene and related groups.

\subsection{Bolometric magnitudes} 

\begin{deluxetable}{lcrc} 
\tablecolumns{4}
\tablewidth{0pt}
\tablenum{9}
\tablecaption{Absolute bolometric magnitudes}
\label{t.mbol}
\tablehead{
  \colhead{Group} & \colhead{Num.} & \colhead{$<M_{\rm bol}>$} & \colhead{Notes}
}
\startdata
Fullerene           & 5 & $-$4.34 $\pm$ 0.52 \\
Mixed               & 7 & $-$4.30 $\pm$ 0.37 \\
Big 11              & 9 & $-$4.36 $\pm$ 0.86 \\
PAH-like            & 6 & $-$4.70 $\pm$ 0.16 & excluding IRAS 00350\\
21 \mum\ (Class D2) & 6 & $-$4.55 $\pm$ 0.25 \\
21 \mum\ (Class D1) & 3 & $-$4.90 $\pm$ 0.20 \\
21 \mum\ (Class B)  & 2 & $-$5.12 $\pm$ 0.07 & 
\enddata
\end{deluxetable}

Table~\ref{t.colopt} also includes absolute bolometric
magnitudes for the sample, which are estimated as described
in Appendix~2.  All of the sources have bolometric 
magnitudes between $-$3.3 (SMP SMC 099) and $-$5.8 (SMP SMC 
011), with the exception of IRAS~00350, which has $M_{\rm 
bol}$ = $-$7.1, right at the AGB limit.  Table~\ref{t.mbol} 
gives the mean bolometric magnitude for each spectral
group (excluding IRAS~00350).  The overlap betwen the
Fullerene, Mixed, and Big-11 groups is complete, suggesting
a common origin, although the scatter in the Big-11 sources
is substantial.  The PAH-like and 21~\mum\ groups are 
brighter, and there is enough separation between the 21~\mum\
subgroups to suggest that the Class D2 spectra come from a
fainter population.

\subsection{Optical spectroscopy} 

\begin{figure} 
\includegraphics[width=3.4in]{\figpath 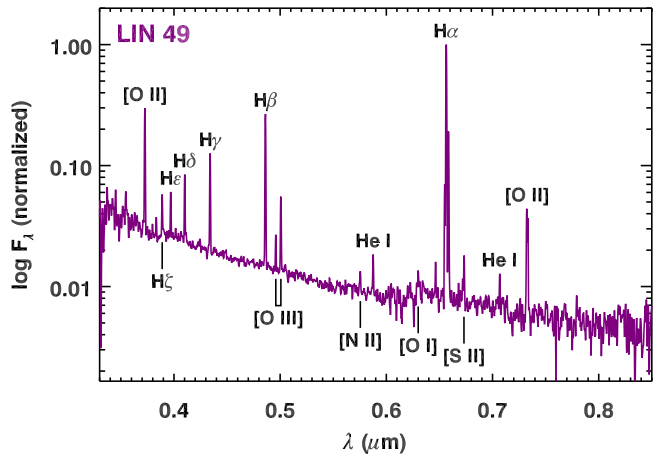}
\caption{An optical spectrum of LIN 49, showing the Balmer
series of hydrogen recombination lines, He I emission lines,
and forbidden lines from [O II], [O III], [N II], and 
[S II].\label{f.lin49}}
\end{figure}

Figure~\ref{f.lin49} shows the optical spectrum for LIN~49,
taken with the Double-Beam Spectrograph on the 2.3 m 
telescope at Siding Springs Observatory on 2008 December 17.
The spectrum shows the Balmer series of hydrogen 
recombination lines and forbidden lines from [O II] at 3727 
and 7323~\AA\ and [O III] at 4958 and 5007~\AA.  Weaker lines 
are also apparent from He I at 5876 and 7065~\AA, [N II] at 
5755~\AA, and [S II] at 6731~\AA.  These relatively 
low-excitation lines indicate that LIN~49 is a relatively 
cool PN, much like the other fullerene sources.

\subsection{Optical variability} 

In the final stages of their evolution on the AGB, stars are
associated with strong pulsational variability as they eject 
their envelopes and evolve into white dwarfs.  One might 
expect some variability in the embedded post-AGB sources in 
our sample, where the central star still has a significant 
envelope and is relatively cool.

A search for sources in our sample in the OGLE-III survey of
the Magellanic Clouds \citep{sos09,sos11} reveals five 
variables.  Four of the five, J004441, IRAS~F05192, J052043, 
and IRAS~Z05259 are 21~\mum\ sources with Class D2 PAH
emission.  The first three are semi-regular variables; the
last is a small-amplitude variable.  The fifth variable
is IRAS~00350, the exceptional source in the PAH-like group.
It is classified as a small-amplitude variable.

The EROS-2 LMC survey uncovered two additional variables, 
both 21~\mum\ sources, IRAS~05092, and IRAS~05413 
\citep{spa11}.  Multiple periods are given for IRAS~05092, 
making it a likely semi-regular.  It is a 21-\mum\ source 
with Class D1 PAH emission.  IRAS~05413 is a PAH-like source, 
and its variability was not characterized.

Both the OGLE and EROS surveys include a Mira only 3\farcs6
away from SMP LMC 048, but this is a different source.

Surprisingly, the four 21~\mum\ variables detected in the
OGLE survey are classified as oxygen-rich, based on their 
Wesenheit index, which is insensitive to reddening and
usually separates carbon-rich and oxygen-rich AGB stars 
reliably.  However, optical spectroscopy of J004441 shows 
that it is clearly a carbon-rich object \citep{des12}, 
casting some doubt on the applicability of the Wesenheit 
index to post-AGB objects.

The variables include three semi-regulars and one 
small-amplitude variable among the Class D2 sources,
one semi-regular among the Class D1 sources, and two
small-amplitude or uncharacterized variables among the
PAH-like sources.  Thus three of the six Class D2 sources
are semi-regular variables, compared to one of the five other
21~\mum\ sources.  This is most likely a selection effect 
caused by the differences in dust extinction in the sources.  
In the Class D2 sources, we can see deeper into the system, 
making the variability easier to detect.

\section{Discussion} 

\subsection{Notes on individual sources} 

Spectral classification involves placing sources which show a
continuous distribution of spectral properties into discrete
bins, usually with the constraint that the number of bins
should be minimized.  Successful classification can
reveal trends in the properties of the sources not otherwise
apparent, which the classification presented here has 
accomplished.  However, a number of sources resisted our 
efforts to pigeon-hole them, either because they are at the 
boundary between two groups, or because they show unusual 
characteristics which do not fit the average properties of
any bin.

Several of the transition sources have been pointed out 
above.  The Mixed source SMP SMC~013 could be placed in the 
Fullerene group, and the Big-11 source SMP LMC~085 could 
belong with the Mixed sources.  As noted in 
Section~\ref{s.new21}, the PAH-like source IRAS~05063 is 
almost a 21~\mum\ source, and IRAS~05073 may be similar.

IRAS~05588 is classified in the Big-11 group despite the
absence of an obvious big-11 feature because it shows the
associated strong emission shoulder at $\sim$18~\mum.  This
approach is analogous to the classification of some sources 
by \cite{kra02} as 21~\mum\ sources because their spectra 
included the associated spectral features but not the 
21~\mum\ feature itself.  However, this source is much cooler 
than the other three Big-11 sources.  Its inclusion in any 
group should not take away from its unique spectrum.

The Big-11 source SMP SMC~020 has an unusual spectrum
dominated by strong broad features with peaks at $\sim$11
and $\sim$19~\mum.  If this spectrum were redshifted silicate
emission, such as has been seen in active galactic nuclei
\citep[e.g.][]{hao05,hon11}, then the 10~\mum\ silicate 
feature (rest wavelength) would be more asymmetric, with a 
sharper rise on the blue side than the red, and the 18~\mum\ 
silicate feature would have to be at 20~\mum\ to match the 
shift at 10~\mum.  The complete lack of a shift in the 
[Ne III] line at 15.6~\mum\ is the nail in the coffin for 
this scenario.  Instead, we have a spectrum showing more 
clearly than any other the 18~\mum\ shoulder associated with 
the big-11 feature.

SMP SMC~006 in the PAH-like group differs in some ways from
the rest of the group.  Its spectral structure at 
11--14~\mum\ fits the rest of the group well, but its 
spectrum also shows an inflection at 18~\mum\ that could be
the shoulder associated with the big-11 feature.

IRAS~00350 is another exception among the PAH-like sources.
First, its spectrum shows nearly normal PAH emission in both 
the 6--9 and 11--14~\mum\ ranges.  Only the width of the 
11.3~\mum\ feature is consistent with Class D1 emission.  
Second, it is 2.4 mag brighter than the mean for the 
remaining PAH-like sources and 1.3 mag brighter than the next 
brightest source in our sample.  Third, its infrared colors 
place it close to or in the range of colors covered by carbon 
stars.  \cite{whi89} first suggested that IRAS~00350 may be 
an interacting binary, although they also noted that it could 
be a post-AGB object evolving to a PN.  The combination of 
its unusual luminosity, colors, and carbon-rich nature are 
suggestive that it is a symbiotic carbon star (Whitelock, 
private communication, 2013)

Two non-Fullerene sources show unusually blue $B-V$ colors,
the PAH-like source IRAS~05127 and the 21~\mum\ source
J010546.  IRAS~05127 is the only Class A PAH source among
the PAH-like spectra, and its blue colors in both $B-V$ and
$U-B$ indicate less dust extinction to the central nebula, 
where the PAHs may be more processed.  

J010546 has a unique spectrum which is difficult to classify,
as noted by \cite{vol11}.  Its infrared spectrum shows a 
strong jump from 10 to 11~\mum\ reminiscent of the larger
jump in the spectrum of the T Tauri star CoKu Tau~4
\citep{for04}.  The latter spectrum arises from a
circumstellar disk with an inner gap.  While one must be
careful about invoking disk geometries, the parallels are
interesting.  In the optical, J010546 is bluer than the
other Class D2 sources by 0.8 mag in $B-V$ and 0.6 mag in 
$U-B$, and such a relatively clear line of sight in a 
relatively red source (at longer wavelengths) is consistent 
with a disk or torus-like geometry.

We included SMP SMC~025 in the sample because the 11~\mum\ 
feature might have been SiC, but this source is oxygen-rich.  
The spectrum shows emission peaks at 18--19, 23--24, 27--28, 
and 33~\mum, and these reveal the presence of crystalline 
silicates.  The shape of the spectrum in the 25--35~\mum\ 
region is too structured to be consistent with any of the MgS 
features in the other spectra.  Furthermore, the feature at 
$\sim$11~\mum\ is bluer and broader than the big-11 features.  
Silicate dust would normally show a peak at 10~\mum\ if it 
were amorphous, or structure throughout the 10--11~\mum\ 
region if it were crystalline.  While the longer-wavelength 
features can be ascribed to crystalline silicates with 
confidence, the nature of the 11~\mum\ feature remains 
unclear.  

\subsection{Coated grains and SiC} 
\label{s.coatings}

Section~\ref{s.mgs} explained that grains coated with MgS
would produce the observed 26-30~\mum\ feature while staying
within the abundance limits.  This mechanism may also solve a 
similar problem with SiC as the carrier of the big-11 feature 
first noted by \cite{jbs09}.  Comparisons of carbon stars in 
the Galaxy, LMC, and SMC show that more metal-poor carbon 
stars have less SiC in their spectra \citep[e.g.][and 
references therein]{slo12}.  This relation is likely driven 
by the abundance of Si.  How then can the big-11 features be 
so strong in Magellanic post-AGB objects if the strength of 
the SiC feature is limited by abundances on the AGB?

Grains coated with SiC would produce strong big-11 features,
as observed, despite the lower amounts of available Si.  This 
scenario is consistent with observations of how the strength 
of the SiC feature behaves with increasing mass-loss rate for 
stars of different metallicities.  In the more metal-rich 
Galactic sample, SiC forms first, followed by a layer of 
amorphous carbon and finally, once the dust shell is 
sufficiently optically thick, MgS \citep[see Figure~10 by][as 
an example]{slo12}.  However in the more metal-poor 
Magellanic samples, the SiC strength rises more slowly as the 
dust shells thicken and redden, due presumably to the lower 
abundance of Si.  \cite{lag07} and \cite{lei08} both 
suggested that the metal-poor case is consistent with a layer 
of SiC-rich material around a core of amorphous carbon.

As a star evolves off the AGB and sheds its envelope,
circumstellar dust grains are exposed to an increasingly 
harsh radiation field.  SiC has a higher condensation 
temperature than amorphous carbon, and increased 
photo-processing of the grains would preferentially
remove solid carbon from the outer layers of the grains, 
making the grains more SiC-rich at their surface and
producing stronger SiC features in the dust spectrum.
The condensation sequence in the Galaxy, with SiC grains 
forming first and coatings of amorphous carbon second, would 
explain why the spectra of carbon-rich PNe in the Galaxy 
rarely show SiC dust features \citep{cas01}.  The entire 
mantle would have to be removed to expose the SiC core.


The origin of the 18~\mum\ shoulder associated with the
big-11 feature remains unknown.  With the exception of
SMP SMC~025, none of our sources show any evidence of
oxygen-rich dust at any other wavelength, making it unlikely 
that the 18~\mum\ shoulder could arise from silicates.  In 
the more intense radiation fields expected in post-AGB
environments, some processing of the CO gas is expected,
and while silicates could conceivably form from the freed
oxygen, the lack of a 10~\mum\ silicate emission feature
would be difficult to explain.


\subsection{The Fullerene, Mixed, and Big-11 spectra} 

Compared to the general population of PNe associated with
carbon-rich dust, the properties of the Fullerene, Mixed, 
and Big-11 groups roughly follow two progressions.   First, 
we see a temperature progression, with the fullerenes
appearing only in the vicinity of central stars with 
effective temperatures of $\sim$25,000--40,000 K, and the 
big-11 feature at temperatures up to $\sim$ 70,000 K.  
Second, we see an optical depth progression, with the bluest 
$UBV$ colors associated with the Fullerene group, which is
defined to be those fullerene sources with no PAH emission in 
the 6--9~\mum\ range.  The Mixed spectra, which show both 
fullerene and PAH features, are more enshrouded than the 
Fullerene group, both in the optical and at longer 
wavelengths, like $[8]-[24]$.  However, the photometric data 
complicate this scenario.

The SEDs of the Mixed and Big-11 groups are quite similar
and show substantial overlap in all colors considered.  How
can one group have fullerene emission in its spectra
and not the other?  Invoking a disk or torus and differences
in inclination between the two groups might help, but that
is inconsistent with the similarities in the SEDs and 
photometric colors in the two groups.  Explaining the
differences between the Fullerene and Mixed groups is also
a challenge.  A pole-on inclination would give a direct line 
of sight to the central star and its immediate environment,
explaining the presence of fullerene emission and the blue
optical colors of the Fullerene group, but our large beam 
would also include the surrounding disk which presumably 
would show the PAH emission evident in the Mixed group.

It is also difficult to explain why the Fullerene group has 
such tight distributions in infrared colors, starting with 
$K_s-[3.6]$ and continuing to $[8]-[24]$.  The Mixed group, 
which is also composed of fullerene sources, shows broader
color distributions which overlap the Fullerene group.  It is 
the combination of the presence of fullerenes and absence of 
PAHs that somehow constrains the possible shape of the SED 
from 2 to 24~\mum.

In the sequence from Big-11 to Mixed to Fullerene, the
center of the 26--30~\mum\ feature shifts from 30.7 to 
29.9~\mum.  If MgS is the carrier, then this shift
corresponds to a rise in temperature of the MgS dust from
200 to 250 K, and its temperature traces a relatively 
shielded or more distant dust component.  The optical colors 
of the Fullerene group suggest less optically thick dust, 
which would lead to higher MgS temperatures.  The lower MgS 
temperatures of the Big-11 group imply more optically thick 
dust, but as already noted the broad-band colors of the Mixed 
and Big-11 groups show little difference.

Several authors have suggested that fullerenes result from 
the photo-processing of larger grains, either large PAHs 
\citep{ber12} or grains or clusters composed of mixtures of 
aromatic and aliphatic hydrocarbons \citep{gh12,mic12,ots13}.
For the cooler PNe, fullerenes appear to be the final
survivor of the hydrocarbon mix, but in PNe with hotter
central stars, if the dust is carbon-rich, we either see 
PAHs or no solid-state component at all.  In these harsher
radiation fields, the fullerenes either cannot form as the
parent hydrocarbon population is destroyed, or they cannot
survive long enough to be detected.


\subsection{The 21~\mum\ and PAH-like spectra} 


The 21~\mum\ and PAH-like groups are more embedded in their
own dust than the Fullerene, Mixed, and Big-11 groups, and
their central stars are even cooler.  Together these point to 
an earlier evolutionary phase closer to the AGB.  Their dust
properties are suggestive of less processed carbonaceous
material more like amorphous carbon than the classic PAHs,
fullerenes, and possibly SiC-coated grains seen in more
evolved sources.

Spectra with the 21~\mum\ feature almost always show PAH
or PAH-related emission features, a point first noted by
\cite{kwo89}.  This association suggests that the carrier of 
the 21~\mum\ feature is either a hydrocarbon itself, or 
closely related.  \cite{jus96} raised this possibility, 
based on their study of Galactic 21~\mum\ sources, and a 
more recent study based on IRS spectra of Galactic 
sources reached a similar conclusion \citep{cer11}.  
\cite{hil98} showed that nano-diamonds could be the 
responsible hydrocarbon, based on fits to the shape of the 
emission feature.

Our spectral sample reveals that over half of the 21~\mum\ 
sources are associated with the newly defined Class D2 PAH 
emission at 11--14~\mum, and all but three show Class D PAH 
structure at 7--9~\mum.  The 6.9~\mum\ feature from aliphatic 
hydrocarbons appears in all of the 21~\mum\ spectra, in some 
cases quite strongly, and in over half the 21~\mum\ spectra, 
the 7.3~\mum\ aliphatic feature is visible as well.  

The spectra of all of the 21~\mum\ sources include an 
emission complex at 16~\mum, which we have separated into
components at 15.8 and 17.1~\mum.  The 15.8~\mum\ feature
aligns well with a C--H mode in a terminal C$\equiv$CH bond 
on a longer aliphatic chain, and it further strengthens the 
connection of the 21~\mum\ feature to aliphatic hydrocarbons.  
The 17.1~\mum\ feature may also arise from aliphatic chains
with alkyne bonds, but the supporting evidence is weaker.
To produce triple bonds in aliphatics chains, it is necessary 
to partially dehydrogenate the methyl and methylene groups
at the ends and in the middle of the chains, respectively.
This suggests that the aliphatic component of the hydrocarbon 
mix is being photo-processed.

The carrier of 21~\mum\ feature remains unknown, but the case 
for a hydrocarbon-based carrier is growing, even if it still 
circumstantial.  Some caution is advisable.  Numerous 
candidates for the 21~\mum\ carrier have been proposed over 
the years.  \cite{zha09a} examined several inorganic 
candidates and ruled out some because they produced 
additional spectral features which were not observed and some 
because they violated abundance constraints.  The latter 
group included TiC, fullerenes with Ti atoms, SiS$_2$ and 
unusual forms of SiC.  Given the possibility that coated 
grains could bypass the abundance constraints, some of these 
candidates may still be viable.

\subsection{Non-spherical geometries} 

Figure~\ref{f.sed} shows that the 21~\mum\ sources associated
with Class B or Class D1 PAHs are the most embedded sources
in our sample (after the Red sources).  The Class D2 sources
differ, in that they have more warm dust, indicating a 
clearer line of sight through the circumstellar material 
toward the center of the nebula.  This could result from 
either a thinner spherical shell or an inclination effect in 
a disk or torus-like geometry.  In the latter case, the Class 
D2 sources are more pole-on.  The other sources could be 
either more edge-on or less asymmetric.

Invoking a disk-like geometry might also help explain the 
origin of the 21~\mum\ feature and its associated dust 
features.  If the dust were trapped in a disk in these 
systems instead of simply flowing outward, then it would 
remain near the central sources and be subject to more 
intense photo-processing for a longer period.  An outflow 
velocity of only 5~km/s corresponds to outward motion of 
1~AU/yr.\footnote{A 10~km/s outflow velocity is a common 
assumption \citep[e.g.]{gro09}; we have halved that value to 
be conservative.}  Thus the outflow timescales are short even
compared to the short evolutionary timescales from the AGB to 
the PN stage.  Perhaps the carrier of the 21~\mum\ feature is 
an unusual hydrocarbon only produced when hydrocarbons are 
trapped in a disk in a post-AGB system.
  
The 26--30~\mum\ feature in the PAH-like and 21~\mum\ spectra 
has a central wavelength between 27.9 and 28.9~\mum, which is
shifted significantly from the position seen in the
Fullerene-Big-11 family, 29.9--30.7~\mum.  The corresponding 
MgS temperatures range from 350 to 400 K.  The Class D2 
sources are in the middle of the range for the PAH-like and
21~\mum\ family.  If the SEDs are indicating a clearer line 
of sight to the center of Class D2 sources, then one would 
expect them to have warmer dust, and while the near-IR colors 
reflect that, the MgS temperatures do not.



The spectra in this sample are not always self-consistent.
For most of the trends we have uncovered, there exists at
least one spectrum that is an exception.  The coming and
going of the 6.9~\mum\ aliphatic feature is a good
example.  These spectral disparities can be explained by
complex geometries, either disks or tori, and possibly 
clumping as well.  The result would be multiple regions in
our single beam, each with its own physical conditions and 
spectral signature.  If a disk or torus is viewed 
edge-on, then extinction will block our view of the inner 
disk and central star more effectively at shorter 
wavelengths.  Conversely, if scattering gives us an indirect 
view of the interior, that view is better at shorter 
wavelengths.  Follow-up polarimetric observations would test 
these hypotheses and clarify the geometry.

%
%


\section{Summary} 

We have investigated a sample of carbon-rich post-AGB
objects in the Magellanic Clouds by first classifying the
mid-IR spectra into self-consistent groups and then 
investigating their ancillary properties.  This approach had 
led to a number of findings.
\begin{itemize}
\item The width and position of the narrow emission feature 
at $\sim$18.8~\mum\ can distinguish between fullerene 
emission at 18.9~\mum\ and [S III] emission at 18.7~\mum.
\item Combining this information with an assessment of the
17.4~\mum\ fullerene feature and the structure of the
6--9~\mum\ region can determine which spectra show fullerenes
and which do not.  This analysis adds three sources to the
family of fullerene emitters:  LIN~49, SMP LMC~008, and SMP
SMC~001.  It also also removes three previously identified 
fullerene candidates:  SMP SMC~020, SMP SMC~027, and SMP 
LMC~099.
\item We confirm that the big-11 feature is due primarily to
SiC dust.  In order to stay within abundance constraints, it
is likely that the SiC is coating grains consisting of 
amorphous carbon.
\item We have divided the sources showing Class D PAH
emission at 7--9~\mum\ into two groups based on their
spectra at 11--14~\mum.  Class D1 spectra show an unusual
profile with a strong 12.0~\mum\ feature compared to the
12.7~\mum\ feature.  In Class D2, the features in the
11--14~\mum\ range are shifted to $\sim$11.4, 12.4, and
13.2~\mum.
\item All of the PAH-like and some of the 21~\mum\ spectra
show Class D1 PAH emission.  Removing SiC from the big-11
feature leaves a residual that resembles Class D1 PAHs.
Not all Class D1 PAHs are associated with Class D PAHs 
at 6--9~\mum.
\item Over half of the 21~\mum\ sources show Class D2 PAH
emission at 11--14~\mum, and all of these show Class D PAHs
at 6--9~\mum.  Most of these sources are also associated with
strong aliphatic emission at 6.9~\mum.
\item The 21~\mum\ sources can show Class B, D1, or D2 PAH
emission.  The Class D2 PAH sources show the least 
extinguished line of sight to the central source.
\item The 15.8~\mum\ feature associated with 21~\mum\ 
sources is fitted well with laboratory data from aliphatic
hydrocarbon chains terminated with an alkyne CH group.  The
carrier of the 17.1~\mum\ feature may be related.
\item We attribute the 26--30~\mum\ feature to MgS.  This
feature shifts from a central wavelength of $\sim$30 to 
$\sim$28~\mum\ in the sequence Big-11 --- Mixed --- Fullerene 
--- 21~\mum, consistent with a shift in temperature of MgS
from $\sim$200 to $\sim$400 K
\item The Fullerene group shows the bluest $UBV$ colors in
our sample, indicating a relatively clear line of sight to 
the $\sim$30,000~K stars at their center.  From $K$ to [24],
their infrared colors exhibit a remarkably narrow
distribution.  The colors of the Mixed group overlap those
of the Fullerene group, but show a broader range.
\item The optical spectrum of LIN 49, one of the newly
detected fullerene sources, reveals a low-excitation PN and
a cool central star, which is consistent with the other 
Fullerene sources.
\end{itemize}

The Fullerene, Mixed, and Big-11 groups form one family.
They show fullerenes, mixtures of fullerenes and PAHs, or
just PAHs in their spectra (respectively).  Most show the 
big-11 and 26--30~\mum\ features from what we believe to 
be grains coated with SiC and MgS, and the MgS feature
typically peaks to the red of 29~\mum, indicating lower
dust temperatures.  Previous studies have shown that
fullerene emission is limited to those PNe with relatively
cool central stars, and we confirm that result and extend it 
to the Big-11 spectra.  The most significant new clue is the 
relatively clear view of the central star in the Fullerene 
group.

The 21~\mum\ and PAH-like groups also form a family.  Two
of the PAH-like sources may in fact be 21~\mum\ sources.
All of the 21~\mum\ spectra and half of the PAH-like group
show clear evidence of aliphatic hydrocarbons in their
spectra, and most show unusual Class D1 or D2 PAH emission 
profiles, with the Class D2 spectra associated with clearer
lines of sight to the central nebula.  Our identification of 
the 15.8~\mum\ feature, which usually accompanies the
21~\mum\ feature, with alkyne bonds in aliphatic chains, adds 
to the growing body of circumstantial evidence that the 
21~\mum\ feature is related to hydrocarbons.  Aliphatics
suggest less photo-processing, which is consistent with the
earlier evolutionary phase suggested by the cooler central
stars and thicker dust shells of the 21~\mum\ sources.

\vspace{0.3cm}

\acknowledgements

We thank the referee, A.~P.\ Jones, for a thorough report
which led to substantial improvements in this paper.
We are grateful to P.~R.\ Wood for providing the spectrum of 
LIN~49.  G.~C.~S.\ was supported by NASA through Contract
Number 1257184 issued by the Jet Propulsion Laboratory,
California Institute of Technology under NASA contract 1407.
F.~K.\ received support from the National Science Council of
Taiwan, grant NSC100-2112-M-001-023-MY3.
This research relied on the following resources:
NASA's Astrophysics Data System, the Infrared Science Archive 
at the Infrared Processing and Analysis Center, operated 
by JPL, and the SIMBAD and VizieR databases, operated at the 
Centre de Donn\'{e}es astronomiques de Strasbourg.

\section*{Appendix 1  Spectral analysis}  

\begin{deluxetable}{lcc} 
\tablecolumns{3}
\tablewidth{0pt}
\tablenum{10}
\tablecaption{Continuum-fitting wavelengths}
\label{t.wave}
\tablehead{
  \colhead{ } & \colhead{Blue continuum} & \colhead{Red continuum} \\
  \colhead{Feature} & \colhead{(\mum)} & \colhead{(\mum)}
}
\startdata
6--9~\mum\ combined flux    &  5.90--5.99  &  9.00--9.12 \\
\\
6.2~\mum\ PAH               &  5.90--5.96  &  6.56--6.62 \\
6.9~\mum\ aliphatic         &  6.68--6.75  &  7.05--7.11 \\
7.3~\mum\ aliphatic         &  7.11--7.17  &  7.38--7.44 \\
8.6~\mum\ PAH               &  8.21--8.33  &  8.82--8.94 \\
7--9 \mum\ base PAH feature &  6.65--6.74  &  9.00--9.12 \\
\\
6.5~\mum\ fullerene-related &  6.29--6.35  &  6.74--6.80 \\
7.0~\mum\ fullerene         &  6.80--6.86  &  7.26--7.32 \\
7.6~\mum\ fullerene-related &  7.29--7.38  &  7.90--8.03 \\
8.1~\mum\ fullerene-related &  7.84--7.97  &  8.21--8.33 \\
8.5~\mum\ fullerene         &  8.21--8.33  &  8.82--8.94 \\
\\
17.4~\mum\ fullerene        & 16.82--16.98 & 17.75--17.92 \\
18.7~\mum\ [S III]          & 18.34--18.52 & 19.02--19.20 \\
18.9~\mum\ fullerene        & 18.17--18.34 & 19.45--19.61 \\
18.7--18.9~\mum\ gaussian   & 18.25--18.43 & 19.28--19.45
\enddata
\end{deluxetable}

Our spectral analysis follows a standard sequence.  We first 
simplified the overall structure in the spectra by fitting 
cubic splines and subtracting them.  We chose the spline 
points so that the fitted continuum would pass through minima 
in the spectra; therefore, the spline points vary from one 
spectrum to the next.  They are visible as diamonds in 
Figures~\ref{f.spfull} to \ref{f.spred}.  In cases where 
forbidden lines would interfere with estimates of the 
continuum around spectral features of interest, we replaced 
them with lines interpolated from the spectrum to either side 
before measuring feature strengths and positions.  
We removed the following lines:  [Ar III] at 8.99~\mum, 
[S IV] at 10.52~\mum, [Ne II] at 12.81~\mum, [Ne III] at 
15.56~\mum, [S III] at 33.5~\mum, and [Si II] at 34.8~\mum.  

To extract central wavelengths and feature strengths we 
estimated a continuum by fitting line segments to the 
wavelengths given in Table~\ref{t.wave} and integrating the 
flux density in between.  The wavelengths used to fit line
segments were fixed for all spectra, except the 21~\mum\
sources (Appendix 1.3).  The central wavelength $\lambda_C$ 
is defined as the position in the spectrum that bisects the 
integrated flux from the feature.  Fitting a linear continuum 
is essential for accurate central wavelengths because they 
are sensitive to any residual slopes in the spectrum.  The 
main purpose of the spline is to remove the curvature 
underneath the broader features to facilitate the fitting of 
line segments.  

The use of line segments when measuring the PAH features
assumes the presence of PAH plateaus, which have long been 
hypothesized to arise from larger PAH clusters and underlie 
the emission features \citep[e.g.][]{atb89,bre89}.  Whether 
or not this assumption is correct, the line-segment method 
generates central wavelengths which are easily measured and 
reproduced, and which can be compared to similarly determined 
values from previous (and future) work.  We rely on these 
wavelengths to identify and track the various spectral 
features.  For the most part, our measurements of
feature strengths first serve to check the validity of the
central wavelength, and we have included them in the 
following tables even when we use them for only that purpose.

Other methods could also be used, most notably fitting of
gaussians or other assumed feature profiles, from which the
apparent plateaus can be built.  Central wavelengths found
using this method depend on the shape of the adopted profile
shape, and because the method assumes that the plateaus 
do not exist, it forces the creation of a series of spectral
components which can modify the apparent positions of the
stronger features on which we are focused.

\subsection*{Appendix 1.1.  17--19~\mum}  

\begin{deluxetable*}{lcccccc} 
\tablecolumns{7}
\tablewidth{0pt}
\tablenum{11}
\tablecaption{Fullerene features at 17.4 and 18.9~\mum}
\label{t.fulldat}
\tablehead{
  \colhead{ } & \multicolumn{2}{c}{17.4~\mum\ fullerene feature} &
                \multicolumn{2}{c}{18.9~\mum\ fullerene feature} &
                \multicolumn{2}{c}{Gaussian fit to 18.9~\mum} \\
  \colhead{Source} & \colhead{$\lambda_C$(\mum)} &
  \colhead{$F$ (10$^{-18}$ W m$^{-2}$)} &
  \colhead{$\lambda_C$ (\mum)} &
  \colhead{$F$ (10$^{-18}$ W m$^{-2}$)} &
  \colhead{$\lambda_0$ (\mum)} & \colhead{FWHM (\mum)}
}
\startdata
SMP LMC 002 & 17.44 $\pm$ 0.05 & 35.4 $\pm$  4.7 & 18.88 $\pm$ 0.03 & 49.6 $\pm$  4.3 & 18.85   & 0.32 \\
SMP LMC 056 & 17.39 $\pm$ 0.11 & 22.4 $\pm$  6.7 & 18.86 $\pm$ 0.02 & 80.7 $\pm$  2.8 & 18.86   & 0.41 \\
SMP SMC 024 & 17.38 $\pm$ 0.08 & 18.8 $\pm$  4.2 & 18.84 $\pm$ 0.03 & 63.6 $\pm$  4.3 & 18.82   & 0.38 \\
LIN 49      & 17.41 $\pm$ 0.08 & 12.1 $\pm$  2.4 & 18.93 $\pm$ 0.02 & 28.8 $\pm$  1.3 & 18.93   & 0.40 \\
SMP SMC 016 & 17.44 $\pm$ 0.13 & 18.4 $\pm$  5.2 & 18.90 $\pm$ 0.02 & 74.4 $\pm$  2.8 & 18.90   & 0.45 \\
\\
SMP SMC 013 & 17.41 $\pm$ 0.10 &  8.6 $\pm$  2.4 & 18.82 $\pm$ 0.06 & 37.8 $\pm$  4.5 & 18.83   & 0.47 \\
SMP SMC 018 & 17.37 $\pm$ 0.18 & 10.7 $\pm$  4.1 & 18.86 $\pm$ 0.03 & 66.9 $\pm$  4.0 & 18.85   & 0.38 \\
SMP SMC 015 & \nodata          & 12.3 $\pm$ 12.1 & 18.84 $\pm$ 0.15 & 32.5 $\pm$ 10.9 & 18.83   & 0.31 \\
SMP SMC 001 & \nodata          &  9.5 $\pm$  5.1 & 18.88 $\pm$ 0.08 & 70.6 $\pm$  9.4 & 18.89   & 0.55 \\
SMP LMC 008 & \nodata          & 22.4 $\pm$ 18.8 & 18.86 $\pm$ 0.08 & 59.6 $\pm$  7.4 & 18.81   & 0.38 \\
SMP LMC 048 & \nodata          &  6.1 $\pm$  6.9 & 18.78 $\pm$ 0.05 & 42.0 $\pm$  7.7 & 18.76   & 0.22 \\
SMP LMC 025 & 17.23 $\pm$ 0.18 & 13.1 $\pm$  5.6 & 18.79 $\pm$ 0.01 & 90.8 $\pm$  3.3 & 18.77   & 0.28 \\
\\
SMP LMC 085 & \nodata          & 23.4 $\pm$ 16.4 & \nodata          & 24.4 $\pm$ 12.8 & \nodata & \nodata \\
SMP LMC 076 & \nodata          &  7.0 $\pm$  6.0 & 18.78 $\pm$ 0.08 & 25.8 $\pm$  6.2 & 18.76   & 0.23 \\
SMP LMC 058 & 17.42 $\pm$ 0.21 & 20.8 $\pm$  6.6 & \nodata          & \nodata         & \nodata & \nodata \\
SMP SMC 027 & 17.48 $\pm$ 0.16 & 55.9 $\pm$  2.2 & 18.75 $\pm$ 0.04 & 14.4 $\pm$  2.6 & 18.76   & 0.18 \\
SMP LMC 051 & 17.21 $\pm$ 0.11 & 16.3 $\pm$  5.1 & 18.86 $\pm$ 0.21 & 21.5 $\pm$  7.6 & 18.79   & 0.30 \\
IRAS 05370  & \nodata          & \nodata         & \nodata          & 14.0 $\pm$ 13.2 & \nodata & \nodata \\
IRAS 05537  & \nodata          & 11.1 $\pm$  8.8 & \nodata          & 27.8 $\pm$ 12.8 & \nodata & \nodata \\
SMP SMC 020 & \nodata          &  0.8 $\pm$  5.8 & \nodata          &  7.0 $\pm$  3.9 & \nodata & \nodata \\
IRAS 05588  & \nodata          &  6.9 $\pm$ 13.9 & \nodata          & \nodata         & \nodata & \nodata 
\enddata
\end{deluxetable*}

\begin{deluxetable*}{lcccc} 
\tablecolumns{5}
\tablewidth{0pt}
\tablenum{12}
\tablecaption{[S III] emission in the PNe sample}
\label{t.siiidat}
\tablehead{
  \colhead{ } & \multicolumn{2}{c}{[S III] emission line} &
                \multicolumn{2}{c}{Gaussian fit to 18.9~\mum} \\
  \colhead{Source} & \colhead{$\lambda_C$ (\mum)} & 
  \colhead{$F$ (10$^{-18}$ W m$^{-2}$)} &
  \colhead{$\lambda_0$ (\mum)} & \colhead{FWHM (\mum)}
}
\startdata

SMP SMC 005 & 18.74 $\pm$ 0.04 & 28.2 $\pm$  5.0 & 18.73 & 0.19 \\
SMP SMC 008 & 18.70 $\pm$ 0.04 &  8.0 $\pm$  2.2 & 18.72 & 0.15 \\
SMP LMC 004 & 18.74 $\pm$ 0.09 &  9.8 $\pm$  3.4 & 18.73 & 0.18 \\
SMP LMC 009 & 18.76 $\pm$ 0.07 & 17.8 $\pm$  5.4 & 18.76 & 0.09 \\
SMP LMC 010 & 18.73 $\pm$ 0.04 & 15.7 $\pm$  4.9 & 18.73 & 0.12 \\
SMP LMC 016 & 18.73 $\pm$ 0.03 & 26.3 $\pm$  5.4 & 18.74 & 0.15 \\
SMP LMC 019 & 18.72 $\pm$ 0.03 & 38.5 $\pm$  9.8 & 18.73 & 0.13 \\
SMP LMC 021 & 18.73 $\pm$ 0.02 & 87.9 $\pm$ 13.1 & 18.73 & 0.15 \\
SMP LMC 034 & 18.73 $\pm$ 0.04 & 22.3 $\pm$  5.6 & 18.75 & 0.16 \\
SMP LMC 038 & 18.75 $\pm$ 0.05 & 52.3 $\pm$ 11.5 & 18.73 & 0.14 \\
SMP LMC 040 & 18.68 $\pm$ 0.07 &  9.1 $\pm$  3.6 & 18.72 & 0.17 \\
SMP LMC 045 & 18.74 $\pm$ 0.03 & 66.2 $\pm$ 12.8 & 18.74 & 0.13 \\
SMP LMC 053 & 18.72 $\pm$ 0.04 & 36.6 $\pm$ 10.4 & 18.73 & 0.12 \\
SMP LMC 061 & 18.76 $\pm$ 0.03 & 70.3 $\pm$ 10.9 & 18.76 & 0.16 \\
SMP LMC 066 & 18.77 $\pm$ 0.07 &  7.0 $\pm$  2.8 & 18.76 & 0.08 \\
SMP LMC 062 & 18.72 $\pm$ 0.02 & 66.4 $\pm$  7.8 & 18.72 & 0.15 \\
SMP LMC 071 & 18.73 $\pm$ 0.03 & 34.3 $\pm$  6.3 & 18.73 & 0.15 \\
SMP LMC 079 & 18.74 $\pm$ 0.03 & 26.3 $\pm$  6.4 & 18.75 & 0.14 \\
SMP LMC 080 & 18.73 $\pm$ 0.02 & 24.7 $\pm$  2.9 & 18.73 & 0.16 \\
SMP LMC 087 & 18.73 $\pm$ 0.04 & 52.9 $\pm$ 11.8 & 18.74 & 0.13 \\
SMP LMC 095 & 18.72 $\pm$ 0.08 & 15.4 $\pm$  5.3 & 18.72 & 0.21 \\
SMP LMC 099 & 18.74 $\pm$ 0.03 & 59.7 $\pm$  9.3 & 18.74 & 0.16 \\
SMP LMC 100 & 18.72 $\pm$ 0.04 & 14.7 $\pm$  4.7 & 18.73 & 0.13
\enddata
\end{deluxetable*}

For the spectra with possible fullerene emission, we analyzed
the 17--19~\mum\ spectral region using our standard approach
and a second method fitting gaussians to deblend the [S III] 
and fullerenes at 18.7--18.9~\mum.  To better understand the
behavior of the [S III] line, we added a control sample of PNe
with [S III] lines detected at a SNR of 2.5~$\sigma$ or better,
but no fullerenes.

Table~\ref{t.fulldat} presents the results for the 
program sources, while Table~\ref{t.siiidat} presents the
comparison sample of PNe.  SMP LMC~099 appears in the latter 
table.  In this and all data tables, $F$ is the integrated 
flux from a feature, $\lambda_C$ is the central wavelength of 
the feature, and $\lambda_O$ is the centroid wavelength based 
on a fitted gaussian.  Wavelengths are not reported if the 
SNR of the extracted feature is less than 2.5.  The gaussian 
centroids ($\lambda_O$) generally differ slightly from 
the central wavelengths determined by integration due to 
noise in the spectra and asymmetries in the features.

The results of the gaussian test appear in 
Section~\ref{s.full18} and Figure~\ref{f.pos19}.  The
relation between the width and position of the blended 
feature is tight, with the unambiguous fullerene spectra 
showing broader, redder features and the PNe showing 
narrower, bluer features.  The PNe have a mean central 
wavelength of 18.73 $\pm$ 0.01~\mum\ and a FWHM of 
0.15 $\pm$ 0.03~\mum, compared to 18.87 $\pm$ 0.04~\mum\ and 
0.39 $\pm$ 0.05~\mum\ for the fullerene sources.\footnote{We 
attribute the shift of 0.02~\mum\ from the nominal position 
of the [S~III] feature to uncertainties in the wavelength 
calibration; the shift amounts to about a quarter pixel.}  
We used the results of the gaussian fitting to determine
which set of continuum wavelengths in Table~\ref{t.wave} to 
use when integrating the 18.9~\mum\ feature.

Even when fullerenes are present, the LL spectra do not have 
the resolution to deblend the 18.9~\mum\ feature from [S III] 
emission, which is most likely present.  Thus the ratios of 
the strengths of the 17.4 and 18.9~\mum\ features based on 
Table~\ref{t.fulldat} will be lower than those determined 
from higher-resolution data or from model calculations 
\citep[e.g.][]{jbs12}.

\subsection*{Appendix 1.2  6--9~\mum}  

\begin{deluxetable*}{lccrcrcccc} 
\tablecolumns{10}
\tablewidth{0pt}
\tablenum{13}
\tablecaption{Fullerene-related features at 6--9~\mum}
\label{t.full69}
\tablehead{
  \colhead{ } & \colhead{Total Feature} & \multicolumn{2}{c}{6.5~\mum\ Feature} & 
  \multicolumn{2}{c}{7.0~\mum\ Feature} & \multicolumn{2}{c}{7.6~\mum\ Feature} & 
  \multicolumn{2}{c}{8.5~\mum\ Feature} \\
  \colhead{ } & \colhead{Flux (6--9~\mum)} & \colhead{$\lambda_C$} & 
  \colhead{Flux/Total} & \colhead{$\lambda_C$} & \colhead{Flux/Total} & 
  \colhead{$\lambda_C$} & \colhead{Flux/Total} & \colhead{$\lambda_C$} & \colhead{Flux/Total} \\
  \colhead{Target} & \colhead{(10$^{-15}$ W m$^{-2}$)} & \colhead{(\mum)} & \colhead{(\%)} & \colhead{(\mum)} & 
  \colhead{(\%)} & \colhead{(\mum)} & \colhead{(\%)} & \colhead{(\mum)} & \colhead{(\%)}
}
\startdata
SMP LMC 002   & 0.33 $\pm$ 0.01 & \nodata          &     1.2 $\pm$ 0.7 & 7.03 $\pm$ 0.01 &   18.9 $\pm$  1.1 & 7.58 $\pm$ 0.02 &  8.9 $\pm$  0.5 & 8.51 $\pm$ 0.02 & 5.4 $\pm$ 0.4 \\
SMP LMC 056   & 0.44 $\pm$ 0.02 &  6.55 $\pm$ 0.05 &     9.9 $\pm$ 1.4 & 7.03 $\pm$ 0.01 &   24.9 $\pm$  1.7 & \nodata         &  3.5 $\pm$  1.5 & 8.50 $\pm$ 0.02 & 3.8 $\pm$ 0.4 \\ 
SMP SMC 024   & 0.57 $\pm$ 0.01 &  6.54 $\pm$ 0.03 &     3.6 $\pm$ 0.3 & 7.03 $\pm$ 0.01 &   11.1 $\pm$  0.7 & 7.58 $\pm$ 0.03 &  8.7 $\pm$  0.8 & 8.51 $\pm$ 0.04 & 2.9 $\pm$ 0.4 \\ 
LIN 49        & 0.59 $\pm$ 0.01 &  6.53 $\pm$ 0.02 &     8.6 $\pm$ 0.6 & 7.04 $\pm$ 0.01 &    8.1 $\pm$  0.5 & 7.57 $\pm$ 0.02 &  5.5 $\pm$  0.5 & 8.52 $\pm$ 0.15 & 1.6 $\pm$ 0.6 \\ 
SMP SMC 016   & 0.84 $\pm$ 0.02 &  6.54 $\pm$ 0.03 &     7.8 $\pm$ 1.1 & 7.05 $\pm$ 0.02 &    9.2 $\pm$  0.9 & 7.59 $\pm$ 0.06 &  5.9 $\pm$  0.9 & 8.52 $\pm$ 0.04 & 1.8 $\pm$ 0.4 \\ 
\\
SMP SMC 018   & 0.97 $\pm$ 0.04 &  6.54 $\pm$ 0.06 &     2.6 $\pm$ 0.8 & 7.04 $\pm$ 0.05 &    4.8 $\pm$  0.8 & 7.59 $\pm$ 0.03 &  9.5 $\pm$  0.8 & \nodata         & \nodata       \\
SMP SMC 015   & 0.60 $\pm$ 0.01 & \nodata          &     1.4 $\pm$ 0.8 & 7.06 $\pm$ 0.04 &    5.1 $\pm$  0.9 & 7.59 $\pm$ 0.02 & 15.0 $\pm$  0.8 & \nodata         & \nodata       \\
SMP LMC 008   & 0.84 $\pm$ 0.03 & \nodata          & $-$11.4 $\pm$ 1.8 & \nodata         & $-$2.4 $\pm$  0.5 & 7.57 $\pm$ 0.11 & 13.4 $\pm$  3.0 & \nodata         & \nodata       \\
SMP LMC 048   & 1.07 $\pm$ 0.05 & \nodata          &  $-$0.3 $\pm$ 2.4 & 7.05 $\pm$ 0.04 &    9.8 $\pm$  1.6 & 7.62 $\pm$ 0.05 & 10.8 $\pm$  1.8 & \nodata         & \nodata       \\
SMP LMC 025   & 0.93 $\pm$ 0.03 & \nodata          &  $-$1.7 $\pm$ 0.8 & 7.03 $\pm$ 0.07 &    2.4 $\pm$  0.7 & 7.58 $\pm$ 0.02 & 10.1 $\pm$  0.6 & \nodata         & \nodata       \\
\\
SMP SMC 027   & 0.04 $\pm$ 0.01 & \nodata          &    13.0 $\pm$ 7.2 & \nodata         &    4.8 $\pm$ 14.5 & 7.56 $\pm$ 0.10 & 43.4 $\pm$ 14.6 & \nodata         & \nodata       \\     
IRAS 05537    & 0.70 $\pm$ 0.07 & \nodata          & $-$16.2 $\pm$ 2.8 & \nodata         & $-$3.7 $\pm$  1.8 & 7.61 $\pm$ 0.04 & 29.3 $\pm$  2.9 & \nodata         & \nodata
\enddata
\end{deluxetable*}

\begin{deluxetable*}{lrcrcrcc} 
\tablecolumns{8}
\tablewidth{0pt}
\tablenum{14}
\tablecaption{PAH features at 6--9~\mum}
\label{t.pah69}
\tablehead{
  \colhead{ } & \colhead{Total Feature} & \multicolumn{2}{c}{6.2~\mum\ Feature} & 
  \multicolumn{2}{c}{8.6~\mum\ Feature} & \multicolumn{2}{c}{7--9~\mum\ Complex} \\
  \colhead{ } & \colhead{Flux (6--9~\mum)} & \colhead{$\lambda_C$} & \colhead{Flux/Total} & 
  \colhead{$\lambda_C$} & \colhead{Flux/Total} & \colhead{$\lambda_C$} & \colhead{Flux/Total} \\
  \colhead{Target} & \colhead{(10$^{-15}$ W m$^{-2}$)} & \colhead{(\mum)} & \colhead{(\%)} & 
  \colhead{(\mum)} & \colhead{(\%)} & \colhead{(\mum)} & \colhead{(\%)}
}
\startdata
SMP SMC 013       &   0.59 $\pm$ 0.02 & 6.25 $\pm$ 0.13 &    3.9 $\pm$  1.5 & 8.58 $\pm$ 0.05 &    3.3 $\pm$ 0.4 & 7.73 $\pm$ 0.05 & 42 $\pm$    2 \\
SMP SMC 018       &   0.97 $\pm$ 0.04 & 6.27 $\pm$ 0.11 &    3.8 $\pm$  1.1 & 8.60 $\pm$ 0.04 &    2.1 $\pm$ 0.3 & 7.59 $\pm$ 0.04 & 47 $\pm$    2 \\
SMP SMC 015       &   0.60 $\pm$ 0.01 & 6.33 $\pm$ 0.05 &    5.4 $\pm$  0.8 & 8.59 $\pm$ 0.03 &    3.5 $\pm$ 0.4 & 7.57 $\pm$ 0.03 & 44 $\pm$    2 \\
SMP SMC 001       &   1.16 $\pm$ 0.03 & 6.25 $\pm$ 0.00 &   18.2 $\pm$  0.5 & 8.61 $\pm$ 0.05 &    4.8 $\pm$ 1.0 & 7.75 $\pm$ 0.02 & 65 $\pm$    2 \\
SMP LMC 008       &   0.84 $\pm$ 0.03 & 6.26 $\pm$ 0.00 &   25.1 $\pm$  0.7 & \nodata         &    4.0 $\pm$ 2.0 & 7.66 $\pm$ 0.06 & 50 $\pm$    4 \\
SMP LMC 048       &   1.07 $\pm$ 0.05 & 6.31 $\pm$ 0.05 &   14.4 $\pm$  2.5 & \nodata         &    0.7 $\pm$ 0.5 & 7.58 $\pm$ 0.08 & 56 $\pm$    4 \\
SMP LMC 025       &   0.93 $\pm$ 0.03 & 6.27 $\pm$ 0.03 &    8.7 $\pm$  1.2 & 8.60 $\pm$ 0.06 &    2.2 $\pm$ 0.5 & 7.60 $\pm$ 0.03 & 48 $\pm$    1 \\
\\
SMP LMC 085       &   1.42 $\pm$ 0.03 & 6.26 $\pm$ 0.00 &   21.8 $\pm$  0.6 & 8.59 $\pm$ 0.02 &    4.5 $\pm$ 0.4 & 7.67 $\pm$ 0.03 & 60 $\pm$    2 \\
SMP LMC 076       &   0.66 $\pm$ 0.02 & 6.26 $\pm$ 0.00 &   17.6 $\pm$  0.4 & 8.63 $\pm$ 0.04 &    3.7 $\pm$ 0.5 & 7.78 $\pm$ 0.02 & 69 $\pm$    2 \\
SMP LMC 058       &   1.67 $\pm$ 0.18 & 6.26 $\pm$ 0.01 &   24.0 $\pm$  1.3 & 8.62 $\pm$ 0.05 &    2.7 $\pm$ 0.6 & 7.91 $\pm$ 0.01 & 70 $\pm$    1 \\
SMP SMC 027       &   0.04 $\pm$ 0.01 & \nodata         &    8.4 $\pm$ 18.4 & \nodata         &   13.6 $\pm$ 5.7 & \nodata         & 32 $\pm$   28 \\
SMP LMC 051       &   2.59 $\pm$ 0.08 & 6.25 $\pm$ 0.01 &   19.8 $\pm$  1.1 & 8.63 $\pm$ 0.03 &    3.8 $\pm$ 0.5 & 7.88 $\pm$ 0.02 & 66 $\pm$    2 \\
IRAS 05370        &   1.21 $\pm$ 1.62 & 6.25 $\pm$ 0.01 &   39.1 $\pm$  1.3 & 8.64 $\pm$ 0.02 &    9.4 $\pm$ 0.9 & 7.73 $\pm$ 0.03 & 51 $\pm$    3 \\
IRAS 05537        &   0.70 $\pm$ 0.07 & 6.25 $\pm$ 0.01 &   40.5 $\pm$  2.0 & 8.59 $\pm$ 0.03 &   12.9 $\pm$ 1.6 & 7.61 $\pm$ 0.07 & 60 $\pm$    8 \\
SMP SMC 020       &   0.26 $\pm$ 0.02 & 6.23 $\pm$ 0.03 &   24.6 $\pm$  3.9 & \nodata         &    1.7 $\pm$ 1.3 & 7.83 $\pm$ 0.08 & 16 $\pm$    4 \\
IRAS 05588        &   8.03 $\pm$ 0.09 & 6.23 $\pm$ 0.00 &   15.4 $\pm$  0.3 & 8.59 $\pm$ 0.03 &    5.3 $\pm$ 0.6 & 7.76 $\pm$ 0.01 & 69 $\pm$    2 \\
\\
SMP SMC 006       &   1.49 $\pm$ 0.04 & 6.25 $\pm$ 0.01 &   16.4 $\pm$  0.9 & 8.63 $\pm$ 0.02 &    2.7 $\pm$ 0.3 & 7.81 $\pm$ 0.02 & 65 $\pm$    2 \\
IRAS 05073        &   1.93 $\pm$ 0.04 & 6.25 $\pm$ 0.01 &   11.9 $\pm$  0.8 & \nodata         &    0.5 $\pm$ 0.5 & 7.83 $\pm$ 0.03 & 65 $\pm$    2 \\
IRAS 05063        &   0.74 $\pm$ 0.04 & 6.26 $\pm$ 0.02 &   20.1 $\pm$  1.8 & \nodata         &    1.1 $\pm$ 1.2 & \nodata         & \nodata       \\ 
IRAS 05413        &   0.78 $\pm$ 0.12 & 6.24 $\pm$ 0.12 &   37.5 $\pm$ 10.2 & \nodata         &    7.2 $\pm$ 5.3 & 7.74 $\pm$ 0.17 & 97 $\pm$   12 \\
IRAS 00350        &   2.25 $\pm$ 0.17 & 6.29 $\pm$ 0.01 &    4.3 $\pm$  0.9 & 8.64 $\pm$ 0.04 &    8.8 $\pm$ 1.4 & 7.84 $\pm$ 0.03 & 90 $\pm$    4 \\
IRAS 05127        &   2.74 $\pm$ 0.09 & 6.25 $\pm$ 0.01 &   16.2 $\pm$  0.6 & 8.60 $\pm$ 0.01 &    4.3 $\pm$ 0.3 & 7.72 $\pm$ 0.02 & 59 $\pm$    1 \\
\\
J010546           &   2.31 $\pm$ 0.04 & 6.23 $\pm$ 0.01 &    9.9 $\pm$  0.5 & 8.59 $\pm$ 0.02 &    4.0 $\pm$ 0.3 & 7.82 $\pm$ 0.02 & 65 $\pm$    1 \\
IRAS F05192       &   4.16 $\pm$ 0.13 & 6.27 $\pm$ 0.04 &    9.2 $\pm$  1.5 & 8.61 $\pm$ 0.03 &    3.8 $\pm$ 0.4 & 7.89 $\pm$ 0.03 & 66 $\pm$    2 \\
J052043           &   0.97 $\pm$ 0.02 & 6.24 $\pm$ 0.01 &    9.3 $\pm$  0.4 & 8.59 $\pm$ 0.09 &    1.9 $\pm$ 0.5 & 7.84 $\pm$ 0.04 & 69 $\pm$    2 \\
IRAS 05110        &   3.79 $\pm$ 0.03 & 6.26 $\pm$ 0.00 &    7.7 $\pm$  0.1 & 8.57 $\pm$ 0.08 &    0.7 $\pm$ 0.2 & 7.80 $\pm$ 0.04 & 61 $\pm$    2 \\
IRAS Z05259       &   1.38 $\pm$ 0.03 & 6.25 $\pm$ 0.01 &    8.1 $\pm$  0.3 & 8.58 $\pm$ 0.04 &    2.5 $\pm$ 0.3 & 7.83 $\pm$ 0.03 & 68 $\pm$    2 \\
J004441           &   1.78 $\pm$ 0.01 & 6.24 $\pm$ 0.00 &    9.7 $\pm$  0.2 & 8.61 $\pm$ 0.05 &    1.0 $\pm$ 0.2 & 7.81 $\pm$ 0.03 & 57 $\pm$    2 \\
NGC 1978 WBT 2665 &   3.24 $\pm$ 0.06 & 6.23 $\pm$ 0.02 &    9.0 $\pm$  1.0 & \nodata         & $-$0.0 $\pm$ 0.2 & 7.87 $\pm$ 0.05 & 50 $\pm$    2 \\
IRAS 06111        &   5.13 $\pm$ 0.04 & 6.25 $\pm$ 0.00 &   10.9 $\pm$  0.2 & \nodata         &    0.4 $\pm$ 0.2 & 7.86 $\pm$ 0.02 & 57 $\pm$    1 \\
IRAS 05092        &   7.48 $\pm$ 0.07 & 6.24 $\pm$ 0.00 &   15.1 $\pm$  0.2 & 8.61 $\pm$ 0.02 &    5.2 $\pm$ 0.3 & 7.79 $\pm$ 0.01 & 59 $\pm$    0 \\
IRAS 05185        &  14.65 $\pm$ 0.16 & 6.25 $\pm$ 0.00 &   15.6 $\pm$  0.2 & 8.62 $\pm$ 0.02 &    5.9 $\pm$ 0.4 & 7.82 $\pm$ 0.01 & 59 $\pm$    1 \\
IRAS 05360        &   9.95 $\pm$ 0.10 & 6.25 $\pm$ 0.00 &   15.2 $\pm$  0.2 & 8.62 $\pm$ 0.02 &    5.9 $\pm$ 0.4 & 7.84 $\pm$ 0.01 & 59 $\pm$    1 \\
\\
SMP SMC 011       &   2.43 $\pm$ 0.10 & 6.23 $\pm$ 0.00 &   32.7 $\pm$  0.6 & 8.58 $\pm$ 0.05 &    4.7 $\pm$ 1.0 & 7.68 $\pm$ 0.03 & 73 $\pm$    2 \\
IRAS 05315        &   0.21 $\pm$ 0.05 & \nodata         & $-$8.9 $\pm$ 14.5 & \nodata         &    0.2 $\pm$ 9.2 & 8.15 $\pm$ 0.26 & 92 $\pm$   27 \\
IRAS 05495        &   0.37 $\pm$ 0.03 & 6.21 $\pm$ 0.10 &   11.1 $\pm$  3.1 & \nodata         &    4.7 $\pm$ 3.9 & 7.98 $\pm$ 0.08 & 91 $\pm$    7 \\
SMP LMC 099       &   0.94 $\pm$ 0.04 & 6.29 $\pm$ 0.03 &   12.7 $\pm$  1.7 & \nodata         &    2.6 $\pm$ 1.1 & 7.87 $\pm$ 0.04 & 68 $\pm$    3
\enddata
\end{deluxetable*}

\begin{deluxetable*}{lrcrcr} 
\tablecolumns{6}
\tablewidth{0pt}
\tablenum{15}
\tablecaption{Aliphatic hydrocarbon features at 6--9~\mum}
\label{t.aliph69}
\tablehead{
  \colhead{ } & \colhead{Total Feature} & \multicolumn{2}{c}{6.9~\mum\ Feature} & \multicolumn{2}{c}{7.3~\mum\ Feature} \\
  \colhead{ } & \colhead{Flux (6--9~\mum)} & \colhead{$\lambda_C$} & \colhead{Flux/Total} & \colhead{$\lambda_C$} & \colhead{Flux/Total} \\
  \colhead{Target} & \colhead{(10$^{-15}$ W m$^{-2}$)} & \colhead{(\mum)} & \colhead{(\%)} & \colhead{(\mum)} & \colhead{(\%)}
}
\startdata
SMP SMC 020       &   0.26 $\pm$ 0.02 &  6.83 $\pm$ 0.08 &     4.8 $\pm$  1.3 & \nodata          &    3.5 $\pm$  2.2 \\
IRAS 05588        &   8.03 $\pm$ 0.09 &  6.91 $\pm$ 0.02 &     2.1 $\pm$  0.2 & \nodata          & $-$0.9 $\pm$  0.3 \\
\\
IRAS 05073        &   1.93 $\pm$ 0.04 &  6.90 $\pm$ 0.01 &    12.5 $\pm$  0.9 & \nodata          & $-$0.0 $\pm$  0.5 \\
IRAS 05063        &   0.74 $\pm$ 0.04 &  6.92 $\pm$ 0.03 &     7.0 $\pm$  1.1 & \nodata          &    0.6 $\pm$  0.7 \\
IRAS 05413        &   0.78 $\pm$ 0.12 &  6.85 $\pm$ 0.05 &    10.0 $\pm$  3.5 & \nodata          & $-$3.9 $\pm$  4.8 \\
IRAS 00350        &   2.25 $\pm$ 0.17 & \nodata          &  $-$3.1 $\pm$  1.2 &  7.24 $\pm$ 0.04 &    2.0 $\pm$  0.6 \\
\\
J010546           &   2.31 $\pm$ 0.04 &  6.91 $\pm$ 0.05 &     0.9 $\pm$  0.2 & \nodata          & $-$0.0 $\pm$  0.3 \\
IRAS F05192       &   4.16 $\pm$ 0.13 &  6.95 $\pm$ 0.05 &     2.1 $\pm$  0.6 & \nodata          &    0.4 $\pm$  0.8 \\
J052043           &   0.97 $\pm$ 0.02 &  6.90 $\pm$ 0.02 &     3.9 $\pm$  0.3 &  7.28 $\pm$ 0.04 &    0.8 $\pm$  0.2 \\
IRAS 05110        &   3.79 $\pm$ 0.03 &  6.89 $\pm$ 0.00 &    12.8 $\pm$  0.3 &  7.28 $\pm$ 0.00 &    2.6 $\pm$  0.1 \\
IRAS Z05259       &   1.38 $\pm$ 0.03 &  6.92 $\pm$ 0.03 &     2.7 $\pm$  0.4 &  7.29 $\pm$ 0.03 &    0.4 $\pm$  0.1 \\
J004441           &   1.78 $\pm$ 0.01 &  6.89 $\pm$ 0.00 &    10.5 $\pm$  0.3 &  7.28 $\pm$ 0.01 &    1.8 $\pm$  0.2 \\
NGC 1978 WBT 2665 &   3.24 $\pm$ 0.06 &  6.91 $\pm$ 0.01 &    15.2 $\pm$  1.2 &  7.27 $\pm$ 0.02 &    2.5 $\pm$  0.4 \\
IRAS 06111        &   5.13 $\pm$ 0.04 &  6.90 $\pm$ 0.01 &    11.4 $\pm$  0.4 &  7.27 $\pm$ 0.01 &    1.5 $\pm$  0.1 \\
IRAS 05092        &   7.48 $\pm$ 0.07 &  6.90 $\pm$ 0.01 &     1.3 $\pm$  0.1 & \nodata          & $-$0.3 $\pm$  0.1 \\
IRAS 05185        &  14.65 $\pm$ 0.16 &  6.91 $\pm$ 0.01 &     1.0 $\pm$  0.1 & \nodata          & $-$0.3 $\pm$  0.1 \\
IRAS 05360        &   9.95 $\pm$ 0.10 &  6.92 $\pm$ 0.02 &     0.8 $\pm$  0.1 & \nodata          & $-$0.3 $\pm$  0.1 \\
\\
SMP SMC 011       &   2.43 $\pm$ 0.10 &  6.92 $\pm$ 0.01 &     1.9 $\pm$  0.2 & \nodata          & $-$2.8 $\pm$  0.6
\enddata
\end{deluxetable*}

We followed our standard method to measure the strengths and
positions of the fullerene-related, PAH, and aliphatic 
hydrocarbon features.  Tables~\ref{t.full69}, \ref{t.pah69}, 
and \ref{t.aliph69} present the results.  For all of the 
spectra, we have integrated the total strength of the 
6--9~\mum\ emission complex and expressed the strengths of 
the individual features as a percentage of the flux from the 
overall complex.  

We extracted features from the spectra after spline-fitting 
and removing a continuum and removing the forbidden lines 
when they were clearly present.  Only [Ar III] at 8.99~\mum\ 
would affect the analysis at 6--9~\mum.  The tables only list 
the sources where at least one feature of the relevant family 
has been detected at a SNR  of at least 2.5~$\sigma$.  When the
fullerene and PAH features are too blended to disentangle, we
do not report a result.

\subsection*{Appendix 1.3.  Longer wavelengths}

\begin{deluxetable*}{lrrccrcrcr} 
\tablecolumns{10}
\tablewidth{0pt}
\tablenum{16}
\tablecaption{Features in the 21~\mum\ spectra}
\label{t.21dat}
\tablehead{
  \colhead{ } & \colhead{Total Feature} & \colhead{6--9~\mum} & \colhead{11--14~\mum} & \multicolumn{2}{c}{16~\mum\ Complex} & 
  \multicolumn{2}{c}{21~\mum\ Feature} & \multicolumn{2}{c}{26--30~\mum\ Feature} \\
  \colhead{ } & \colhead{Flux (5--37~\mum)} & \colhead{Flux/Total} & \colhead{Flux/Total} & \colhead{$\lambda_C$} & 
  \colhead{Flux/Total} & \colhead{$\lambda_C$} & \colhead{Flux/Total} & \colhead{$\lambda_C$} & \colhead{Flux/Total} \\
  \colhead{Target} & \colhead{(10$^{-15}$ W m$^{-2}$)} & \colhead{(\%)} & \colhead{(\%)} & \colhead{(\mum)} & 
  \colhead{(\%)} & \colhead{(\mum)} & \colhead{(\%)} & \colhead{(\mum)} & \colhead{(\%)}
}
\startdata
J010546           &  5.01 $\pm$ 0.03 & 60.9 $\pm$ 0.3 & 30.8 $\pm$ 0.2 & 16.75 $\pm$ 0.17 &    1.5 $\pm$ 0.1 & 20.62 $\pm$ 0.25 &  2.5 $\pm$ 0.2 & 28.07 $\pm$ 0.58 &  3.2 $\pm$ 0.2 \\
IRAS F05192       & 14.68 $\pm$ 0.15 & 42.9 $\pm$ 0.6 & 26.2 $\pm$ 0.3 & 16.83 $\pm$ 0.54 &    0.8 $\pm$ 0.1 & 20.44 $\pm$ 0.05 & 12.9 $\pm$ 0.3 & 28.90 $\pm$ 0.27 & 13.8 $\pm$ 0.5 \\
J052043           &  4.30 $\pm$ 0.04 & 29.4 $\pm$ 0.3 & 23.2 $\pm$ 0.7 & 15.98 $\pm$ 0.05 &    5.6 $\pm$ 0.2 & 20.35 $\pm$ 0.03 & 17.3 $\pm$ 0.2 & 28.72 $\pm$ 0.09 & 23.3 $\pm$ 0.3 \\
IRAS 05110        & 12.99 $\pm$ 0.07 & 31.0 $\pm$ 0.2 & 29.2 $\pm$ 0.4 & 16.23 $\pm$ 0.05 &    2.1 $\pm$ 0.1 & 20.40 $\pm$ 0.04 &  4.4 $\pm$ 0.1 & 28.46 $\pm$ 0.04 & 30.6 $\pm$ 0.1 \\
IRAS Z05259       &  5.95 $\pm$ 0.03 & 29.7 $\pm$ 0.2 & 33.9 $\pm$ 0.3 & 16.07 $\pm$ 0.08 &    3.4 $\pm$ 0.2 & 20.43 $\pm$ 0.01 & 22.8 $\pm$ 0.1 & 28.33 $\pm$ 0.13 &  9.3 $\pm$ 0.2 \\
J004441           &  4.21 $\pm$ 0.03 & 46.9 $\pm$ 0.2 & 31.8 $\pm$ 0.3 & 16.11 $\pm$ 0.11 &    3.2 $\pm$ 0.2 & 20.32 $\pm$ 0.11 &  6.6 $\pm$ 0.3 & 28.63 $\pm$ 0.24 &  8.6 $\pm$ 0.3 \\
NGC 1978 WBT 2665 & 13.37 $\pm$ 0.08 & 23.4 $\pm$ 0.4 & 28.1 $\pm$ 0.1 & 16.69 $\pm$ 0.17 &    1.5 $\pm$ 0.1 & 20.60 $\pm$ 0.23 &  0.4 $\pm$ 0.1 & 28.00 $\pm$ 0.03 & 46.0 $\pm$ 0.2 \\
IRAS 06111        & 14.49 $\pm$ 0.06 & 30.2 $\pm$ 0.2 & 31.2 $\pm$ 0.2 & 16.99 $\pm$ 0.09 &    1.1 $\pm$ 0.1 & 20.55 $\pm$ 0.34 &  0.4 $\pm$ 0.1 & 28.17 $\pm$ 0.04 & 36.7 $\pm$ 0.2 \\
IRAS 05092        & 14.16 $\pm$ 0.04 & 55.9 $\pm$ 0.2 & 29.3 $\pm$ 0.1 & 16.21 $\pm$ 0.11 &    2.1 $\pm$ 0.1 & 20.57 $\pm$ 0.04 &  3.4 $\pm$ 0.1 & 28.60 $\pm$ 0.09 &  8.7 $\pm$ 0.1 \\
IRAS 05185        & 28.44 $\pm$ 0.10 & 63.3 $\pm$ 0.3 & 21.7 $\pm$ 0.0 & 16.80 $\pm$ 0.03 &    1.8 $\pm$ 0.0 & 20.51 $\pm$ 0.04 &  3.1 $\pm$ 0.1 & 28.87 $\pm$ 0.07 &  9.4 $\pm$ 0.1 \\
IRAS 05360        & 21.23 $\pm$ 0.09 & 60.0 $\pm$ 0.2 & 20.9 $\pm$ 0.3 & 16.99 $\pm$ 0.04 &    1.5 $\pm$ 0.1 & 20.41 $\pm$ 0.20 &  0.4 $\pm$ 0.1 & 28.58 $\pm$ 0.05 & 16.9 $\pm$ 0.1 \\
\\
IRAS 05073        & 10.43 $\pm$ 0.07 & 18.8 $\pm$ 0.3 & 28.9 $\pm$ 0.3 & 16.86 $\pm$ 0.39 &    1.6 $\pm$ 0.2 & \nodata          &  0.1 $\pm$ 0.2 & 28.10 $\pm$ 0.05 & 48.1 $\pm$ 0.3 \\
IRAS 05063        &  7.93 $\pm$ 0.05 &  8.9 $\pm$ 0.3 & 40.6 $\pm$ 0.3 & 17.13 $\pm$ 0.22 &    0.9 $\pm$ 0.2 & 20.55 $\pm$ 0.35 &  1.3 $\pm$ 0.2 & 28.04 $\pm$ 0.05 & 46.4 $\pm$ 0.3 \\
IRAS 05413        &  6.82 $\pm$ 0.21 & 14.9 $\pm$ 1.7 & 39.4 $\pm$ 0.8 & \nodata          & $-$0.1 $\pm$ 0.2 & 20.14 $\pm$ 1.53 &  1.3 $\pm$ 0.4 & 27.90 $\pm$ 0.18 & 40.6 $\pm$ 1.0
\enddata
\end{deluxetable*}

\begin{deluxetable}{lrcr} 
\tablecolumns{4}
\tablewidth{0pt}
\tablenum{17}
\tablecaption{The 26--30~\mum\ feature}
\label{t.mgsdat}
\tablehead{
  \colhead{ } & \colhead{Total Feature} & \multicolumn{2}{c}{26--30~\mum\ Feature} \\
  \colhead{ } & \colhead{Flux (5--37~\mum)} & \colhead{$\lambda_C$} & \colhead{Flux/Total} \\
  \colhead{Target} & \colhead{(10$^{-15}$ W m$^{-2}$)} & \colhead{(\mum)} & \colhead{(\%)}
}
\startdata
SMP LMC 002       &  0.95 $\pm$ 0.02 & 30.08 $\pm$ 0.56 &   18.9 $\pm$ 1.2 \\
SMP LMC 056       &  1.00 $\pm$ 0.02 & 30.15 $\pm$ 0.52 &   13.4 $\pm$ 1.1 \\
SMP SMC 024       &  1.43 $\pm$ 0.02 & \nodata          &   18.9 $\pm$ 1.0 \\
LIN 49            &  1.19 $\pm$ 0.01 & 29.32 $\pm$ 0.11 &   21.2 $\pm$ 0.3 \\
SMP SMC 016       &  1.79 $\pm$ 0.03 & 29.96 $\pm$ 0.21 &   11.8 $\pm$ 0.4 \\
\\
SMP SMC 013       &  1.53 $\pm$ 0.03 & \nodata          &    6.7 $\pm$ 0.5 \\
SMP SMC 018       &  3.69 $\pm$ 0.03 & 31.03 $\pm$ 0.17 &    8.4 $\pm$ 0.2 \\
SMP SMC 015       &  4.60 $\pm$ 0.04 & 30.24 $\pm$ 0.17 &    6.0 $\pm$ 0.2 \\
SMP SMC 001       &  5.71 $\pm$ 0.05 & 31.09 $\pm$ 0.29 &    5.1 $\pm$ 0.2 \\
SMP LMC 008       & 10.39 $\pm$ 0.13 & 30.17 $\pm$ 0.14 &   10.9 $\pm$ 0.2 \\
SMP LMC 048       &  4.54 $\pm$ 0.06 & 29.79 $\pm$ 0.07 &   25.6 $\pm$ 0.2 \\
SMP LMC 025       &  4.78 $\pm$ 0.05 & 29.72 $\pm$ 0.07 &   28.0 $\pm$ 0.3 \\
\\
SMP LMC 085       &  9.09 $\pm$ 0.06 & 30.03 $\pm$ 0.09 &   16.0 $\pm$ 0.2 \\
SMP LMC 076       &  3.06 $\pm$ 0.02 & 30.04 $\pm$ 0.39 &    5.2 $\pm$ 0.3 \\
SMP LMC 058       &  8.77 $\pm$ 0.08 & 31.23 $\pm$ 0.38 &    2.2 $\pm$ 0.2 \\
SMP SMC 027       &  0.57 $\pm$ 0.03 & \nodata          &   25.2 $\pm$ 1.4 \\
SMP LMC 051       & 10.21 $\pm$ 0.08 & 30.86 $\pm$ 0.32 &    3.8 $\pm$ 0.3 \\
IRAS 05370        & 16.87 $\pm$ 0.06 & \nodata          &    0.4 $\pm$ 0.2 \\
IRAS 05537        & 11.93 $\pm$ 0.07 & 31.08 $\pm$ 0.33 &    4.8 $\pm$ 0.3 \\
SMP SMC 020       &  3.83 $\pm$ 0.03 & \nodata          &    0.4 $\pm$ 0.1 \\
IRAS 05588        & 18.72 $\pm$ 0.15 & \nodata          &    0.9 $\pm$ 0.1 \\
\\
SMP SMC 006       &  6.42 $\pm$ 0.04 & 31.51 $\pm$ 0.36 &    4.7 $\pm$ 0.2 \\
IRAS 00350        &  8.15 $\pm$ 0.15 & 27.34 $\pm$ 0.10 &   30.8 $\pm$ 0.4 \\
IRAS 05127        &  9.03 $\pm$ 0.05 & 28.83 $\pm$ 0.06 &   44.5 $\pm$ 0.3 \\
\\
SMP SMC 011       & 12.28 $\pm$ 0.17 & \nodata          & $-$0.4 $\pm$ 0.5 \\
IRAS 05315        &  9.52 $\pm$ 0.14 & 27.52 $\pm$ 0.11 &   47.2 $\pm$ 0.6 \\
IRAS 05495        &  7.99 $\pm$ 0.08 & 28.32 $\pm$ 0.10 &   56.5 $\pm$ 0.6 \\
SMP SMC 025       &  0.90 $\pm$ 0.02 & \nodata          &   13.8 $\pm$ 0.8 \\
SMP LMC 099       &  2.81 $\pm$ 0.10 & \nodata          &   22.1 $\pm$ 1.5 
\enddata
\end{deluxetable}

We adjusted our method to measure the strengths and positions 
of the features in the 21~\mum\ spectra because they can 
shift from one spectrum to the next.  Rather than use fixed 
wavelengths to set the continuum, we have allowed the 
wavelengths for integration to vary by setting them to the 
minima in the spline-removed spectra in intervals on either 
side of the features.

Table~\ref{t.21dat} presents the total integrated fluxes for
all features, and for individual features, the central 
wavelength and the strength as a percentage of the total 
feature flux.  We do not report a central wavelength for 
features detected at less that 2.5~$\sigma$.  The features 
are integrated between the minima found in the following 
intervals:  5--6, 9--10.2, 13.8--15.2, 17--19, 22--26, and 
35--37~\mum.  We have also zeroed negative data to the red 
of 35~\mum\ (after finding the position of the minimum) to
reduce the impact of a poorly determined spline at the 
boundary of the data.

Table~\ref{t.21dat} includes a column giving the central
wavelength and fraction of the flux from the 26--30~\mum\ 
emission feature for the eleven 21~\mum\ sources and the 
three 21~\mum\ candidates.  Table~\ref{t.mgsdat} provides the 
same data for the remainder of the sample.  Some spectra do 
not show a 26--30~\mum\ feature or show it with only a low 
SNR.  In others, the spectrum actually climbs at the longest 
wavelengths, possibly due to background subtraction issues.  
In these cases, we have not reported a central wavelength.  

\section*{Appendix 2.  Photometry}  

\begin{deluxetable*}{lccrrrrrr} 
\tablecolumns{9}
\tablewidth{0pt}
\tablenum{18}
\tablecaption{Combined near- and mid-infrared photometry}
\label{t.irphot}
\tablehead{
  \colhead{Target} & \colhead{$J$} & \colhead{$H$} & \colhead{$K_s$} & \colhead{[3.6]} & \colhead{[4.5]} & \colhead{[5.8]} & \colhead{[8.0]} & \colhead{[24]}
}
\startdata
SMP LMC 002       & 16.83 $\pm$ 0.06 & 16.43 $\pm$ 0.09 & 15.39 $\pm$ 0.08 & 13.78 $\pm$ 0.09 & 12.81 $\pm$ 0.02 & 12.05 $\pm$ 0.04 & 10.52 $\pm$ 0.01 &  7.40 $\pm$ 0.02 \\
SMP LMC 056       & 15.74 $\pm$ 1.68 & 15.04 $\pm$ 1.96 & 15.22 $\pm$ 0.21 & 13.47 $\pm$ 0.20 & 12.48 $\pm$ 0.06 & 11.73 $\pm$ 0.04 & 10.13 $\pm$ 0.00 &  6.76 $\pm$ 0.00 \\
SMP SMC 024       & 16.18 $\pm$ 0.13 & 15.73 $\pm$ 0.08 & 14.68 $\pm$ 0.02 & 13.11 $\pm$ 0.18 & 12.26 $\pm$ 0.06 & 11.47 $\pm$ 0.01 &  9.76 $\pm$ 0.04 &  6.06 $\pm$ 0.04 \\
LIN 49            & 16.58 $\pm$ 0.08 & 15.72 $\pm$ 0.07 & 14.51 $\pm$ 0.15 & 12.93 $\pm$ 0.14 & 12.17 $\pm$ 0.03 & 11.33 $\pm$ 0.05 &  9.74 $\pm$ 0.01 &  6.47 $\pm$ 0.04 \\
SMP SMC 016       & 16.05 $\pm$ 0.10 & 15.28 $\pm$ 0.09 & 14.08 $\pm$ 0.15 & 12.57 $\pm$ 0.16 & 11.72 $\pm$ 0.03 & 10.90 $\pm$ 0.04 &  9.29 $\pm$ 0.04 &  5.98 $\pm$ 0.06 \\
\\
SMP SMC 013       & 16.04 $\pm$ 0.10 & 15.73 $\pm$ 0.10 & 14.57 $\pm$ 0.43 & 13.24 $\pm$ 0.17 & 12.37 $\pm$ 0.08 & 11.35 $\pm$ 0.01 &  9.50 $\pm$ 0.04 &  5.71 $\pm$ 0.02 \\
SMP SMC 018       & 15.31 $\pm$ 0.31 & 14.85 $\pm$ 0.24 & 14.01 $\pm$ 0.41 & 12.37 $\pm$ 0.10 & 11.45 $\pm$ 0.03 & 10.32 $\pm$ 0.01 &  8.60 $\pm$ 0.00 &  4.46 $\pm$ 0.02 \\
SMP SMC 015       & 15.56 $\pm$ 0.19 & 15.31 $\pm$ 0.05 & 14.13 $\pm$ 0.13 & 12.56 $\pm$ 0.17 & 11.49 $\pm$ 0.03 & 10.30 $\pm$ 0.00 &  8.53 $\pm$ 0.02 &  4.50 $\pm$ 0.00 \\
SMP SMC 001       & 16.15 $\pm$ 0.03 & 15.48 $\pm$ 0.04 & 14.49 $\pm$ 0.11 & 12.57 $\pm$ 0.12 & 11.56 $\pm$ 0.04 & 10.12 $\pm$ 0.03 &  8.33 $\pm$ 0.00 &  4.13 $\pm$ 0.01 \\
SMP LMC 008       & 15.63 $\pm$ 0.16 & 15.60 $\pm$ 0.17 & 14.40 $\pm$ 0.06 & 12.35 $\pm$ 0.25 & 10.95 $\pm$ 0.01 &  9.54 $\pm$ 0.00 &  7.81 $\pm$ 0.00 &  3.26 $\pm$ 0.02 \\
SMP LMC 048       & 15.39 $\pm$ 0.05 & 14.10 $\pm$ 1.46 & 13.17 $\pm$ 1.45 & 11.82 $\pm$ 1.32 & 11.16 $\pm$ 1.11 & 10.63 $\pm$ 0.03 &  8.88 $\pm$ 0.00 &  4.35 $\pm$ 0.07 \\
SMP LMC 025       & 15.50 $\pm$ 0.04 & 15.25 $\pm$ 0.10 & 14.23 $\pm$ 0.13 & 12.61 $\pm$ 0.02 & 11.76 $\pm$ 0.07 & 10.68 $\pm$ 0.03 &  8.83 $\pm$ 0.01 &  4.24 $\pm$ 0.00 \\
\\
SMP LMC 085       & 15.46 $\pm$ 0.12 & 15.16 $\pm$ 0.07 & 14.01 $\pm$ 0.14 & 12.38 $\pm$ 0.22 & 11.28 $\pm$ 0.01 &  9.75 $\pm$ 0.02 &  7.93 $\pm$ 0.01 &  3.42 $\pm$ 0.04 \\
SMP LMC 076       & 15.85 $\pm$ 0.20 & 15.97 $\pm$ 0.06 & 15.04 $\pm$ 0.02 & 13.46 $\pm$ 0.25 & 12.47 $\pm$ 0.07 & 10.89 $\pm$ 0.06 &  9.01 $\pm$ 0.04 &  4.66 $\pm$ 0.03 \\
SMP LMC 058       & 15.62 $\pm$ 0.17 & 15.53 $\pm$ 0.09 & 14.53 $\pm$ 0.03 & 12.27 $\pm$ 0.24 & 11.08 $\pm$ 0.02 &  9.70 $\pm$ 0.01 &  8.08 $\pm$ 0.01 &  4.00 $\pm$ 0.01 \\
SMP SMC 027       & 16.02 $\pm$ 0.09 & 15.87 $\pm$ 0.48 & 15.37 $\pm$ 0.05 & 14.01 $\pm$ 0.26 & 13.26 $\pm$ 0.04 & 12.73 $\pm$ 0.01 & 10.87 $\pm$ 0.03 &  6.15 $\pm$ 0.05 \\
SMP LMC 051       & 15.84 $\pm$ 0.12 & 15.61 $\pm$ 0.02 & 14.23 $\pm$ 0.07 & 11.87 $\pm$ 0.24 & 10.68 $\pm$ 0.03 &  9.17 $\pm$ 0.02 &  7.40 $\pm$ 0.01 &  3.30 $\pm$ 0.01 \\
IRAS 05370        & 17.50 $\pm$ 0.37 & 16.49 $\pm$ 0.01 & 15.19 $\pm$ 0.03 & 12.04 $\pm$ 0.19 & 10.76 $\pm$ 0.14 &  9.20 $\pm$ 0.04 &  7.39 $\pm$ 0.01 &  2.96 $\pm$ 0.02 \\
IRAS 05537        & 16.38 $\pm$ 0.14 & 15.93 $\pm$ 0.18 & 15.03 $\pm$ 0.08 & 12.97 $\pm$ 0.30 & 11.91 $\pm$ 0.04 &  9.90 $\pm$ 0.00 &  7.74 $\pm$ 0.05 &  3.00 $\pm$ 0.00 \\
SMP SMC 020       & 15.88 $\pm$ 0.14 & 15.98 $\pm$ 0.04 & 15.06 $\pm$ 0.04 & 12.90 $\pm$ 0.28 & 11.48 $\pm$ 0.01 & 10.26 $\pm$ 0.04 &  8.72 $\pm$ 0.03 & \nodata          \\
IRAS 05588        & 14.64 $\pm$ 0.06 & 14.30 $\pm$ 0.22 & 13.41 $\pm$ 0.35 & 11.50 $\pm$ 0.32 & 10.48 $\pm$ 0.21 &  8.83 $\pm$ 0.09 &  6.78 $\pm$ 0.03 &  3.04 $\pm$ 0.01 \\
\\
SMP SMC 006       & 15.91 $\pm$ 0.21 & 15.87 $\pm$ 0.14 & 14.95 $\pm$ 0.07 & 12.75 $\pm$ 0.24 & 11.54 $\pm$ 0.02 &  9.90 $\pm$ 0.04 &  8.06 $\pm$ 0.03 &  3.74 $\pm$ 0.02 \\
IRAS 05073        & 16.57 $\pm$ 0.06 & 15.37 $\pm$ 0.05 & 14.53 $\pm$ 0.03 & 12.42 $\pm$ 0.34 & 11.40 $\pm$ 0.14 &  9.69 $\pm$ 0.01 &  7.57 $\pm$ 0.01 &  3.26 $\pm$ 0.03 \\
IRAS 05063        & 15.08 $\pm$ 0.17 & 13.86 $\pm$ 0.11 & 13.16 $\pm$ 0.16 & 11.45 $\pm$ 0.29 & 10.44 $\pm$ 0.17 &  9.17 $\pm$ 0.03 &  7.56 $\pm$ 0.01 &  3.57 $\pm$ 0.00 \\
IRAS 05413        & 15.59 $\pm$ 0.10 & 14.20 $\pm$ 0.15 & 13.11 $\pm$ 0.14 & 11.27 $\pm$ 0.66 & 10.38 $\pm$ 0.45 &  9.14 $\pm$ 0.04 &  7.67 $\pm$ 0.03 &  3.72 $\pm$ 0.01 \\
IRAS 00350        & 11.37 $\pm$ 0.07 & 10.19 $\pm$ 0.04 &  9.10 $\pm$ 0.03 &  7.68 $\pm$ 0.08 &  7.03 $\pm$ 0.05 &  6.51 $\pm$ 0.00 &  5.80 $\pm$ 0.02 &  3.55 $\pm$ 0.00 \\
IRAS 05127        & 15.52 $\pm$ 0.16 & 15.05 $\pm$ 0.06 & 14.15 $\pm$ 0.25 & 12.58 $\pm$ 0.23 & 11.80 $\pm$ 0.00 & 10.11 $\pm$ 0.02 &  8.25 $\pm$ 0.02 &  3.64 $\pm$ 0.00 \\
\\
J010546           & 14.72 $\pm$ 0.01 & 14.62 $\pm$ 0.03 & 14.47 $\pm$ 0.12 & 13.54 $\pm$ 0.03 & 12.89 $\pm$ 0.07 & 10.62 $\pm$ 0.04 &  8.13 $\pm$ 0.02 &  4.92 $\pm$ 0.03 \\
IRAS F05192       & 14.64 $\pm$ 0.05 & 14.25 $\pm$ 0.03 & 13.92 $\pm$ 0.14 & 12.83 $\pm$ 0.22 & 11.53 $\pm$ 0.04 &  9.43 $\pm$ 0.01 &  7.17 $\pm$ 0.03 &  3.41 $\pm$ 0.01 \\
J052043           & 13.17 $\pm$ 0.30 & 12.73 $\pm$ 0.32 & 12.52 $\pm$ 0.26 & 12.23 $\pm$ 0.49 & 12.22 $\pm$ 0.26 & 11.31 $\pm$ 0.01 &  8.94 $\pm$ 0.00 &  4.46 $\pm$ 0.04 \\
IRAS 05110        & 13.29 $\pm$ 0.01 & 12.83 $\pm$ 0.04 & 12.60 $\pm$ 0.08 & 12.19 $\pm$ 0.10 & 11.70 $\pm$ 0.04 &  9.92 $\pm$ 0.03 &  7.38 $\pm$ 0.00 &  3.56 $\pm$ 0.00 \\
IRAS Z05259       & 13.65 $\pm$ 0.04 & 13.46 $\pm$ 0.06 & 13.33 $\pm$ 0.07 & 13.12 $\pm$ 0.03 & 12.96 $\pm$ 0.03 & 11.24 $\pm$ 0.04 &  8.59 $\pm$ 0.01 &  4.12 $\pm$ 0.01 \\
J004441           & 13.63 $\pm$ 0.06 & 13.25 $\pm$ 0.02 & 13.04 $\pm$ 0.08 & 12.33 $\pm$ 0.07 & 11.87 $\pm$ 0.04 & 10.39 $\pm$ 0.00 &  8.32 $\pm$ 0.02 &  5.13 $\pm$ 0.01 \\
NGC 1978 WBT 2665 & 14.72 $\pm$ 0.07 & 13.63 $\pm$ 0.11 & 12.97 $\pm$ 0.15 & 11.96 $\pm$ 0.10 & 11.23 $\pm$ 0.06 &  9.52 $\pm$ 0.01 &  7.33 $\pm$ 0.00 &  3.33 $\pm$ 0.03 \\
IRAS 06111        & 16.29 $\pm$ 0.04 & 14.93 $\pm$ 0.07 & 14.08 $\pm$ 0.06 & 12.39 $\pm$ 0.02 & 11.04 $\pm$ 0.02 & \nodata          & \nodata          & \nodata          \\
IRAS 05092        & 15.28 $\pm$ 0.13 & 14.80 $\pm$ 0.04 & 14.17 $\pm$ 0.04 & 11.91 $\pm$ 0.26 & 11.22 $\pm$ 0.07 &  8.95 $\pm$ 0.06 &  6.93 $\pm$ 0.02 &  3.90 $\pm$ 0.05 \\
IRAS 05185        & 16.62 $\pm$ 0.23 & 15.63 $\pm$ 0.14 & 14.65 $\pm$ 0.07 & 11.49 $\pm$ 0.23 & 10.80 $\pm$ 0.02 &  8.38 $\pm$ 0.03 &  6.49 $\pm$ 0.01 &  3.33 $\pm$ 0.03 \\
IRAS 05360        & 16.28 $\pm$ 0.20 & 15.15 $\pm$ 0.05 & 14.14 $\pm$ 0.12 & 11.57 $\pm$ 0.23 & 10.46 $\pm$ 0.03 &  8.49 $\pm$ 0.01 &  6.67 $\pm$ 0.00 &  3.31 $\pm$ 0.03 \\
\\
SMP SMC 011       & 16.10 $\pm$ 0.21 & 15.44 $\pm$ 0.15 & 15.27 $\pm$ 0.10 & 14.90 $\pm$ 0.02 & 14.87 $\pm$ 0.06 & \nodata          & \nodata          & \nodata          \\
IRAS 05315        & 18.54 $\pm$ 0.08 & 17.94 $\pm$ 0.11 & 17.23 $\pm$ 0.22 & 13.89 $\pm$ 0.03 & 12.12 $\pm$ 0.14 & 10.22 $\pm$ 0.00 &  7.83 $\pm$ 0.00 &  2.76 $\pm$ 0.02 \\
IRAS 05495        & \nodata          & \nodata          & 15.52 $\pm$ 0.21 & 13.63 $\pm$ 0.12 & 11.31 $\pm$ 0.03 &  8.45 $\pm$ 0.03 &  2.42 $\pm$ 0.04 & \nodata          \\
SMP SMC 025       & 17.89 $\pm$ 0.20 & 17.56 $\pm$ 0.40 & 16.84 $\pm$ 0.10 & 15.95 $\pm$ 0.13 & 14.67 $\pm$ 0.14 & 13.67 $\pm$ 0.07 & 11.68 $\pm$ 0.03 &  5.88 $\pm$ 0.02 \\
SMP LMC 099       & 16.07 $\pm$ 0.01 & 15.92 $\pm$ 0.04 & 14.95 $\pm$ 0.12 & 13.51 $\pm$ 0.03 & 12.49 $\pm$ 0.02 & \nodata          & \nodata          & \nodata         
\enddata
\end{deluxetable*}

\begin{deluxetable*}{lccccl} 
\tablecolumns{6}
\tablewidth{0pt}
\tablenum{19}
\tablecaption{Combined UBVI photometry and bolometric magnitudes}
\label{t.optphot}
\tablehead{
  \colhead{Target} & \colhead{$U$} & \colhead{$B$} & \colhead{$V$} & \colhead{$I$} & \colhead{$M_{\rm bol}$}
}
\startdata
SMP LMC 002       & 15.97 $\pm$ 0.06 & 17.13 $\pm$ 0.03 & 17.01 $\pm$ 0.00 & 17.26 $\pm$ 0.07 & $-$3.99\tablenotemark{a} \\
SMP LMC 056       & 16.56 $\pm$ 0.09 & 17.36 $\pm$ 0.04 & 17.11 $\pm$ 0.03 & 17.39 $\pm$ 0.15 & $-$3.61\tablenotemark{a} \\
SMP SMC 024       & 15.67 $\pm$ 0.04 & 16.66 $\pm$ 0.04 & 15.90 $\pm$ 0.03 & 17.04 $\pm$ 0.07 & $-$4.81\tablenotemark{a} \\
LIN 49            & 15.86 $\pm$ 0.05 & 17.10 $\pm$ 0.02 & 17.23 $\pm$ 0.03 & 17.04 $\pm$ 0.12 & $-$4.58\tablenotemark{a} \\
SMP SMC 016       & 15.78 $\pm$ 0.05 & 16.94 $\pm$ 0.02 & 16.81 $\pm$ 0.02 & 16.94 $\pm$ 0.03 & $-$4.73\tablenotemark{a} \\
\\
SMP SMC 018       & 15.83 $\pm$ 0.03 & 16.64 $\pm$ 0.04 & 15.99 $\pm$ 0.05 & 16.35 $\pm$ 0.35 & $-$4.33 \\
SMP SMC 013       & 16.12 $\pm$ 0.04 & 16.30 $\pm$ 0.04 & 15.34 $\pm$ 0.02 & 16.89 $\pm$ 0.17 & $-$3.61 \\
SMP SMC 015       & 15.72 $\pm$ 0.05 & 16.24 $\pm$ 0.10 & 15.33 $\pm$ 0.03 & 16.51 $\pm$ 0.12 & $-$4.44 \\
SMP SMC 001       & \nodata          & \nodata          & \nodata          & 16.91 $\pm$ 0.07 & $-$4.42 \\
SMP LMC 008       & 16.65 $\pm$ 0.04 & 16.98 $\pm$ 0.03 & 15.95 $\pm$ 0.04 & 16.58 $\pm$ 0.06 & $-$4.84 \\
SMP LMC 048       & 15.38 $\pm$ 0.08 & 16.45 $\pm$ 0.04 & 15.10 $\pm$ 0.01 & 16.11 $\pm$ 0.02 & $-$4.21 \\
SMP LMC 025       & 15.26 $\pm$ 0.06 & 15.83 $\pm$ 0.03 & 14.82 $\pm$ 0.11 & 16.33 $\pm$ 0.13 & $-$4.24 \\
\\
SMP LMC 085       & 15.83 $\pm$ 0.04 & 16.53 $\pm$ 0.02 & 15.71 $\pm$ 0.02 & 16.39 $\pm$ 0.04 & $-$4.73 \\
SMP LMC 076       & 15.85 $\pm$ 0.04 & 16.42 $\pm$ 0.02 & 15.51 $\pm$ 0.04 & 16.85 $\pm$ 0.06 & $-$3.66 \\
SMP LMC 058       & 15.94 $\pm$ 0.04 & 16.24 $\pm$ 0.05 & 15.35 $\pm$ 0.03 & 16.55 $\pm$ 0.04 & $-$4.39 \\
SMP SMC 027       & \nodata          & \nodata          & \nodata          & 16.75 $\pm$ 0.09 & $-$2.65 \\
SMP LMC 051       & 17.95 $\pm$ 0.06 & 17.56 $\pm$ 0.04 & 16.26 $\pm$ 0.10 & 17.10 $\pm$ 0.15 & $-$4.80 \\
IRAS 05370        & 20.44 $\pm$ 0.37 & 21.23 $\pm$ 0.24 & 18.78 $\pm$ 0.17 & 18.08 $\pm$ 0.03 & $-$4.99 \\
IRAS 05537        & \nodata          & \nodata          & \nodata          & 17.67 $\pm$ 0.24 & $-$4.93 \\
SMP SMC 020       & \nodata          & \nodata          & \nodata          & 16.73 $\pm$ 0.08 & $-$3.71 \\
IRAS 05588        & 15.98 $\pm$ 0.05 & 16.04 $\pm$ 0.02 & 15.51 $\pm$ 0.13 & 14.92 $\pm$ 0.10 & $-$5.35 \\
\\
SMP SMC 006       & 16.98 $\pm$ 0.04 & 16.98 $\pm$ 0.13 & 15.95 $\pm$ 0.04 & 17.13 $\pm$ 0.20 & $-$4.82 \\
IRAS 05073        & \nodata          & \nodata          & \nodata          & \nodata          & $-$4.81 \\
IRAS 05063        & \nodata          & \nodata          & \nodata          & \nodata          & $-$4.78 \\
IRAS 05413        & \nodata          & \nodata          & \nodata          & \nodata          & $-$4.64 \\
IRAS 00350        & 17.04 $\pm$ 0.07 & 15.95 $\pm$ 0.05 & 14.59 $\pm$ 0.02 & 13.06 $\pm$ 0.07 & $-$7.07 \\
IRAS 05127        & 15.60 $\pm$ 0.07 & 16.59 $\pm$ 0.04 & 16.30 $\pm$ 0.21 & 16.05 $\pm$ 0.11 & $-$4.44 \\
\\
J010546           & 15.44 $\pm$ 0.03 & 15.82 $\pm$ 0.03 & 15.44 $\pm$ 0.06 & 15.02 $\pm$ 0.11 & $-$4.39 \\
IRAS F05192       & 20.13 $\pm$ 0.16 & 19.21 $\pm$ 0.06 & 17.67 $\pm$ 0.26 & 15.71 $\pm$ 0.11 & $-$4.82 \\
J052043           & 17.62 $\pm$ 0.07 & 16.52 $\pm$ 0.06 & 15.13 $\pm$ 0.08 & 13.87 $\pm$ 0.08 & $-$4.43 \\
IRAS 05110        & 20.53 $\pm$ 0.19 & 18.42 $\pm$ 0.05 & 16.63 $\pm$ 0.03 & 14.51 $\pm$ 0.02 & $-$4.91 \\
IRAS Z05259       & 17.56 $\pm$ 0.04 & 16.57 $\pm$ 0.03 & 15.37 $\pm$ 0.04 & 14.92 $\pm$ 1.16 & $-$4.34 \\
J004441           & 19.25 $\pm$ 0.15 & 17.52 $\pm$ 0.02 & 15.96 $\pm$ 0.18 & 14.45 $\pm$ 0.01 & $-$4.42 \\
NGC 1978 WBT 2665 & \nodata          & \nodata          & \nodata          & 16.31 $\pm$ 0.07 & $-$4.94 \\
IRAS 06111        & \nodata          & \nodata          & \nodata          & 18.04 $\pm$ 0.17 & $-$5.07 \\
IRAS 05092        & \nodata          & 20.18 $\pm$ 0.05 & 18.41 $\pm$ 0.05 & 16.39 $\pm$ 0.11 & $-$4.67 \\
IRAS 05185        & \nodata          & \nodata          & \nodata          & 18.04 $\pm$ 0.19 & $-$5.17 \\
IRAS 05360        & \nodata          & \nodata          & \nodata          & \nodata          & $-$5.07 \\
\\
SMP SMC 011       & \nodata          & 20.73 $\pm$ 0.09 & 19.03 $\pm$ 0.13 & 16.93 $\pm$ 0.23 & $-$5.79 \\
IRAS 05315        & \nodata          & \nodata          & \nodata          & \nodata          & $-$4.94 \\
IRAS 05495        & \nodata          & \nodata          & \nodata          & \nodata          & $-$5.30 \\
SMP SMC 025       & 17.84 $\pm$ 0.05 & 18.33 $\pm$ 0.05 & 17.70 $\pm$ 0.10 & 18.65 $\pm$ 0.06 & $-$2.41 \\
SMP LMC 099       & \nodata          & \nodata          & \nodata          & 17.02 $\pm$ 0.12 & $-$3.30 
\enddata
\tablenotetext{a}{Modified by adding a 30,000 K Planck function.}
\end{deluxetable*}

To assemble optical and infrared photometry for our sample,
we began with the observed IRS coordinates, as determined by 
the optimal extraction, then found the counterpart point 
source as observed by IRAC in either the SAGE or SAGE-SMC 
catalogs \citep{mei06,gor11}.\footnote{IRAC = Infrared Array 
Camera on {\it Spitzer}; SAGE = Surveying the Agents of 
Galactic Evolution.}  The SAGE-SMC catalog includes updated 
photometry from the earlier S$^3$MC survey of the core of the 
SMC \citep{bol07}.  IRAC observed all sources except 
IRAS~06111 and SMP LMC~099, which lie outside the SAGE 
footprint.  The SMC sources LIN~49 and IRAS~00350 are 
$\sim$2\arcsec\ and 4$\farcs$5, respectively, away from their 
position as given by SIMBAD.

Using the new coordinates, we searched the 2MASS catalog
\citep{skr06}, and where we found a source, again updated
the coordinates.  We then searched several additional
catalogs.  The 2MASS-6X catalog covers the Magellanic Clouds
with deeper sensitivity and in multiple epochs \citep{cut06}.
The Infrared Survey Facility (IRSF) also surveyed the 
Magellanic Clouds at $JHK_s$ and provides additional temporal 
coverage \citep{kat07}.  The Third Release of the 
DENIS catalog extended our temporal coverage at $J$ and 
$K_s$.\footnote{DENIS = Deep Near-Infrared Survey.}
\cite{cio00} introduced the original DENIS point-source 
catalog of the Magellanic Clouds.   We also used the WISE
survey for additional coverage at 3.4 and 4.6~\mum\ 
\citep{wri10}.\footnote{WISE = Wide-field Infrared Survey
experiment.}

We used the IRSA website to access all of the catalogs except 
the IRSF catalog, which we obtained through 
VizieR.\footnote{IRSA = Infrared Science Archive, maintained 
by NASA and the Infrared Processing and Analysis Center 
(IPAC) at Caltech, and available at 
http://irsa.ipac.caltech.edu.} 
We assumed equivalence in closely overlapping photometric 
filters.  Thus we did not distinguish for differences in 
near-IR filters, and we combined WISE data at 3.4 and 
4.6~\mum\ with IRAC data at 3.6 and 4.5~\mum, respectively.  

Table~\ref{t.irphot} presents the combined infrared 
photometry for our sample.  The coordinates reported in 
Table~\ref{t.sample} are an average of the sources reported 
in the 2MASS, 2MASS 6X, and IRSF surveys, with equal weight 
assigned for each epoch in which the source was observed.  
None of those surveys detected IRAS~05495, and for it we
report the mean IRAC position.  We estimate, based on
discrepancies between positions for a given source among the
catalogs, that the positions we report are good to 
$\sim$0$\farcs$2 in most cases.


Table~\ref{t.optphot} presents the combined optical 
photometry.  We obtained $UBVI$ photometry from the 
Magellanic Clouds Photometric Survey of the SMC \citep{zar02} 
and LMC \citep{zar04}, supplemented with $I$-band data from 
DENIS and data from the OGLE-III shallow survey of the LMC 
\citep{ula12}.\footnote{OGLE = Optical Gravitational Lens 
Experiment.}  Again, we did not distinguish between 
differences in filters and simply combined all available
data for each source in each filter.

Table~\ref{t.optphot} also contains a column of absolute
bolometric magnitudes, which are found by integrating 
through the IRS data and the available photometry to the 
blue of 5~\mum.  We assumed a Wien distribution to the blue 
of the available data and a Rayleigh-Jeans distribution to 
the red of the IRS data.  The absolute bolometric magnitude 
was determined assuming a distance modulus of 18.50 to the 
LMC and 18.90 to the SMC, and $M_{bol} = -4.75$ for the Sun.  
$M_{bol}$ = $-$5.2 corresponds to $\approx$10$^4$ 
$L_{\odot}$.

For the Fullerene group, we modified the procedure for
determining bolometric magnitude.  The blue $U-B$ colors
point to the presence of emission from the central star
in the spectrum, and to account for this additional 
emission, we appended a 30,000 K blackbody to the spectrum,
scaling it to the $U$ magnitude.  The bolometric magnitudes
in Table~\ref{t.optphot} reflect this correction, which 
raised the brightness of the sources 1.5 mag on average.  
Before the correction, the bolometric magnitudes
ranged from $-$2.1 to $-$3.4, compared to $-$3.6 to $-$4.8
after.



\end{document}